\documentclass[USenglish]{article}	

\usepackage[utf8]{inputenc}				
\usepackage[big,online]{dgruyter}	
\usepackage{lmodern} 
\usepackage{microtype}
\usepackage[numbers,square,sort&compress]{natbib}

\theoremstyle{dgthm}

\theoremstyle{dgdef}

\newcommand{\Real}{\mathbb{R}}
\begin{document}

\articletype{Research Article}
\received{Month	DD, YYYY}
\revised{Month	DD, YYYY}
\accepted{Month	DD, YYYY}
\journalname{International Journal of Biostatistics}
\journalyear{YYYY}
\journalvolume{XX}
\journalissue{X}
\startpage{1}
\aop
\DOI{10.1515/sample-YYYY-XXXX}

\title{Regression Trees and Ensembles for Cumulative Incidence Functions}
\runningtitle{Regression Trees and Ensembles for Cumulative Incidence Functions}

\author[1]{Youngjoo Cho}
\author[2]{Annette M. Molinaro}
\author[3]{Chen Hu} 
\author*[4]{Robert L. Strawderman}
\runningauthor{Cho et al.}
\affil[1]{\protect\raggedright 
	The University of Texas at El Paso, Department of Mathematical Sciences, El Paso, TX, United States, e-mail: ycho@utep.edu}
\affil[2]{\protect\raggedright 
	University of California San Francisco, Department of Neurological Surgery, San Francisco, CA, United States, e-mail: annette.molinaro@ucsf.edu}
\affil[3]{\protect\raggedright 
	Johns Hopkins University School of Medicine, Department of Oncology, Baltimore, MA, United States, e-mail: chu22@jhmi.edu}
\affil[4]{\protect\raggedright 
	University of Rochester, Department of Biostatistics \& Computational Biology, Rochester, NY, United States, email: robert\_strawderman@urmc.rochester.edu}

\abstract{The use of cumulative incidence functions for characterizing the risk of one type of event in the presence of others has become increasingly popular over the past decade. The problems of modeling, estimation and inference have been treated using parametric, nonparametric and semi-parametric methods. Efforts to develop suitable extensions of machine learning methods, such as regression trees and related ensemble methods, have begun comparatively recently. In this paper, we propose a novel approach to estimating 
	cumulative incidence curves in a competing risks setting using regression trees and associated ensemble estimators. The proposed methods employ augmented estimators of the Brier score risk as the primary basis for building and pruning trees, and lead to 
	methods that are easily implemented using existing {\tt R} packages. Data from the Radiation Therapy Oncology Group 
	(trial 9410) is used to illustrate these new methods.}

	
	\maketitle

	\section{Introduction} 
A subject being followed over time may experience several types of events, possibly even fatal. For example, in a Phase III trial of concomitant versus sequential chemotherapy and thoracic radiotherapy for patients with inoperable non-small cell lung cancer (NSCLC) conducted by the Radiation Therapy Oncology Group (RTOG), patients were followed up to 5 years, where both the occurrence of  disease progression and death are of particular interest. Such ``competing risks'' data are often encountered in cancer and other biomedical follow-up studies, in addition to the potential complication of right-censoring on the event time(s) of interest. 

Two quantities are often used when analyzing competing risks data: the cause-specific hazard function (CSH) and the cumulative incidence function (CIF). For a given event, the former describes the instantaneous risk of this event at time $t$, given that no events have yet occurred; the latter describes the probability of occurrence, or absolute risk, of that event across time and can be derived 
directly from the subdistribution hazard function \citep{fine1999proportional}.
 \citet{dignam} provides a review of methods for handling competing risks data as of 2012, where parametric and semi-parametric approaches to modeling both the CSH and CIF using hazard-type regression modeling are considered. The literature on tree-based methods, including ensemble approaches like random forests
\citep[RF;][]{Breiman2001}, for estimating the CIF remains comparatively under-developed. 
%
%
In particular, to our knowledge, there is no software package currently available that specifically focuses on estimating
the CIF using regression tree methods, and ensemble-based methods for estimating the CIF 
are currently limited to the work of \cite{ishwaran2014random} and \cite{mogensen2013random}. 
\cite{ishwaran2014random} implement their methods in the {\sf randomForestSRC} package \citep{rfsrc}, where the 
unpruned regression trees that make up the bootstrap ensemble are built using logrank-type splitting rules
appropriate for competing risks \citep[e.g.,][]{gray1988class}.

In its most general form, the original CART algorithm, and by extension RF, relies on the specification of a loss function
that (i) informs all decision-making processes (e.g., what covariate to split on and when/where; when to stop tree growth)
and (ii) induces a particular estimator that minimizes the empirical loss. 
Motivated by the recent work of \cite{steingrimsson2016doubly, steingrimsson2019censoring} for right-censored survival
data, this paper proposes a direct extension of CART and RF for estimating the CIF in the presence
of right-censored competing risks. Specifically, starting with an appropriate version of the Brier loss function \citep[cf.,][]{brier1950verification}, we first develop a simple nonparametric estimate of the CIF for a single event by minimizing this loss function
when there is no loss to follow-up (i.e., with full data) and one has specified a fixed partition structure for the covariate space. 
Estimation in this case may be viewed as a form of binomial regression, where the mean function (i.e., CIF) is piecewise
constant on the covariate space. For the case where there is loss to follow-up, we then construct several observed data loss 
functions that target the same expected loss as the (unobserved) full data Brier loss function. The simplest of these
approaches employs inverse probability of censoring weighted estimation (IPCW).
Finally, we explain how the development of these new loss functions 
leads to new splitting and decision rules that can be used by CART and RF algorithms for estimating the CIF, and importantly,
show how these new methods can be easily implemented using existing software in combination with a certain form of imputation. 
The resulting methods may be viewed as nonparametric alternatives to the 
semiparametric binomial regression approach proposed in \cite{scheike2008predicting}
for estimating a CIF, differing in the approach to estimation (i.e., through minimizing the Brier loss instead of employing 
estimating equations).
Simulation studies are used to investigate performance of these new methods. In addition, we use these new methods to conduct 
some secondary analyses for the RTOG 9410 Phase III lung cancer trial mentioned at the beginning of this section. 
The paper concludes with comments on future work.

\section{Estimating a CIF by Minimizing Squared Error Loss}
\label{S: Estimation}
\subsection{Relevant Data Structures}
\label{S:data}
Let $T^{(m)}$ be the time to event for the event type $m = 1,\ldots, K$ where $K \geq 2$ is fixed. Let $W$ be a vector of $p$ covariates, where $W \in {\cal S} \subset \Real^p$. Let $T = \min(T^{(1)},\ldots, T^{(K)})$ be the minimum of all latent event times; it is assumed 
that $T$ is observed and has a continuous distribution function. Then, in the absence of other loss to follow-up, $F= (T,W,M)$ is assumed to be the fully observed (or full) data for a subject, where $M$ is the observed event type that corresponds to $T.$ The definition of $T$ therefore implies that  $(T^{(M)}, M, W)$ is 
observed and, in addition, that $T^{(m)} > T^{(M)}$ but is otherwise not observed for $m \neq M.$ 
Define $\mathcal{F} = (F_1,\ldots,F_n)$ to be the full data observed on $n$ independent subjects,  where $F_i=(T_i,W_i,M_i), i=1,\ldots,n$  are assumed to be identically distributed (i.i.d.). 

In the case where there is also potential random loss to follow-up, we suppose that $C$ is a continuous random 
variable that, given $W$, is statistically independent of  $(T,M).$  Then, for a given subject, we instead observe 
$O = \{\tilde{T},\Delta, M \Delta, W  \},$ where $\tilde{T} = \min(T,C)$ and $\Delta = I(T \leq C)$ is the (any) event indicator. 
The observed data on $n$ i.i.d.\ subjects is  $\mathcal{O} = (O_1,\ldots,O_n).$ Similarly to the case where $K=1,$ random 
censoring on $T$ permits estimation of the CIF from  the data ${\cal O}$.  We remark here that the notational set-up
intentionally excludes $C$ from the set of possible event times $(T^{(1)},\ldots, T^{(K)})$; the reason for setting the 
problem up in this way will become clear in Section \ref{brier-LTF}.

\subsection{CIF estimation via the Brier Loss: no loss to follow-up}
\label{CIF-full}
Let $\psi_{0m}(t;w) = P(T \leq t, M = m | W=w)$ and define $\Psi_0 = \{ \psi_{0m}(t;w), t > 0; w \in {\cal S}, m=1,\ldots,K\}.$
The set of CIFs $\Psi_0$ can be estimated from the data ${\cal F}$ using any suitable parametric or semiparametric methods 
without further assumptions on the data (e.g., independence of $T^{(1)},\ldots, T^{(K)}$).  This section describes a simple method for 
estimating $\psi_{0m}(t;w)$ for a fixed cause $m$ and time point $t >0$ using the Brier (i.e., squared error) loss function.
As preparation for Section \ref{data}, $\psi_{0m}(t;w)$ is assumed to be piecewise constant as a function of $W;$
however, the basic estimation ideas extend to more complex modeling assumptions
in a straightforward manner \citep[e.g.,][]{scheike2008predicting}.

Let $\mathcal{N}_1,\ldots,\mathcal{N}_L$ form a known partition of ${\cal S}.$ In this section and also in 
Section \ref{brier-LTF}, we assume this partition is given and, consistent with the assumption
that $\psi_{0m}(t;w)$ is a piecewise constant function of $w,$
that $\psi_{0m}(t;w) = \sum_{l=1}^{L}\beta_{0lm}(t)I\{W \in \mathcal{N}_l\},$
where the conditional CIF $\beta_{0lm}(t) = P(T \leq t, M=m | W \in \mathcal{N}_l)$ 
is the same function of $t$ for each $W \in \mathcal{N}_l.$ 
Define $Z_m(t) = I(T \leq t, M=m)$
and let  
\begin{gather}
	\label{psi mod}			
	\psi_m(t;w) = \sum_{l=1}^{L}\beta_{lm}(t)I\{w \in \mathcal{N}_l\}
\end{gather}  
be a model for $\psi_{0m}(t;w),$ $w \in {\cal S}.$ Then, 
fixing both $t > 0$ and $m,$ 
the so-called Brier loss is given by
$
L_{m,t}^{full}(F,\psi_m) = \{Z_m(t) - \psi_m(t;w)\}^2
= \sum_{l=1}^{L} I\{W \in \mathcal{N}_l\} \{Z_{m}(t) - \beta_{lm}(t)\}^2.
$
Assuming that ${\cal F}$ is observed, the corresponding empirical Brier loss is given by
\begin{gather}
	\label{eq:0}
	L_{m,t}^{emp}({\cal F},\psi_m) = \frac{1}{n} \sum_{i=1}^n L_{m,t}^{full}(F_i,\psi_m)
	= 
	\frac{1}{n} \sum_{i=1}^n \sum_{l=1}^{L} I\{W_i \in \mathcal{N}_l\} \{Z_{im}(t) - \beta_{lm}(t)\}^2.
\end{gather}
With $t$ and $m$ fixed and under the assumptions of Section \ref{S:data}, $L_{m,t}^{full}(F,\psi_m)$ is 
an unbiased estimator of the risk $\Re(t,\psi_m) = E[\sum_{l=1}^{L} I\{W \in \mathcal{N}_l\}\{Z_{m}(t) - \beta_{lm}(t)\}^2],$
or equivalently, $\Re(t,\psi_m) = \sum_{l=1}^{L} P\{W \in \mathcal{N}_l\}\{ \beta_{0lm}(t) - \beta_{lm}(t)\}^2;$
hence, so is \eqref{eq:0}. 
Considered as a function of $\beta_{lm}(t), l =1,\ldots,L,$
the risk $\Re(t,\psi_m)$ is minimized when
$\beta_{lm}(t) = \beta_{0lm}(t)$ for each $l$; 
the loss \eqref{eq:0} is minimized when
$\psi_m(t;w) = \hat{\psi}_m(t;w) = \sum_{l=1}^{L} I\{W_i \in \mathcal{N}_l\} \hat{\beta}_{lm}(t),$ where
\begin{gather}
	\label{beta-full}
	\hat{\beta}_{lm}(t) = \frac{ \sum_{i=1}^n  I\{W_i \in \mathcal{N}_l\} Z_{im}(t)}{\sum_{i=1}^n  I\{W_i \in \mathcal{N}_l\} }
\end{gather}
is a nonparametric estimate for $\beta_{0lm}(t)$. By contrast, \cite{scheike2008predicting} use
a semiparametric binomial regression model to estimate $\psi_{0m}(t,w)$ from 
$(Z_{im}(t), W_i),~i=1,\ldots n.$

\subsection{CIF estimation via the Brier Loss: random loss to follow-up}
\label{brier-LTF}

In follow-up studies with competing risks outcomes, the full data 
${\cal F}$ might not be observed due to loss to follow-up. In this case, estimating $\psi_{0m}(t;w)$ for a specified 
$m$ under the loss function \eqref{eq:0}  is not possible. One way to overcome this challenge  is to use a modified 
loss function that  (i) depends only on the observed data ${\cal O}$ and (ii) has the same risk as the (unobserved) 
full data loss 
\citep[c.f.,][]{molinaro2004tree,lostritto2012partitioning,steingrimsson2016doubly}.
Following \cite{steingrimsson2016doubly, steingrimsson2019censoring}, we propose an appropriate class of 
inverse probability of censoring weighted (IPCW),  and subsequently augmented IPCW (AIPCW),  
loss functions that share the  same risk $\Re(t,\psi_m)$ as the (unobservable) empirical loss \eqref{eq:0}.
This allows us to derive a new observed data estimator of the CIF with both $t$ and $m$ fixed. We then extend this class of 
losses to the setting of a composite loss function, where the goal is to simultaneously estimate $\psi_{0m}(t_j;w)$ at time 
$t_j,$ $j = 1,\ldots,J.$ As in the previous section, we assume that
$\psi_{0m}(t;w) = \sum_{l=1}^{L}\beta_{0lm}(t)I\{W \in \mathcal{N}_l\},$
where the partition $\{\mathcal{N}_1,\ldots,\mathcal{N}_L\}$ 
of ${\cal S}$ is known.

\subsubsection{CIF estimation via the IPCW and AIPCW Brier Losses }
\label{CIF-obs}

Fix $t > 0,$ define $G_0(s|W) = P(C \geq s|W)$ for any $s \geq 0$ and suppose that $G_0(T_i|W_i) \geq \epsilon$  
almost surely for some $\epsilon > 0$ ($i=1,\ldots,n$).   
Define $\tilde{Z}_{im}(t) = I(\tilde{T}_i \leq t, M_i=m),$ $i=1,\ldots,n;$ easy calculations then show 
\[
E\left[ \frac{\Delta_i }{G_0(\tilde T_i |W_i)} (\tilde{Z}_{im}(t) - \psi_m(t;W_i) ) ^2 \right] = E\left[ (Z_{im}(t) - \psi_m(t;W_i))^2 \right] = \Re(t,\psi_m)
\]
for a fixed $\psi_m(t;w).$ This risk equivalence motivates the construction of an IPCW-type loss function.
In particular, define for any suitable survivor function $G(\cdot|\cdot)$ 
\begin{gather}
	\label{eq:2}
	L_{m,t}^{ipcw}({\cal O},\psi_m;  G)  =	
	\frac{1}{n}\sum_{i=1}^{n}\sum_{l=1}^L I\{W_i \in \mathcal{N}_l\}\bigg{[}\frac{\Delta_i \{\tilde{Z}_{im}(t) - \beta_{lm}(t)\}^2}{G(\tilde{T}_i |W_i)}\bigg{]};
\end{gather}
then, it is easy to see that
\eqref{eq:2} is minimized by
\begin{gather}
	\label{betaIPCW}
	\hat{\beta}^{ipcw}_{lm}(t;  G) = \frac{  \sum_{i=1}^n I\{W_i \in \mathcal{N}_l\} \frac{\Delta_i \tilde{Z}_{im}(t)}{G(\tilde{T}_i |W_i)}}
	{\sum_{i=1}^n I\{W_i \in \mathcal{N}_l\} \frac{\Delta_i }{G(\tilde{T}_i  |W_i)}}, l=1,\ldots,L,
\end{gather}
implying that
$\hat{\psi}_m(t;w) = \sum_{l=1}^{L} I\{W_i \in \mathcal{N}_l\} \hat{\beta}^{ipcw}_{lm}(t;  G)$
is the corresponding estimator for the CIF at time $t$ for cause $m$. Moreover, 
$L_{m,t}^{ipcw}({\cal O},\psi_m;  G_0)$ is an unbiased estimate of $\Re(t,\psi_m)$
Observe that  \eqref{eq:2} and \eqref{betaIPCW}
respectively reduce to \eqref{eq:0} and \eqref{beta-full} if censoring is absent.

When $K=1,$ the loss \eqref{eq:2} is just a special case of that considered in 
\citet{molinaro2004tree}; see also \citet{lostritto2012partitioning}.
In practice, an estimator $\hat G(\cdot|\cdot)$ for $G_0(\cdot|\cdot)$ is used in \eqref{eq:2}; popular approaches 
here include product-limit estimators
derived from the Kaplan-Meier and Cox regression estimation procedures. Of course, other methods could be used, 
such as regression trees or ensembles for right-censored survival data 
\citep[e.g.,][]{ishwaran2008, steingrimsson2016doubly, steingrimsson2019censoring}.

%
%

\label{CIF-obs3}
As in \cite{steingrimsson2016doubly}, one can employ semiparametric estimation theory for missing data to 
construct an improved estimator of the full data risk $\Re(t,\psi_m)$ by augmenting the IPCW loss function 
\eqref{eq:2} with additional information on censored subjects. In particular, consider the loss function 
$L_{m,t}^{ipcw}({\cal O},\psi_m;  G).$  
Recall that $\Psi_0$ defines the set of
CIFs of interest and let $\Psi$ denote a corresponding model that may or
may not contain  $\Psi_0$.
Define $V_{lm}(u;t,w,\Psi) = 
E_{\Psi}[(Z_{m}(t) - \beta_{lm}(t))^2|T \geq u,W= w]$
for any $t, u \geq 0$  and $w \in {\cal S};$ 
it is shown later how this expression specifically depends on $\Psi.$
Then, fixing $\beta_{1m}(t),\ldots,\beta_{Lm}(t),$ the augmented estimator of 
$\Re(t,\psi_m)$ having the smallest variance
that can be constructed from the unbiased estimator $L_{m,t}^{ipcw}({\cal O},\psi_m;  G_0)$ is given by 
$L_{m,t}^{dr}({\cal O},\psi_m; G_0,\Psi_0) 
= L_{m,t}^{ipcw}({\cal O},\psi_m; G_0) 
+ L_{m,t}^{aug}({\cal O},\psi_m; G_0, \Psi_0)$ 
where
\begin{gather}
	\label{eq:4a1}
	L_{m,t}^{aug}(\mathcal{O},\psi_m; G, \Psi) =
	\frac{1}{n}\sum_{l=1}^{L}\sum_{i=1}^{n}I\{W_i \in \mathcal{N}_l\}
	\int_0^{\tilde T_i} \frac{V_{lm}(u;t,W_i,\Psi)}{G(u|W_i)}dM_G(u|W_i)
\end{gather}
is defined for suitable choices of $\Psi$ and $G(\cdot|\cdot)$ 
and $M_G(t|w) = I(\tilde{T} \leq t,\Delta = 0) - \int_0^t I(\tilde{T} \geq u)d\Lambda_G(u|w),$
where $\Lambda_G(\cdot|\cdot)$ denotes the cumulative hazard function corresponding 
to the model $G(\cdot|\cdot)$ \citep[cf.\ ][Sec.\ 9.3, 10.4]{tsiatis2007semiparametric}.
The ``doubly robust'' loss 
$L_{m,t}^{dr}({\cal O},\psi_m; G,\Psi)$ reduces to a special case of the class of 
loss functions proposed in \citet{steingrimsson2016doubly} when $K=1.$


The loss function  $L_{m,t}^{dr}({\cal O},\psi_m; G,\Psi)$ can be simplified further: because $Z_{m}(t)$ is binary, 
\begin{equation}
	\label{simple-V}
	V_{lm}(u;t,w,\Psi) = y_{m}(u;t,w,\Psi) - 2 y_{m}(u;t,w,\Psi) \beta_{lm}(t) + \beta^2_{lm}(t)
\end{equation}
for any suitable $\Psi$ (e.g., $\Psi_0$), where $y_{m}(u;t,w,\Psi) = E_{\Psi}\{Z_{m}(t)|T \geq u, W=w\}$
reduces to
\begin{gather}
	\label{eq:2b}
	y_{m}(u;t,w,\Psi) =
	\begin{cases}
		\dfrac{P_{\Psi}(u \leq T \leq t, M=m|W=w)}{P_{\Psi}(T \geq u | W=w)} & \text{if} \quad u \leq t \\
		0 & \text{otherwise}
	\end{cases}.
\end{gather}
The notation $E_{\Psi}$ and $P_{\Psi}$ means that these quantities are calculated under the CIF
model specification $\Psi$. Hence, under a model $\Psi$, the calculation of $L_{m,t}^{dr}({\cal O},\psi_m; G,\Psi)$
requires estimating both the CIF for cause $m$ and the  all-cause probability $P_{\Psi}(T \geq u| W= w).$

Considering $L_{m,t}^{dr}({\cal O},\psi_m; G,\Psi)$  as a function of the $L$ scalar parameters 
$\beta_{1m}(t),\ldots,\beta_{Lm}(t)$ only and differentiating with respect to each one, it can be shown that
\begin{gather}
	\label{eq:5}
	\tilde{\beta}_{lm}^{dr}(t; G,\Psi) = \dfrac{\displaystyle \sum_{i=1}^n I\{W_i \in \mathcal{N}_l\}[\widetilde{TS}_{1,im}^1(t) + \widetilde{TS}_{2,im}^1(t) ] }{\displaystyle \sum_{i=1}^n I\{W_i \in \mathcal{N}_l\} [\widetilde{TS}_{1,im}^0 + \widetilde{TS}_{2,im}^0 ]}, ~l=1,\ldots,L
\end{gather}
minimize  $L_{m,t}^{dr}({\cal O},\psi_m; G,\Psi),$ 
where 
\begin{gather}
	\nonumber
	\widetilde{TS}_{1,im}^0  = \frac{\Delta_i}{G(\tilde{T}_i | W_i)} \quad \quad
	\widetilde{TS}_{2,im}^0  = \int_0^{\tilde{T}_i} \frac{dM_G(u|W_i)}{G(u|W_i)} \\
	\label{TSfuns}
	\widetilde{TS}_{1,im}^1(t)  = \frac{\tilde{Z}_{im}(t)\Delta_i}{G(\tilde{T}_i| W_i)} \quad \quad 
	\widetilde{TS}_{2,im}^1(t)  = \int_0^{\tilde{T}_i} \frac{y_m(u;t,W_i,\Psi)}{G(u|W_i)}dM_G(u|W_i).
\end{gather}
The validity of this result relies on the assumption that $G(\tilde{T}_i|W_i) \geq \epsilon > 0$ for
some $\epsilon$ and each $i=1,\ldots,n.$ Under this same assumption, Lemma 1 of \cite{strawderman2000estimating}
implies
\begin{gather*}
	\widetilde{TS}_{1,im}^0  + \widetilde{TS}_{2,im}^0 =
	\dfrac{\Delta_i}{G(\tilde{T}_i | W_i)} + \dfrac{1-\Delta_i}{G(\tilde{T}_i|W_i)} - \displaystyle \int_{0}^{\tilde{T}_i}\dfrac{d \Lambda_G(u|W_i)}{G(u|W_i)} = 1;
\end{gather*}
letting $N_l = \sum_{i=1}^n I\{W_i \in \mathcal{N}_l\}, ~l=1,\ldots,L,$
it follows that \eqref{eq:5} can be rewritten as
\begin{gather}
	\label{eq:5a}
	\hat{\beta}_{lm}^{dr}(t;G,\Psi) =  \frac{1}{N_l}  \sum_{i=1}^n I\{W_i \in \mathcal{N}_l\}[\widetilde{TS}_{1,im}^1(t)  + \widetilde{TS}_{2,im}^1(t) ] , ~l=1,\ldots,L.
\end{gather}
Similarly to Section \ref{CIF-obs}, $\hat{\psi}_m(t;w) = \sum_{l=1}^{L} I\{W_i \in \mathcal{N}_l\} \hat{\beta}_{lm}^{dr}(t;G,\Psi)$
now generates the corresponding CIF estimate at time $t$ for cause $m$ and, in addition,
$L_{m,t}^{dr}({\cal O},\psi_m; G,\Psi)$  and \eqref{eq:5a} respectively reduce to \eqref{eq:0} and \eqref{beta-full} when censoring is absent.

The specification $G(t|w) = \tilde G(t|w) = 1$ for all $t \geq 0$ and 
$w \in {\cal S}$ generates an interesting special case of $L_{m,t}^{dr}({\cal O},\psi_m; G,\Psi)$
despite $\tilde G(\cdot|\cdot)$ being incorrectly modeled in the presence of censoring.
In particular, for suitable $\Psi$,  
(i) $L_{m,t}^{dr}({\cal O},\psi_m; \tilde G,\Psi) = \sum_{l=1}^L L_{ml,t}^{bj}({\cal O},\psi_m; \Psi)$ where 
\[
L_{ml,t}^{bj}({\cal O},\psi_m; \Psi)  =  
\frac{1}{n} \sum_{i=1}^n 
I\{W_i \in \mathcal{N}_l\} [\Delta_i\{\tilde Z_{im}(t) - \beta_{lm}(t)\}^2 + (1-\Delta_i)V_{lm}(\tilde{T}_i;t,W_i,
\Psi)];
\]
and, (ii) for $\Psi = \Psi_0,$ $L_{m,t}^{dr}({\cal O},\psi_m; \tilde G,\Psi_0)$ is an
unbiased estimator of the risk  $\Re(t,\psi_m)$.
Noting that \eqref{simple-V} implies $V_{lm}(\tilde{T}_i;t,W_i,
\Psi)$ can be rewritten in terms of $y_m(\tilde{T}_i; t,w, \Psi)$
for every $i,$ the minimizer of $L_{ml,t}^{bj}({\cal O},\psi_m; \Psi)$ is given by
\begin{gather*}
	\tilde{\beta}_{lm}^{bj}(t;\Psi) = 
	\frac{1}{N_l} 
	\sum_{i=1}^n I\{W_i \in \mathcal{N}_l\} [\Delta_i \tilde{Z}_{im}(t) + (1-\Delta_i)y_m(\tilde{T}_i; t,W_i,\Psi)].
\end{gather*}
That is, under the loss $L_{ml,t}^{bj}({\cal O},\psi_m; \Psi),$
the estimator for $\beta_{lm}(t)$ is the Buckley-James (BJ) estimator 
of the mean response within the partition ${\cal N}_l$
\citep{buckley1979linear},
an estimator that can also be derived directly from \eqref{eq:5a} by setting $G = \tilde G.$
For this reason, we refer to $L_{m,t}^{dr}({\cal O},\psi_m; \tilde G,\Psi)$ as the Buckley-James loss function.
For a fixed value of $m$ and $l$, the function $\beta_{lm}(t)$ (i.e., the cumulative incidence for type $m$ within node ${\cal N}_l$) is
monotone increasing in $t.$ In contrast to the doubly robust loss, the Buckley-James loss function therefore preserves monotonicity; 
this property is useful when considering multiple time points, as considered in the next section.

\subsubsection{Composite AIPCW loss functions: the case of multiple time points}
\label{composite loss}

Under the piecewise constant model \eqref{psi mod}, the quantity being estimated within 
each partition depends on  $t;$
however, the set of partitions remains the same across time. As a result, for a given $m$, we can
further reduce variability when estimating $\psi_{0m}(t|w)$ by considering losses constructed from 
$L_{m,t}^{dr}({\cal O},\psi_m; G,\Psi)$ that incorporate information over several time points. 

Recall that
$L_{m,t}^{dr}({\cal O},\psi_m; G,\Psi) = L_{m,t}^{ipcw}({\cal O},\psi_m; G) 
+ L_{m,t}^{aug}({\cal O},\psi_m; G, \Psi)$
where $L_{m,t}^{ipcw}({\cal O},\psi_m; G)$ is given by \eqref{eq:2} 
and $L_{m,t}^{aug}({\cal O},\psi_m; G, \Psi)$ is given by \eqref{eq:4a1}.
For a given set of time points $0 < t_1 < t_2 < \cdots < t_J < \infty,$ a simple composite loss function 
for a given event type $m$ can be formed by calculating 
\begin{gather}
	\label{eq:7a}
	L_{m,\mathbf{t}}^{mult,dr}(\mathcal{O},\psi_m;G,\Psi) = \sum_{j=1}^J \alpha_j L_{m,t_j}^{dr}({\cal O},\psi_m; G,\Psi),
\end{gather}
where $\alpha_j > 0, j=1,\ldots,J$ are pre-specified weights such that $\sum_{j=1}^J \alpha_j = 1.$
Minimizing \eqref{eq:7a} with respect to $\beta_{lm}(t_j)$
gives
\begin{gather}
	\label{eq:7b}
	\tilde{\beta}_{lm}^{mult,dr}(t_j; G,\Psi) = \frac{1}{N_l} 
	\sum_{i=1}^{n} I\{W_i \in \mathcal{N}_l\}[\widetilde{TS}_{1,im}^1(t_j) + \widetilde{TS}_{2,im}^1(t_j)],
\end{gather}
for $j=1,\ldots,J; l = 1,\ldots, L.$
In the absence of censoring, the indicated composite loss function and partition-specific estimators
reduce to that which would be computed by extending the loss function introduced in Section
\ref{CIF-full} in the manner described above.

Thus far, we have assumed the existence of a fixed partition $\{ {\cal N}_1,\ldots,N_L \}$
of ${\cal S}.$  In this situation, the use of a composite loss like \eqref{eq:7b} yields no extra efficiency gain for estimating the CIF 
for cause $m$ within each partition. This can be seen from \eqref{eq:7b}, which is exactly equal to (\ref{eq:5a}) computed 
for $t=t_j;$ that is, the partition-specific estimators for $t_1,\ldots,t_J$ do not depend on $\alpha_1,\ldots,\alpha_J$. Importantly,
this occurs due to the absence of parametric or semiparametric modeling assumptions 
that restrict the relationship between $\beta_{lm}(t)$ (i.e., the CIF when $W \in {\cal N}_l$) 
and $\beta_{l'm}(t)$ (i.e., the CIF when $W \in {\cal N}_{l'}$) when $l \neq l'.$ 

However, in the case of regression trees, and by extension ensembles of trees (e.g., RF), the 
partition $\{ {\cal N}_1,\ldots,N_L \}$ for every tree is estimated adaptively
from the data, and the use of a weighted composite loss \eqref{eq:7b} 
influences both the selection of $L$ and the chosen partition boundaries. Consequently, performance gains 
may still be expected when estimating $\psi_{m0}(t_j;w), j=1,\ldots,J$ using a composite
loss function whether one uses trees or ensembles of trees. We consider such 
methods further in the next section.


\section{CIF Regression Trees and Ensembles}
\label{data}

The developments in Section \ref{S: Estimation} provide an important building block 
for developing new splitting and evaluation procedures when using
CART to build regression trees for estimating the CIF, with or without loss to follow-up.
Because RF relies on bootstrapped ensembles of CART trees, the loss-based
estimation procedures have similarly important implications for RF.
In the coming sections, we propose several variants on CART and RF for 
competing risks data that use the loss functions introduced in previous section. 

\subsection{Estimating a CIF via CART or RF: no loss to follow-up}
\label{implement0}

Given a specified loss function, CART \citep{breiman1984classification} fits a regression tree as follows:
\begin{enumerate}
	\item Using recursive binary partitioning, grow a maximal tree by selecting a (covariate, cutpoint) combination at every
	stage that minimizes the chosen loss function;
	\item Using cross-validation, select the best tree from the sequence of candidate trees generated by Step 1
	via cost complexity pruning (i.e., using penalized loss).
\end{enumerate}
In its most commonly used form for regression problems with a continuous outcome, 
CART estimates the conditional mean response as a 
piecewise constant function on ${\cal S},$ making all decisions on the basis of minimizing squared error loss. 
The resulting tree-structured regression function estimates the predicted response within each terminal node 
(i.e., partition of ${\cal S}$) using the sample mean of the observations falling into that node.  The set of terminal nodes 
(i.e., the partition structure) is determined adaptively from the data as a result of steps 1 and 2 above. 

The random forests algorithm \citep[RF;][]{Breiman2001} is a simple extension of CART:
\begin{enumerate}
	\item Bootstrap the data; that is, draw $B$ random samples with replacement from $\cal F$.
	\item {For each bootstrapped dataset, run Step 1 of the CART algorithm above, possibly 
		randomly selecting a set of $p^* \leq p$ candidate covariates when determining 
		a (covariate, cutpoint) combination at each possible splitting stage.}
	\item {Compute the terminal node estimators for each subject for each of the $B$ trees and 
		average these to obtain an ensemble predictor.} 
\end{enumerate}

The critical step that underpins both CART and RF is Step 1 of the CART algorithm, where
connections to the developments of Section \ref{S: Estimation} should now be evident. 
In particular,  in the absence of censoring and under the piecewise constant model \eqref{psi mod}
for $\psi_{0m}(t;w),$ 
Section \ref{CIF-full} shows that a nonparametric 
estimate for $\psi_{0m}(t;w)$ at $t$
can be obtained by minimizing the loss \eqref{eq:0}.
This basic estimation problem is equivalent to estimating the conditional mean 
response using the modified dataset ${\cal F}_{red,t} = \{ (Z_{im}(t), W'_i)',i=1,\ldots n \}$ 
by minimizing the squared error loss \eqref{eq:0}.
Therefore, any implementation of CART or RF for squared error loss 
applied to ${\cal F}_{red,t}$ will produce a corresponding CART- or RF-based estimate of $\psi_{0m}(t;w).$ 
For example, CART estimates $L$ and the associated set of terminal nodes  $\{\mathcal{N}_1,\ldots,\mathcal{N}_L\}$ 
from the data ${\cal F}_{red,t},$ and within each terminal node, estimates $\psi_{0m}(t|w)$ by \eqref{beta-full}. 

For the case of multiple time points, the relevant loss function is a special case of \eqref{eq:7a}:
\begin{eqnarray}
	\nonumber
	L_{m,\mathbf{t}}^{emp}({\cal F},\psi_m) & = & 
	\sum_{l=1}^{L}   \sum_{i=1}^n I\{W_i \in \mathcal{N}_l\}  \left( \sum_{j=1}^J   \frac{w_j}{n} \{Z_{im}(t_j) - \beta_{lm}(t_j)\}^2  \right)  \\
	& = & 
	\label{eq:0a}
	\sum_{l=1}^{L} \sum_{i=1}^n I\{W_i \in \mathcal{N}_l\}   \{ \mathbf{Z}_{im} - \boldsymbol{\beta}_{lm}\}^{\top} \mathbf{D}^{-1}   \{ \mathbf{Z}_{im} - \boldsymbol{\beta}_{lm}\}
\end{eqnarray}
where $\mathbf{Z}_{im} = (Z_{im}(t_1),\ldots,Z_{im}(t_J))^\top,$ 
$\boldsymbol{\beta}_{lm} = (\beta_{lm}(t_1),\ldots,\beta_{lm}(t_J))^\top,$ and $\mathbf{D}$ is a diagonal matrix with $D_{jj} = n/w_j, j=1,\ldots,J$. 
One can therefore estimate the desired CIF either by a tree or random forest directly from the data 
$\{ (\mathbf{Z}_{im}, W'_i)',i=1,\ldots n \}$ using the {\sf MultivariateRandomForest} package \citep{MRF}, which
builds regression trees using a Mahalanobis loss function of the form \eqref{eq:0a}; see also \cite{Segal}.  
The {\sf randomForestSRC} package also accommodates multivariate response data, but uses an alternative loss
function in the case of squared-error regression that involves repeatedly standardizing the outcomes
falling into each parent node prior to determining where and when to split. Since this process of repeated
standardization has implications for certain equivalences on which we later rely, we use
the {\sf MultivariateRandomForest} package to implement our methods in the next subsection.

\subsection{Estimating a CIF via CART or RF: loss to follow-up}
\label{implement1}

The CART and RF algorithms as outlined in the previous subsection extend easily to more general loss functions, 
where decisions and predictions are instead derived from minimizing the chosen loss function. 
In particular, the  loss  $L_{m,t}^{dr}({\cal O},\psi_m; G,\Psi)$ or its composite extension 
$L_{m,\mathbf{t}}^{mult,dr}(\mathcal{O},\psi_m;G,\Psi)$ could be used in place of \eqref{eq:0} in 
either algorithm in the presence of censoring.  A detailed description of such an algorithm in the case of CART 
can be found in \cite{choArxiv}.

Existing software may not be able to easily accommodate such changes, particularly so for
general loss functions. However, for the class of augmented Brier 
loss functions considered in Section \ref{brier-LTF},
algorithms that use the loss function $L_{m,t}^{dr}({\cal O},\psi_m; G,\Psi)$
or $L_{m,\mathbf{t}}^{mult,dr}(\mathcal{O},\psi_m;G,\Psi)$ 
can be implemented easily with existing software using a certain form
of response imputation; see  \cite{steingrimsson2019censoring} for
related results in the case where $K=1.$

Recall that 
$L_{m,t}^{dr}({\cal O},\psi_m; G,\Psi) 
= L_{m,t}^{ipcw}({\cal O},\psi_m; G) 
+ L_{m,t}^{aug}({\cal O},\psi_m; G, \Psi)$ 
where $L_{m,t}^{ipcw}({\cal O},\psi_m; G)$
is given by \eqref{eq:2} 
and $L_{m,t}^{aug}({\cal O},\psi_m; G, \Psi)$ 
is given by \eqref{eq:4a1}. Using the results
in \eqref{simple-V} and \eqref{eq:2b} and notation defined
in \eqref{TSfuns}, calculations similar to 
\cite{steingrimsson2019censoring} show that
\begin{gather*}
	L_{m,t}^{dr}(\mathcal{O},\psi_m;G,\Psi) 
	= \frac{1}{n}\sum_{i=1}^n \sum_{l=1}^L I(W_i \in \mathcal{N}_l)[\beta_{ml}^2(t) - 2H^{dr}_m(t;O_i,G,\Psi)\beta_{ml}(t)+H_m^{dr}(t;O_i,G,\Psi)]
\end{gather*} 
where $H_m^{dr}(t;O_i,G,\Psi) = \widetilde{TS}_{1,im}^1(t) + \widetilde{TS}_{2,im}^1(t).$
Define the modified ``imputed'' loss function 
\begin{align*}
	L_{m,t}^{dr,*}(\mathcal{O},\psi_m;G,\Psi) & = \sum_{i=1}^n \sum_{l=1}^L I(W_i \in \mathcal{N}_l) (H_m^{dr}(t;O_i,G,\Psi) - \beta_{ml}(t))^2 \\
	& = \sum_{i=1}^n \sum_{l=1}^L I(W_i \in \mathcal{N}_l) \{\beta_{ml}^2(t)  - 2 H_m^{dr}(t;O_i,G,\Psi) \beta_{ml}(t) + [H_m^{dr}(t;O_i,G,\Psi)]^2 \}.
\end{align*}
Importantly, it can be seen that $L_{m,t}^{dr}(\mathcal{O},\psi_m;G,\Psi)-L_{m,t}^{dr,*}(\mathcal{O},\psi_m;G,\Psi)$ does not depend on
$\beta_{ml}(t), l= 1,\ldots,L$ if $H_m^{dr}(t;O_i,G,\Psi), i=1,\ldots,n$ does not depend on these terms.
Hence, a CART tree or RF built using $L_{m,t}^{dr}(\mathcal{O},\psi_m;G,\Psi)$ 
will be identical to that built using $L_{m,t}^{dr,*}(\mathcal{O},\psi_m;G,\Psi);$ see \citet[Thm. 4.1]{steingrimsson2019censoring}. 

A similar correspondence can be established between the composite loss 
$L_{m,\mathbf{t}}^{mult,dr}(\mathcal{O},\psi_m;G,\Psi)$ 
in \eqref{eq:7a} and the Mahalanobis-type loss function
\begin{gather*}
	L_{m,\mathbf{t}}^{mult,dr,*}(\mathcal{O},\psi_m;G,\Psi) = \sum_{l=1}^L \sum_{i=1}^n I(W_i \in \mathcal{N}_l) 
	(\mathbf{H}_{im}^{dr} - \boldsymbol{\beta}_{lm})^\top \mathbf{D}^{-1}
	(\mathbf{H}_{im}^{dr} - \boldsymbol{\beta}_{lm}),
\end{gather*}
where $\mathbf{H}_{im}^{dr} = (H_{m}^{dr}(t_1;O_i,G,\Psi),\ldots,H_{m}^{dr}(t_J;O_i,G,\Psi))^\top,$ $j=1,\ldots,J;$
compare with \eqref{eq:0a}. Hence, a CART tree or RF built using \eqref{eq:7a} is identical 
to that built using  $L_{m,\mathbf{t}}^{mult,dr,*}(\mathcal{O},\psi_m;G,\Psi).$

Criticially, these results imply that a CART or RF algorithm that uses the loss function 
$L_{m,\mathbf{t}}^{mult,dr}(\mathcal{O},\psi_m;G,\Psi)$  can be implemented  
by applying a version of that algorithm designed for squared error loss to the modified dataset
$(\mathbf{H}_{im}^{dr}, W_i),=1\ldots,n,$ where
$\mathbf{H}_{im}^{dr}$ is an imputed univariate ($J=1$) or multivariate ($J > 1$) response.
This includes the case of Buckley-James loss, which results as a special case
upon setting $G(t|w) = 1$ for all $t > 0$ and $w \in {\cal S}.$ 
Specifically, for a fixed set of times $0 < t_1 < \cdots < t_J < \infty$ and event type $m,$ 
the relevant RF estimation algorithm is as follows:

\noindent \underline{Algorithm $M_0$}:
\begin{enumerate}
	\item Compute $\hat{G}$ and $\hat{\Psi}$ by appropriate modeling;
	\item Compute $H_m^{dr}(t_j;O_i,\hat{G},\hat{\Psi})$ for $i=1,\ldots,n,$ $j=1,\ldots,J$;
	\item Run {\sf MultivariateRandomForest} on the imputed dataset $(\hat{\mathbf{H}}_{im}^{dr}, W_i),=1\ldots,n,$ where
	\[
	\hat{\mathbf{H}}_{im}^{dr} = (H_{m}^{dr}(t_1;O_i,\hat{G},\hat{\Psi}), \ldots, H_{m}^{dr}(t_J;O_i,\hat{G},\hat{\Psi}))^\top,
	~j=1,\ldots,J.
	\]
\end{enumerate}
The above procedure extends in an obvious way 
to other tree- and forest-based algorithms that make all decisions on the basis of minimizing squared error loss.

\subsubsection{A modified imputation approach for doubly robust losses}
\label{sec:modimp}

Recall that
$
\mathbf{H}_{im}^{dr} = (H_{m}^{dr}(t_1;O_i,G,\Psi),\ldots,H_{m}^{dr}(t_J;O_i,G,\Psi))^\top
$
where $H_m^{dr}(t;O_i,G,\Psi) = \widetilde{TS}_{1,im}^1(t) + \widetilde{TS}_{2,im}^1(t)$
and $\widetilde{TS}_{r,im}^1(t), r=1,2$ are defined as \eqref{TSfuns}.
Observing that
\[
\widetilde{TS}_{2,im}^1(t) 
=
(1-\Delta_i) \frac{y_m(\tilde{T}_i;t,W_i,\Psi)}{G(\tilde{T}_i |W_i)}
-
\int_0^{\tilde{T}_i} \frac{y_m(u;t,W_i,\Psi)}{G^2(u|W_i)} Y_i(u) d \bar{G}(u|W_i), 
\]
it can be seen that this term has the potential to be estimated with undesirably high variability due to the presence
of $G^2(\cdot|W_i)$ in the denominator of the second term. 

In the context of devising testing procedures for the Cox regression model,
\cite{Lin93} proposed approximating certain martingale integrals using a
simple but effective simulation technique. 
The basic idea, applied here, involves replacing
$M_G(u|W_i)$ by $M^*_G(u) = \xi_i  I( \tilde{T}_i \leq u, \Delta_i = 0 ),$ where $\xi \sim N(0,1).$
In particular,
suppose that $\widetilde{TS}_{2,im}^1(t)$ in the term $H_{m}^{dr}(t;O_i,G,\Psi)$ is replaced by 
\[
\widetilde{TS}_{2,im}^{1}(t,\xi_i) = \int_0^{\tilde{T}_i} \frac{y_m(u;t,W_i,\Psi)}{G(u|W_i)} dM^*_G(u)
=
\xi_i (1-\Delta_i) \frac{y_m(\tilde{T}_i;t,W_i,\Psi)}{G(\tilde{T}_i |W_i)}.
\]
Defining
$
H_m^{dr}(t,\xi_i;O_i,G,\Psi) = \widetilde{TS}_{1,im}^1(t) + \widetilde{TS}_{2,im}^{1}(t,\xi_i),
$
we obtain the alternative loss function
\begin{align*}
	L_{m,t}^{dr,*}(\mathcal{O},\psi_m,\mathbf{\xi};G,\Psi) & = \sum_{i=1}^n \sum_{l=1}^L I(W_i \in \mathcal{N}_l) (H_m^{dr}(t,\xi_i;O_i,G,\Psi) - \beta_{ml}(t))^2.
\end{align*}
Using a straightforward conditioning argument, it is easy to show that 
$E[ L_{m,t}^{dr}(\mathcal{O},\psi_m;G,\Psi)],$ $E[ L_{m,t}^{dr,*}(\mathcal{O},\psi_m;G,\Psi)]$
and $E[ L_{m,t}^{dr,*}(\mathcal{O},\psi_m,\mathbf{\xi};G,\Psi)]$ have the same minimizers; however, 
a CART tree or RF built using $L_{m,t}^{dr,*}(\mathcal{O},\psi_m,\mathbf{\xi};G,\Psi)$ is no longer
guaranteed to be identical to that built using either $L_{m,t}^{dr}(\mathcal{O},\psi_m;G,\Psi)$ or $L_{m,t}^{dr,*}(\mathcal{O},\psi_m;G,\Psi)$.
This is because $L_{m,t}^{dr,*}(\mathcal{O},\psi_m,\mathbf{\xi};G,\Psi)$ contains an extra mean zero term that involves
$\beta_{ml}(t).$

Define the vector
$
\mathbf{H}_{im}^{dr}(\xi_i) = (H_{m}^{dr}(t_1,\xi_i;O_i,G,\Psi),\ldots,H_{m}^{dr}(t_J,\xi_i;O_i,G,\Psi))^\top.
$
Then, for a fixed set of times $0 < t_1 < \cdots < t_J < \infty$ and event type $m,$
we obtain a modified version of the algorithm presented in Section 3.2: 

\noindent \underline{Algorithm $M_1$}:
\begin{enumerate}
	\item Compute $\hat{G}$ and $\hat{\Psi}$ by appropriate modeling;
	\item Loop over $r = 1,\ldots,R,$ where $R \geq 1$ is set by the user:
	\begin{enumerate}
		\item Generate $\xi^{(r)}_i \sim N(0,1)$ for $i=1,\ldots,n$.
		\item Compute $H_m^{dr}(t_j,\xi^{(r)}_i;O_i,\hat{G},\hat{\Psi})$ for $i=1,\ldots,n,$ $j=1,\ldots,J$.
		\item Run {\sf MultivariateRandomForest}  
		on the modified imputed dataset $(\hat{\mathbf{H}}_{im}^{dr}(\xi^{(r)}_i), W_i),=1\ldots,n,$ where
		$
		\hat{\mathbf{H}}_{im}^{dr} (\xi^{(r)}_i) = (H_{m}^{dr}(t_1, \xi^{(r)}_i;O_i,\hat{G},\hat{\Psi}), \ldots, H_{m}^{dr}(t_J, \xi^{(r)}_i;O_i,\hat{G},\hat{\Psi}))^\top,
		~j=1,\ldots,J.
		$
		\item Record $r^{th}$ result.
	\end{enumerate}
	\item Average the $R$ ensemble estimates to obtain a final ensemble predictor.
\end{enumerate}

Analogously to Algorithm $M_0,$ Step 2(c) of the above algorithm involves bootstrapping the $r^{th}$ modified version of the input dataset $B$ times to obtain a RF predictor for the $r^{th}$ modified dataset. Step 3 then averages these $R$ different RF estimates to produce a single ensemble predictor. As a computationally efficient version of this algorithm, Step 2(c) could be run with $B =1$ only; that is, instead of generating a full RF at this stage, one builds a single tree using random feature selection without pruning.  
The resulting algorithm then reduces to a RF-type algorithm based on $R$ bootstrap samples, but where there is an extra component of randomization used in the generation of each bootstrap sample.

\section{Simulation Study: CIF Estimation via RF}
\subsection{{Main simulation setting}}
\label{simsetting}
In this section, we will evaluate the performance of estimators derived using Algorithms $M_0$ and $M_1$ and 
compare the prediction errors to the RF procedure for CIF estimation proposed by \cite{ishwaran2014random}, 
which is implemented in the {\sf R} package {\tt randomForestSRC}.

Let $W_i \sim N(0,1), i=1,\ldots, 20$ be independent predictor variables. Define the true CIFs $\psi_{0m}(t;\boldsymbol{W}), m=1,2$ as follows:
\begin{gather}
	\label{true sim CIF}
	\psi_{01}(t;\boldsymbol{W}) = 1 - (1-p(1-e^{-t}))^{\exp{(\boldsymbol{\beta}_{1}^T \boldsymbol{Z}(\boldsymbol{W}) )}} \\
	\psi_{02}(t;\boldsymbol{W}) = (1-p)^{\exp{(\boldsymbol{\beta}_{1}^T \boldsymbol{Z}(\boldsymbol{W}))}} \times (1-\exp{(-t\exp{(\boldsymbol{\beta}_{2}^T \boldsymbol{Z}(\boldsymbol{W}))})}),
\end{gather} 
where $\boldsymbol{Z}(\boldsymbol{W}) = (\sin(\pi W_1W_2), W_3^2, W_{10}, I(W_{11} > 0), W_{12}, \exp(W_{15}))$ and the regression coefficients 
are given by $\boldsymbol{\beta}_1 = (0.5,0.5,0.5,0.5,0.6,-0.3)^\top$ and $\boldsymbol{\beta}_2 = (0,-0.5,-0.5,-0.5,0.5,0.1)^\top$. 
Random censoring is generated from log normal distribution with mean $0.1+0.1 \cdot |W_1+W_3+W_5| + 0.1 \cdot |W_{11} + W_{13}+W_{15}|$ and variance 1. In this setting, the overall censoring rate is approximately 28.1\%.


\subsection{{Simulation results}}

\subsubsection{Algorithms to be compared}
We focus on estimation of the CIF at the 25th, 50th and 75th time points of the marginal failure time 
distribution $T;$ these are approximated outside the main simulation using 
a single, very large random sample. These time points are also used
in the computation of all composite loss functions.
CIF estimates are obtained using both Algorithms $M_0$ (Section \ref{implement1}) 
and $M_1$ (Section \ref{sec:modimp}), and compared to those produced by  
{\tt rfsrc} in the {\tt randomForestSRC} package \citep{rfsrc}. 
For Algorithm $M_1,$ we set $R=1$ and $B=500;$ for each simulated dataset;
this corresponds to generating a single set of $n$ independent standard normal 
random variables to be used in the computation of the modified imputed loss 
$L_{m,t}^{dr,*}(\mathcal{O},\psi_m,\mathbf{\xi};G,\Psi),$
and then using 500 bootstrap samples to generate a RF predictor.

For calculating $y_{m}(u;t,w,\Psi)$ in \eqref{eq:2b}, we estimate $\Psi$ using {\tt rfsrc}. We denote the 
resulting Buckley-James (BJ-RF) and doubly robust (DR-RF) transformations 
$H_m^{bj*}(t,O_i;\hat{\Psi})$ and $H_m^{dr*}(t,O_i;\hat{G},\hat{\Psi}), i = 1,\ldots, n$, 
where the censoring distribution estimate $\hat{G}$ is obtained using the methods of 
\cite{leblanc1992relative}. Specifically, 
$\hat G$ is estimated using the {\tt rpart} package \citep{therneau2015introduction} with the 
minimum number of observations in each node (i.e., {\tt minbucket}) set to 30. For comparison, 
we also compute (a) versions of these same estimators using correctly specified parametric models 
derived directly from \eqref{true sim CIF}, with relevant parameters estimated 
using the maximum likelihood approach detailed in \citet{jeong2007parametric}; these results for
the parametric Fine-Gray-type model for the CIF are denoted \textit{BJ-FG (true)} and \textit{DR-FG (true)}, 
respectively, and, (b) the RF estimator of the CIF obtained using {\tt rfsrc}.

Tuning parameters play an important role in the performance of ensemble estimators. In the case of {\tt rfsrc}, two 
key tuning parameters are (i) the minimum number of observations in each terminal node (\textit{nodesize}) and 
(ii) the number of candidate variables selected for consideration at each split (\textit{mtry}). There are identical 
parameters with different names to be selected for use with the package {\tt MultivariateRandomForest}, specifically 
through the use of the required function {\tt build\_single\_tree}; for simplicity, we present and
summarize results using the labels \textit{nodesize} and \textit{mtry}. For each algorithm, we calculate the relevant 
ensemble estimators by setting these tuning parameters as follows: 
\begin{itemize}
	\item Tuning Set 1 : \textit{nodesize} = 20 and \textit{mtry} = $\lfloor \sqrt{p} \rfloor$.
	\item Tuning Set 2 : \textit{nodesize} and \textit{mtry} are selected to minimize the out-of-bag (OOB) error. 
\end{itemize}
Results with the suffix \textit{-opt} correspond to parameters selected under Tuning Set 2.
Hence, results are reported for 4 cases:
\begin{itemize}
	\item[(i)] Fixed \textit{mtry} and \textit{nodesize} with Algorithm $M_0$; 
	\item[(ii)] Optimized \textit{mtry} and \textit{nodesize} with Algorithm $M_0$; 
	\item[(iii)] Fixed \textit{mtry} and \textit{nodesize} with Algorithm $M_1$; 
	\item[(iv)] Optimized \textit{mtry} and \textit{nodesize} with Algorithm $M_1$.
\end{itemize}

\subsubsection{Summary of results}

\begin{figure}[!htb]
	\centering
	\includegraphics[width=0.8\textwidth, height=0.325\textheight]{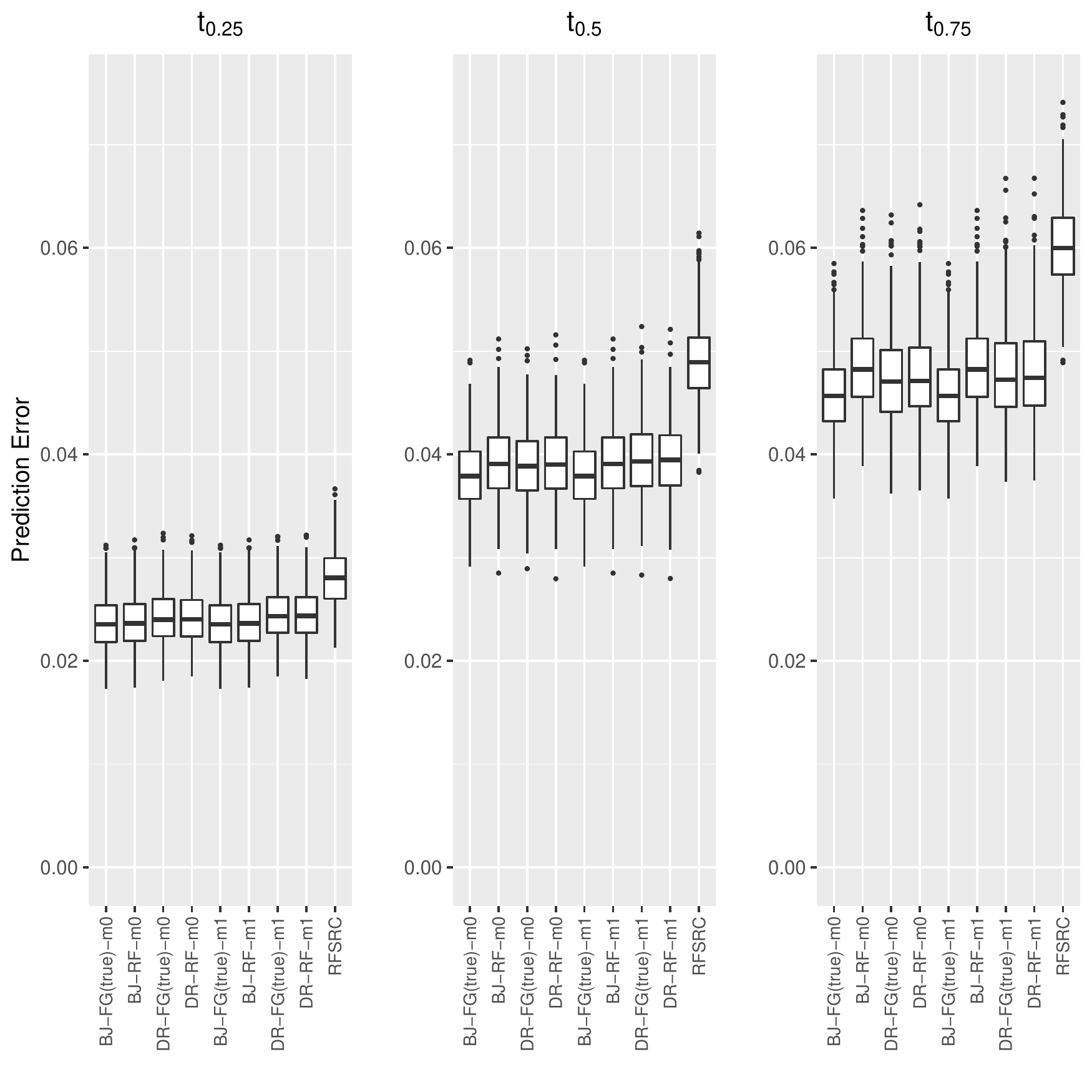}
	\caption{Comparing Algorithms $M_0$ and $M_1$ with Fixed Tuning Parameters for Event 1. Results for {\sf rfsrc} are labeled  RFSRC. Results with Buckley-James and doubly robust losses are respectively prefaced by BJ- and DR-; the methods for $\hat \Psi$ as required by Algorithms $M_0$ and $M_1$ are respectively denoted by RF- and FG(true)-; and, the use of the imputation algorithms $M_0$ and $M_1$ in generating the ensemble estimator is respectively denoted by $m0$ and $m1$.}
	\label{fig:figm1}
\end{figure}

\begin{figure}[!htb]
	\centering
	\includegraphics[width=0.8\textwidth, height=0.325\textheight]{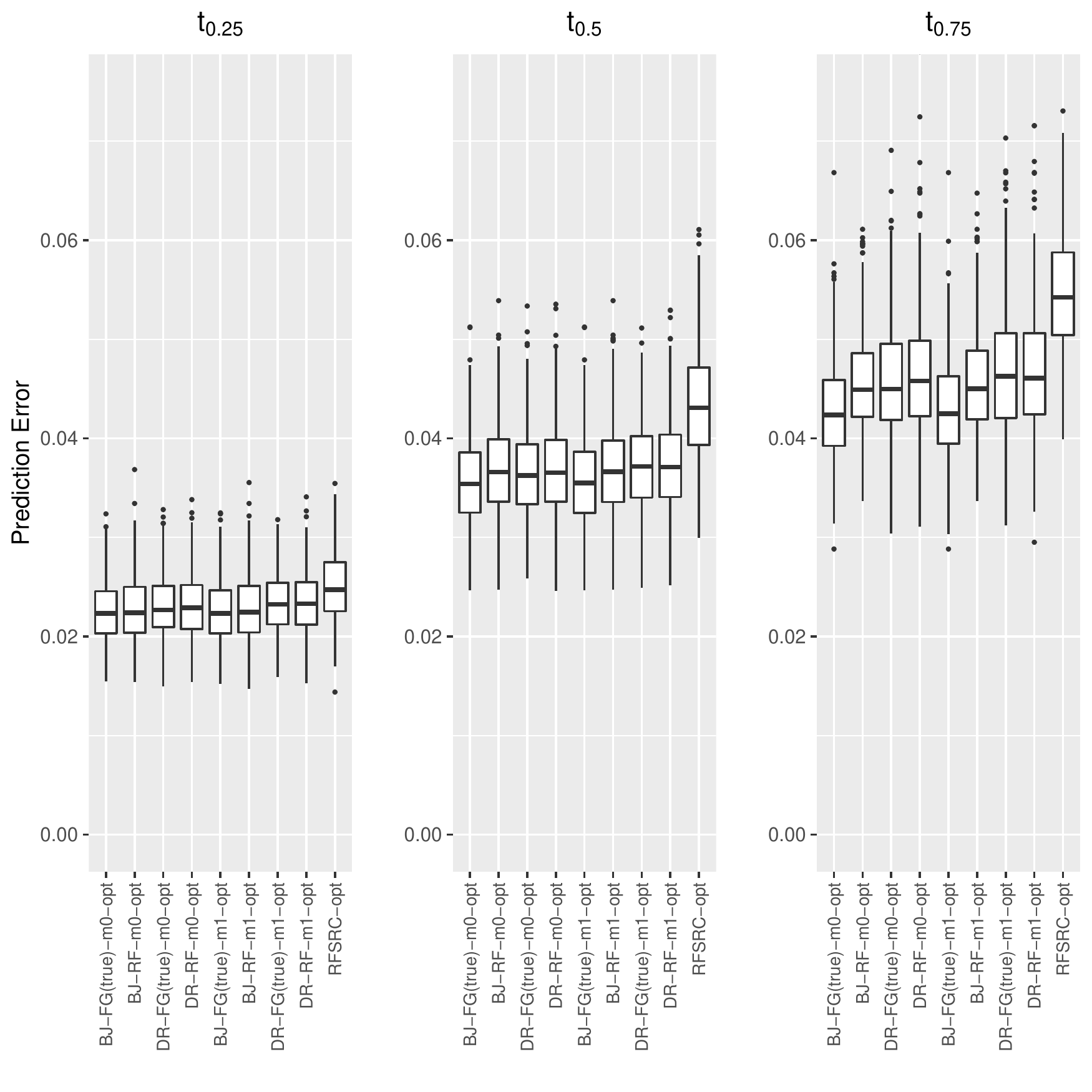}
	\caption{Comparing Algorithms $M_0$ and $M_1$ with Optimized Tuning Parameters for Event 1. Results with Buckley-James and doubly robust losses are respectively prefaced by BJ- and DR-; the methods for $\hat \Psi$ as required by Algorithms $M_0$ and $M_1$ are respectively denoted by RF- and FG(true)-; and, the use of the imputation algorithms $M_0$ and $M_1$ in generating the ensemble estimator is respectively denoted by $m0-$opt and $m1-$opt, the opt indicating use of optimized tuning parameters.}
	\label{fig:figm2}
\end{figure}

\begin{figure}[!htb]
	\centering
	\includegraphics[width=0.8\textwidth, height=0.325\textheight]{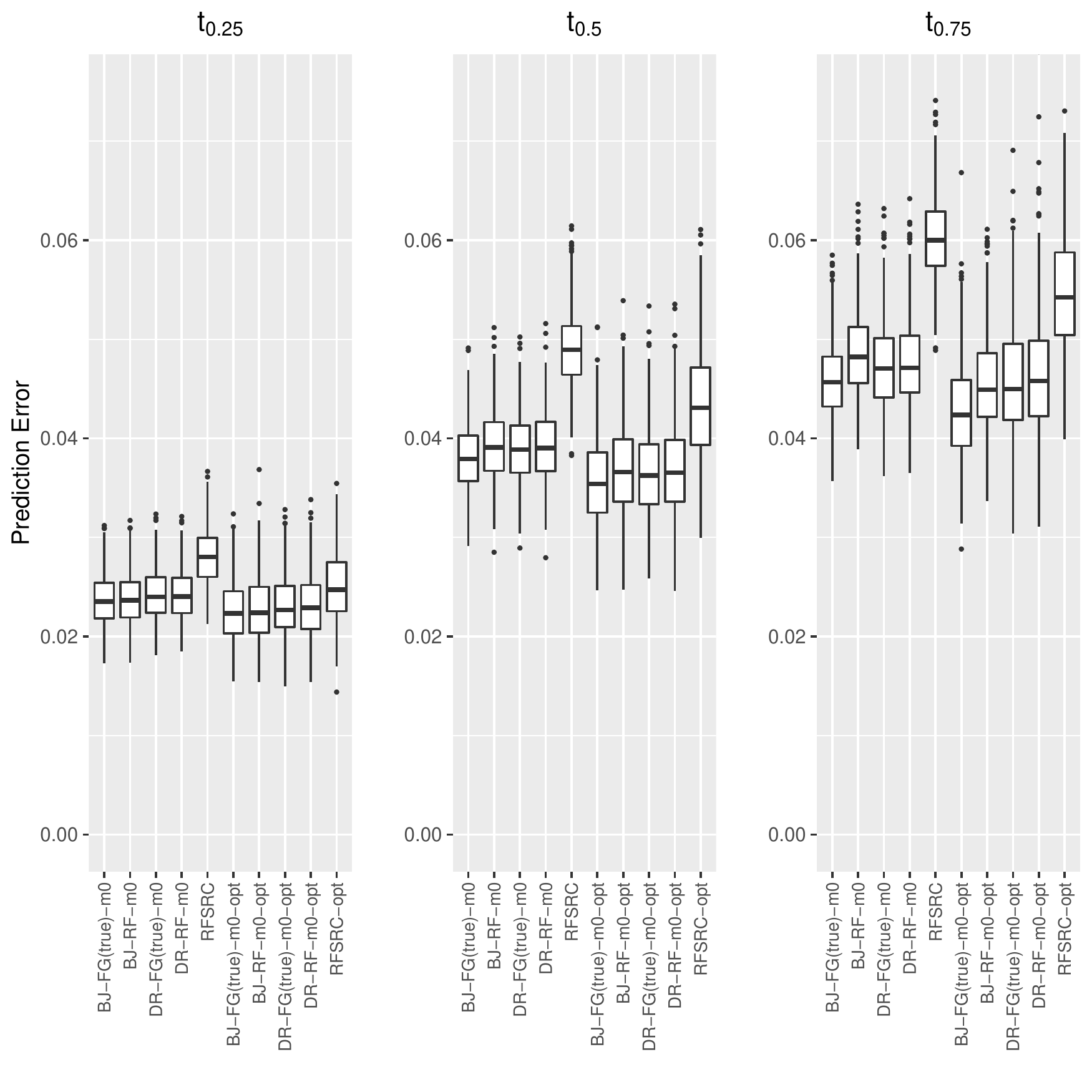}
	\caption{Comparing Fixed and Optimized Tuning Parameters for Algorithm $M_0$ for Event 1. Results with Buckley-James and doubly robust losses are respectively prefaced by BJ- and DR-; the methods for $\hat \Psi$ as required by Algorithm $M_0$ are respectively denoted by RF- and FG(true)-; and, the use of an indicated algorithm with optimized tuning parameters 
		is denoted by $-$opt.}
	\label{fig:figm3}
\end{figure}

\begin{figure}[!htb]
	\centering
	\includegraphics[width=0.8\textwidth, height=0.325\textheight]{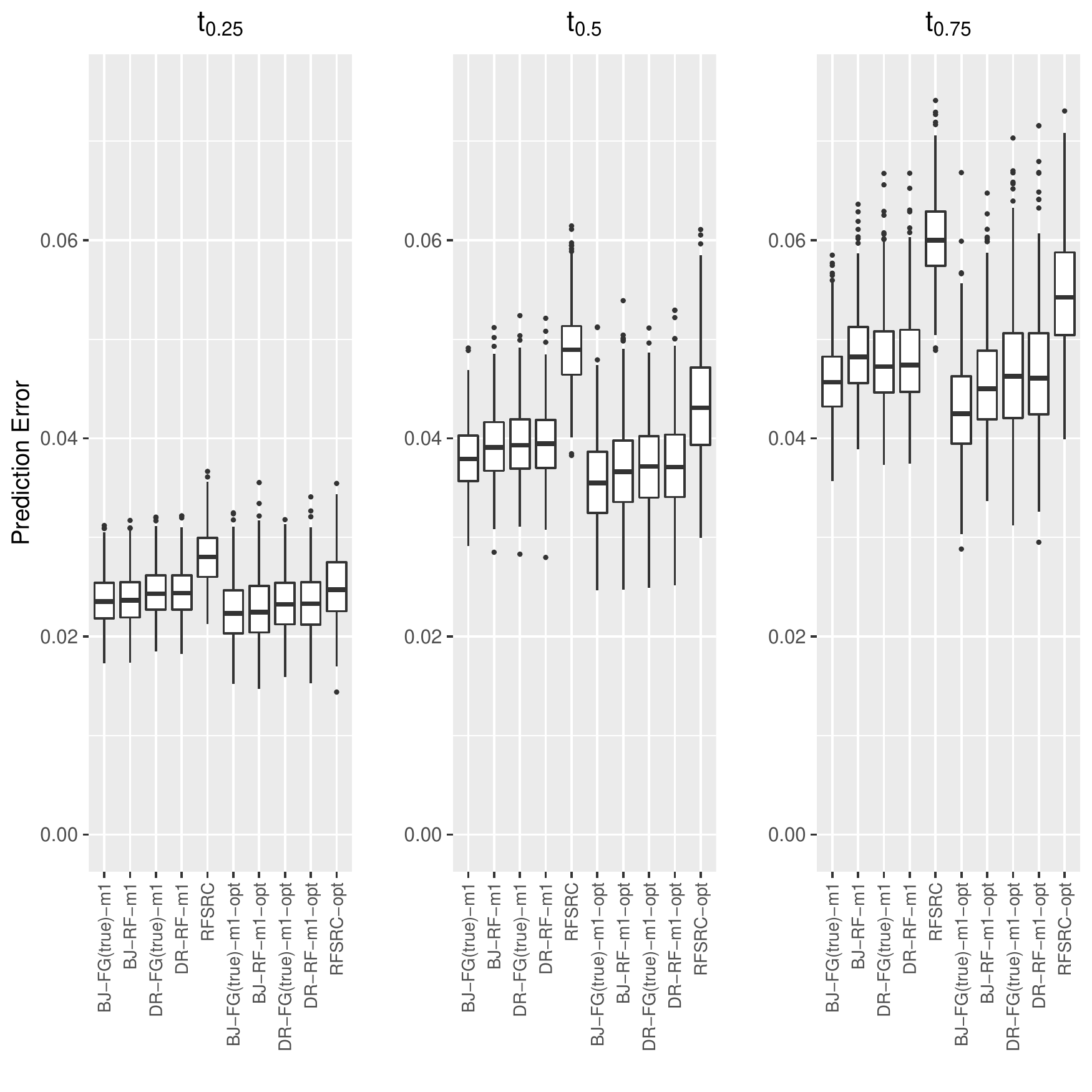}
	\caption{Comparing Fixed and Optimized Tuning Parameters for Algorithm $M_1$ for Event 1. Results with Buckley-James and doubly robust losses are respectively prefaced by BJ- and DR-; the methods for $\hat \Psi$ as required by Algorithm $M_1$ are respectively denoted by RF- and FG(true)-; and, the use of an indicated algorithm with optimized tuning parameters 
		is denoted by $-$opt.}
	\label{fig:figm4}
\end{figure}

\begin{figure}[!htb]
	\centering
	\includegraphics[width=0.8\textwidth, height=0.325\textheight]{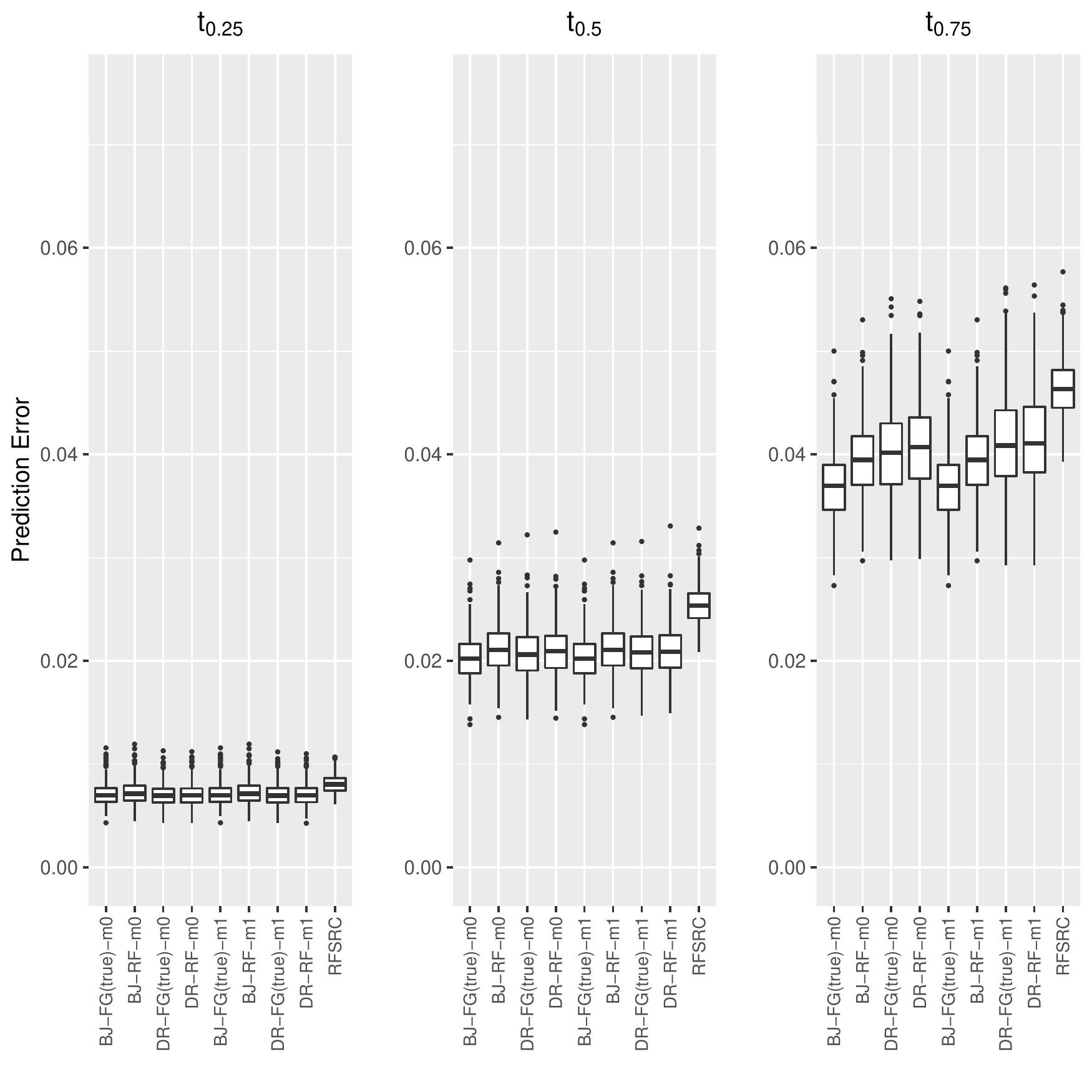}
	\caption{Comparing Algorithms $M_0$ and $M_1$ with Fixed Tuning Parameters for Event 2. Results for {\sf rfsrc} are labeled  RFSRC. Results with Buckley-James and doubly robust losses are respectively prefaced by BJ- and DR-; the methods for $\hat \Psi$ as required by Algorithms $M_0$ and $M_1$ are respectively denoted by RF- and FG(true)-; and, the use of the imputation algorithms $M_0$ and $M_1$ in generating the ensemble estimator is respectively denoted by $m0$ and $m1$.}
	\label{fig:figm1_e2}
\end{figure}

\begin{figure}[!htb]
	\centering
	\includegraphics[width=0.8\textwidth, height=0.325\textheight]{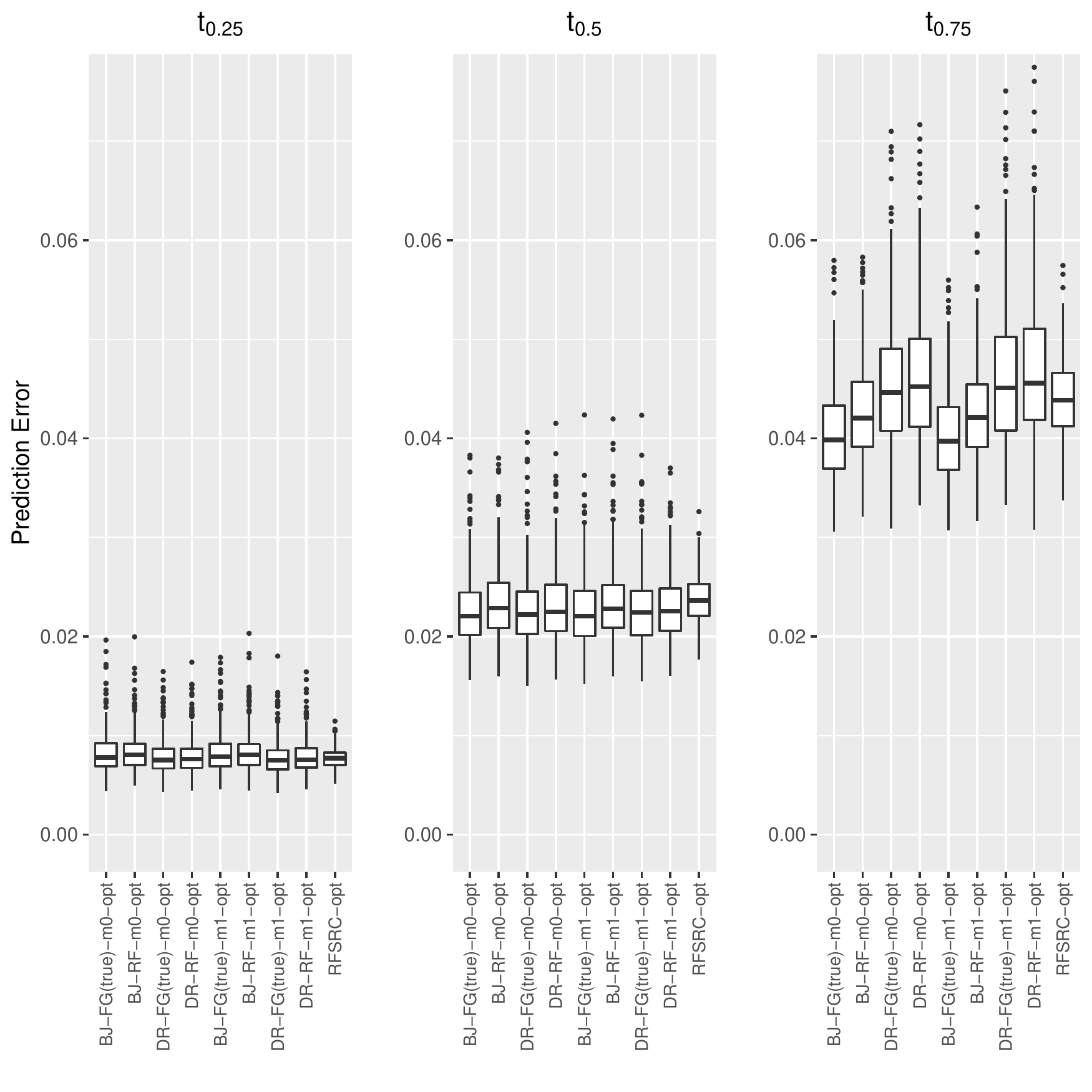}
	\caption{Comparing Algorithms $M_0$ and $M_1$ with Optimized Tuning Parameters for Event 2. Results with Buckley-James and doubly robust losses are respectively prefaced by BJ- and DR-; the methods for $\hat \Psi$ as required by Algorithms $M_0$ and $M_1$ are respectively denoted by RF- and FG(true)-; and, the use of the imputation algorithms $M_0$ and $M_1$ in generating the ensemble estimator is respectively denoted by $m0-$opt and $m1-$opt, the opt indicating use of optimized tuning parameters.}
	\label{fig:figm2_e2}
\end{figure}

\begin{figure}[!htb]
	\centering
	\includegraphics[width=0.8\textwidth, height=0.325\textheight]{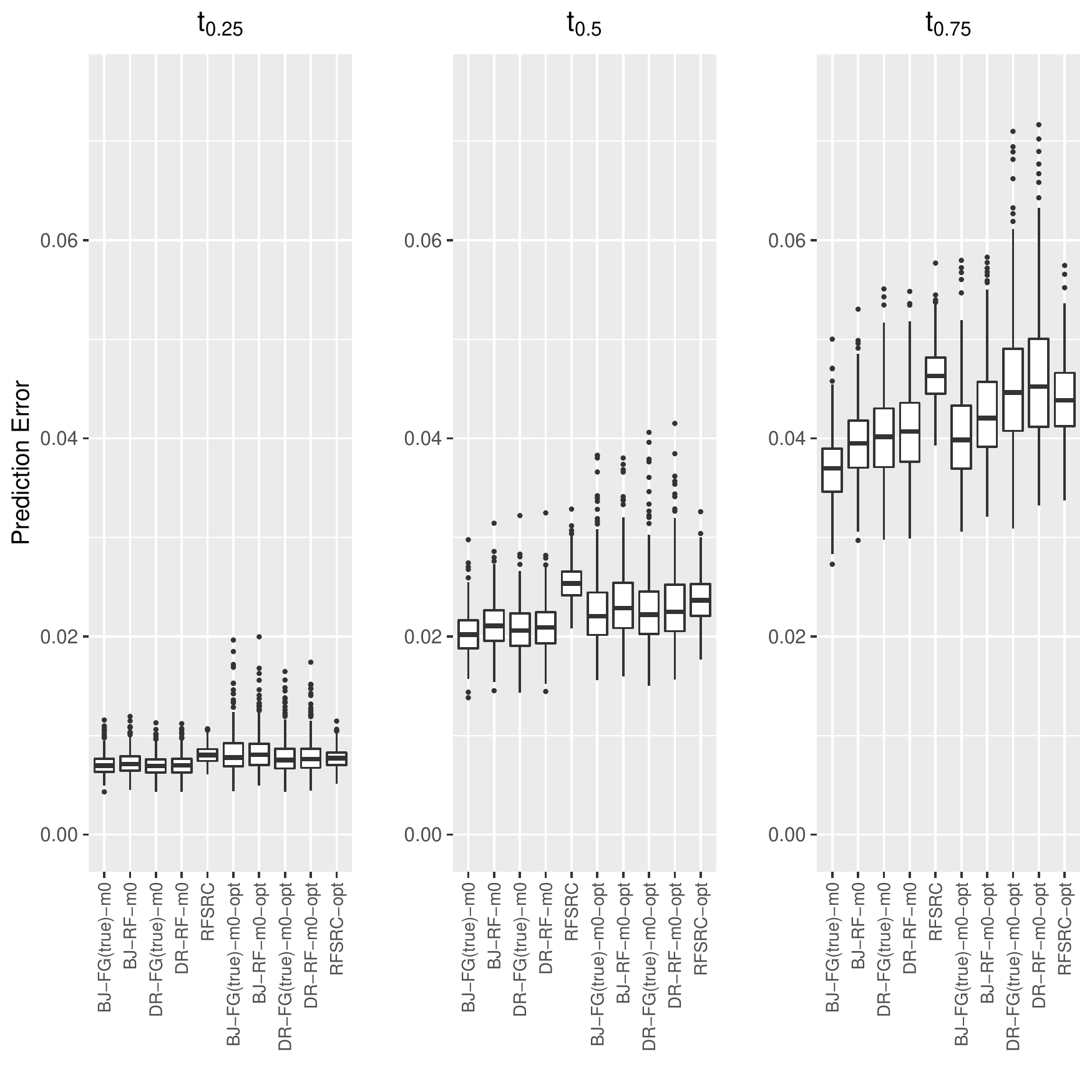}
	\caption{Comparing Fixed and Optimized Tuning Parameters for Algorithm $M_0$ for Event 2. Results with Buckley-James and doubly robust losses are respectively prefaced by BJ- and DR-; the methods for $\hat \Psi$ as required by Algorithm $M_0$ are respectively denoted by RF- and FG(true)-; and, the use of an indicated algorithm with optimized tuning parameters 
		is denoted by $-$opt.}
	\label{fig:figm3_e2}
\end{figure}

\begin{figure}[!htb]
	\centering
	\includegraphics[width=0.8\textwidth, height=0.325\textheight]{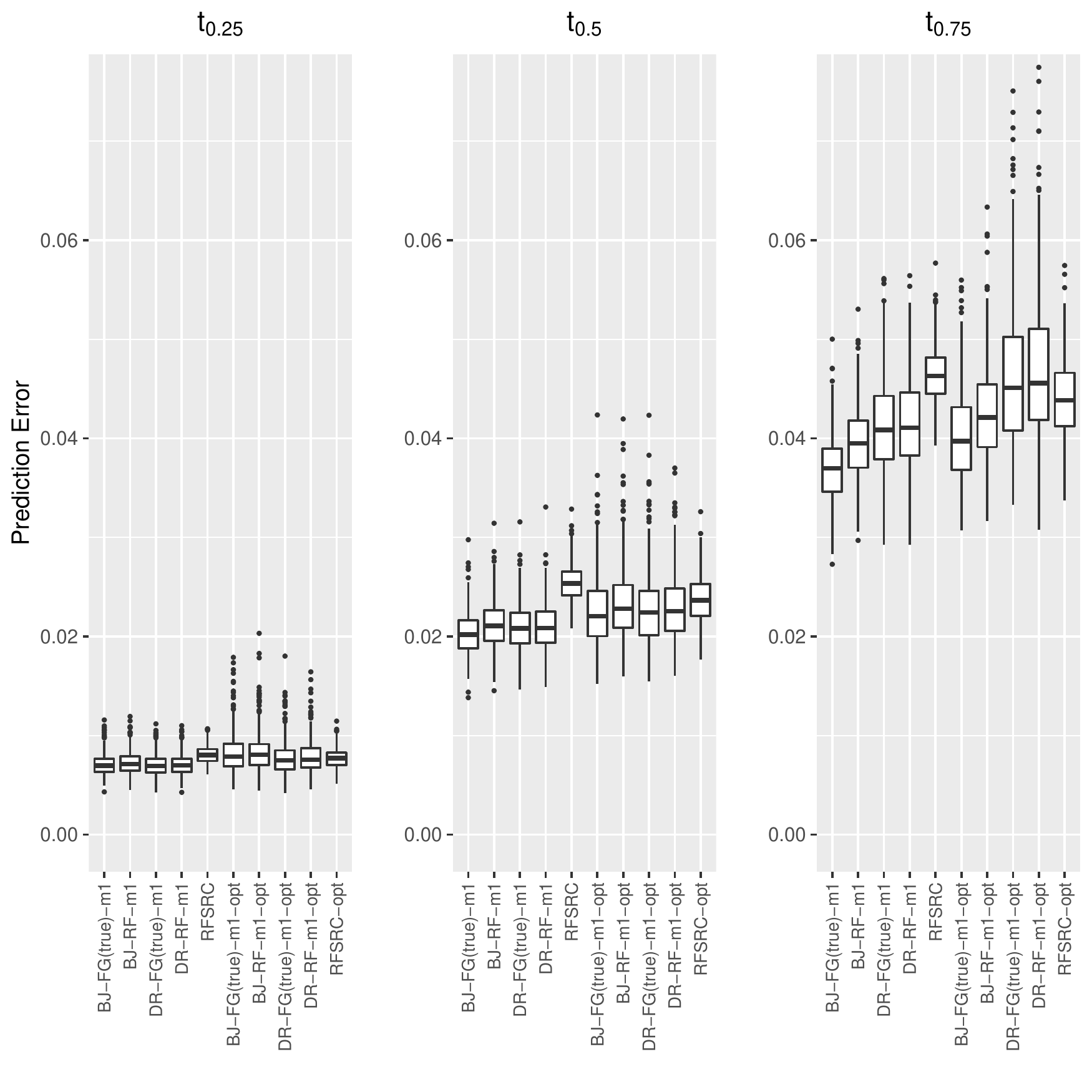}
	\caption{Comparing Fixed and Optimized Tuning Parameters for Algorithm $M_1$ for Event 2. Results with Buckley-James and doubly robust losses are respectively prefaced by BJ- and DR-; the methods for $\hat \Psi$ as required by Algorithm $M_1$ are respectively denoted by RF- and FG(true)-; and, the use of an indicated algorithm with optimized tuning parameters 
		is denoted by $-$opt.}
	\label{fig:figm4_e2}
\end{figure}

For all simulation settings, results are obtained for 400 independent (estimation, test) dataset pairs.
The estimation dataset is generated as in Section \ref{simsetting}, with $n=250;$ 
an independent test dataset of size $n_{test} = 2000$ consisting only of the covariates is 
generated similarly. For all simulation settings, we compute the mean square error 
\begin{gather*}
	\frac{1}{n_{test}} \sum_{r=1}^{n_{test}} \{\hat{\psi}_{m}(t|\boldsymbol{W}_{\!\!r}) - \psi_{0m}(t|\boldsymbol{W}_{\!\!r})\}^2
\end{gather*}
to compare the performance of different algorithms at different values of $t$. Figures \ref{fig:figm1} to \ref{fig:figm4} 
show the results from the simulation setting with 
forests with multiple time points and comparisons with results obtained using {\tt rfsrc} using the same approaches to
selecting \textit{mtry} and \textit{nodeside}. Figures \ref{fig:figm1} to \ref{fig:figm4} focus on $m=1$ and provide comparisons
between fixed and optimized tuning parameters for each algorithm, as well as comparing the algorithms to each other.  
Figures \ref{fig:figm1_e2} to \ref{fig:figm4_e2} repeat these results for $m=2$.
We give an overall summary of these results below:
\begin{itemize}
	\item Algorithms $M_0$ and $M_1$ exhibit similar performance for  the 25th and 50th percentile time points; however, 
	Algorithm $M_1$ tends to perform better for the 75th percentile, where the impact of the censoring rate is higher. 
	
	\item With optimization of tuning parameters, all methods demonstrate similar or slightly improved performance than the same approach using fixed choices of \textit{nodesize} and \textit{mtry}, at least for the 25th and 50th percentile. For the 75th percentile, the effects are somewhat less evident, and in the case of the event $m=2,$ 
	slightly worse for algorithms $M_0$ and $M_1$.
	
	\item The best overall performance is observed for BJ (FG-true) with optimization of tuning parameters, followed by BJ-RF with optimization of tuning parameters.
	The BJ approach has the advantage of not needing to estimate the censoring distribution at all. In general, algorithms
	that use the approach of  \cite{jeong2007parametric}
	for estimating the $\Psi$ required for computing the augmentation term 
	(i.e., a parametric model that agrees with data generating mechanism) perform somewhat better than those that use {\tt rfsrc} for this same limited purpose; however, the results are not dramatically different. 
	
	\item The proposed algorithms perform as well, and often somewhat better, than {\tt rfsrc} in terms of 
	minimizing $MSE(t)$, whether or not tuning parameters are optimized. 
\end{itemize}

\hspace*{-10mm}

\hspace*{-10mm}

\hspace*{-10mm}

\clearpage

The Appendix contains additional simulation results
for the model as described in Section \ref{simsetting}, but where the covariates are 
correlated with each other; see Figures \ref{Afig:figm1} to \ref{Afig:figm4},
which repeat Figures \ref{fig:figm1} to \ref{fig:figm4} in this alternative setting. 
The results using  Algorithm $M_1$ were also re-run with $R=500$ and $B=1$, and were 
indistinguishable from those summarized here (results not shown).

\section{Example: Lung Cancer Treatment Trial}
\label{RTOG9410}

We illustrate our methods using data from the RTOG 9410, a randomized trial of patients with locally advanced inoperable 
non-small cell lung cancer.  The motivation for this trial was to ascertain whether sequential or concurrent delivery of chemotherapy 
and thoracic radiotherapy  (TRT) is a better treatment strategy. 
The original RTOG 9410 study randomized 610 patients to three treatment arms: sequential chemotherapy followed by radiotherapy (RX=1);
once-daily chemotherapy concurrent with radiotherapy (RX=2); and, twice-daily chemotherapy concurrent with radiotherapy (RX=3).
The primary endpoint of interest was overall survival and the main trial analysis results were published in \cite{curran2011sequential}, demonstrating 
a survival benefit of concurrent delivery of chemotherapy and TRT compared with sequential delivery. Secondary analyses of the data using the time 
from randomization to the first occurrence of three possible outcomes are considered: in-field failure (cancer recurrence within the treatment field for TRT); 
out-field failure (cancer recurrence and distant metastasis outside of the treatment field for TRT); and, 
death without documented in-field or out-field failure (i.e., without observed cancer progression). 
Among these event types, those that first experienced out-field failures are of particular interest
since these patients typically have suboptimal prognosis and may be candidates for more intensified 
treatment regimens intended to prevent distant metastasis, including but not limited to consolidative 
chemotherapy, prophylactic cranial irradiation (for brain metastases), and so on.
As such, patients that experienced both in-field failure and out-field failure were 
considered to be out-field failures for purposes of this analysis.
%

At the time the study database was last updated in 2009, there were 577 patients, with approximately 4\% 
censoring on the aforementioned outcomes. Our methods could be applied to directly analyze this
final dataset. However, because the censoring rate is so low, we have decided to take a more illustrative
approach. Specifically, we first create a ``fully observed'' dataset by removing these 23 censored observations.
We then compare the results of analyses of this uncensored dataset of 554 patients to analyses of data 
that were created from this uncensored dataset using an artificially induced censoring mechanism.  
The main purpose of doing this analysis is two-fold; first, the results for the uncensored dataset should largely 
reflect an analysis that would be done for the full dataset of 577 patients; second, we are able to study how 
the introduction of (artificial) censoring affects the results and, in particular, illustrate how well the various 
procedures recover the estimator that would be obtained had outcomes been fully observed (i.e., no random 
loss to follow-up).

We focus on building forests for each outcome using a composite loss function with 3 time points (5.2, 8.5, 15.9 months), 
selected as the 25th, 50th and 75th percentiles of the observed ``all cause" event time (i.e., $T$). Some related analyses
using regression trees alone may be found in \cite{choArxiv}.
Baseline covariates included in this analysis are RX (Treatment),
Age, Stage (American Joint Committee on Cancer [AJCC] stage IIIB vs.\  IIIA or II), 
Gender, KPS (Karnofsky performance score of either 70, 80, 90 or 100), 
Race (White vs.\ non-White), and Histology (Squamous vs.\ non-Squamous).
Censoring is created according to a Uniform $[0, 50]$ distribution, generating approximately 29\% 
censoring on $T$. In addition to building forests using the uncensored version of the dataset using the methods described in 
Section \ref{implement0}, we consider the methods \textit{BJ-RF} and \textit{DR-RF} based on Algorithm $M_1$ 
using optimally tuned parameters as described in the Simulation results, with $R=500$ and $B=1$ (i.e., 500 bootstrap
samples). For comparison, we also report results obtained using {\tt rfsrc} 
using the same approaches to setting the indicated parameters.

\begin{figure}[!htb]
	\centering
	\begin{minipage}{.5\textwidth}
		\centering
		\includegraphics[width=0.9\textwidth, height=0.9\textwidth]{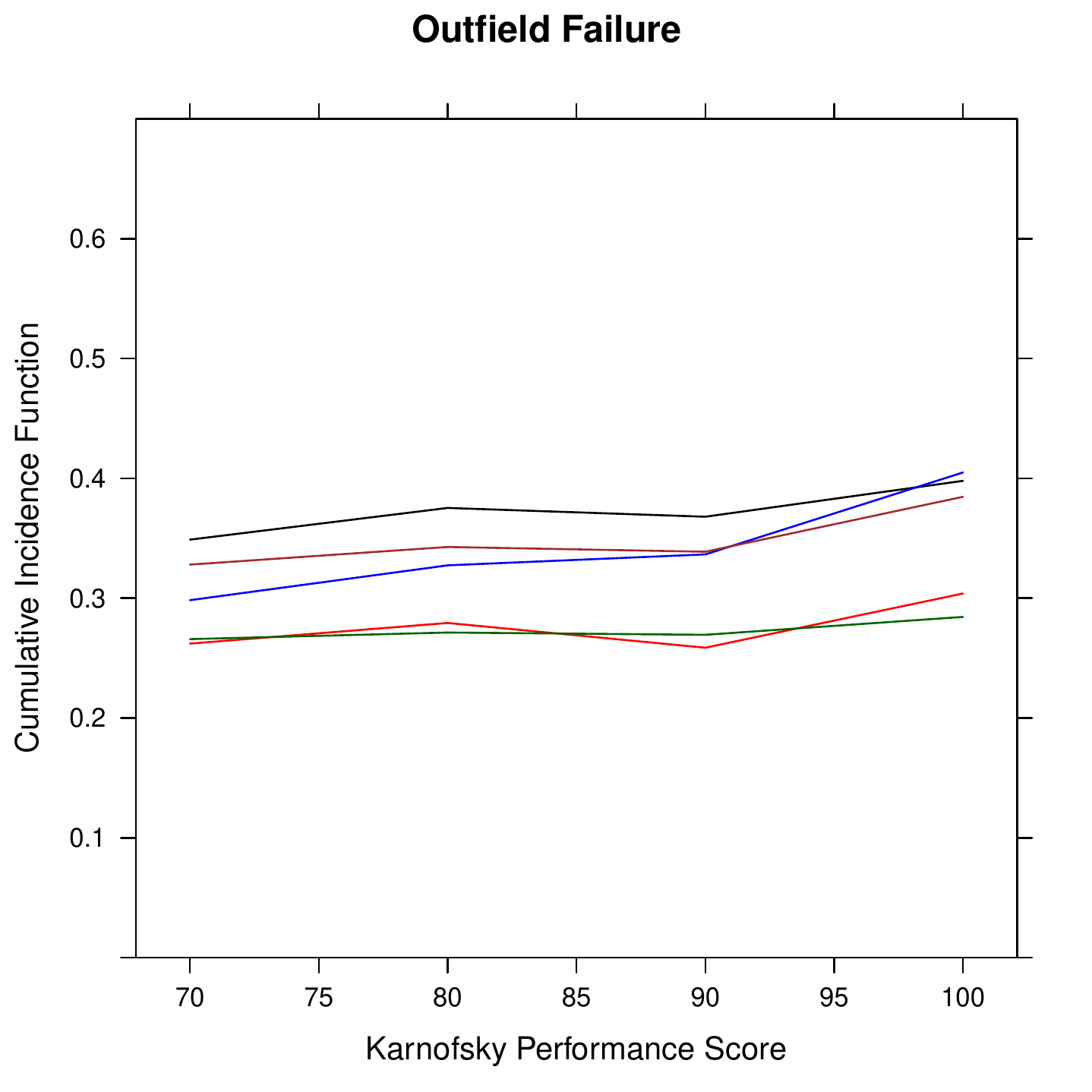}
	\end{minipage}%
	\begin{minipage}{0.5\textwidth}
		\centering
		\includegraphics[width=0.9\textwidth, height=0.9\textwidth]{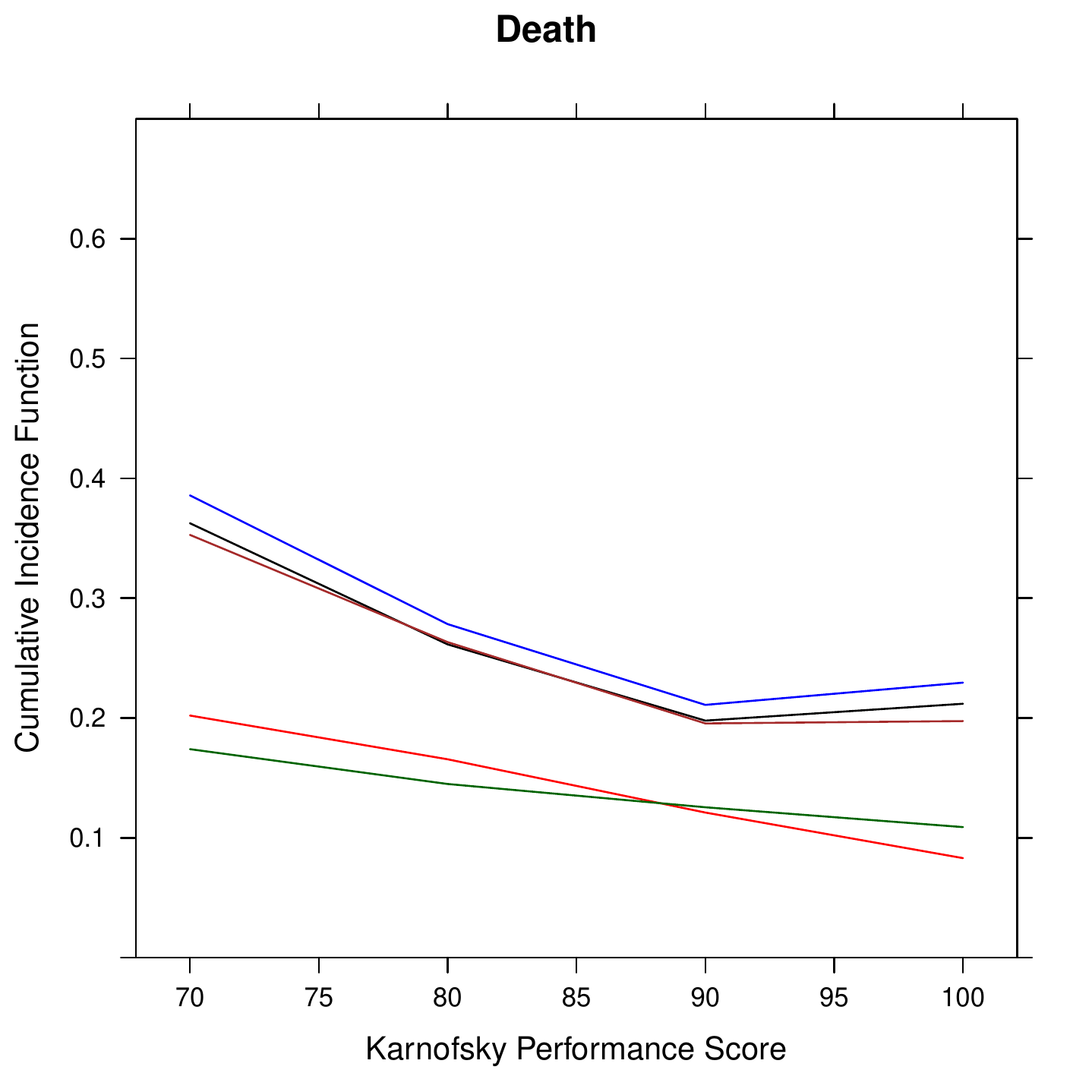}
	\end{minipage}
	\begin{minipage}{.5\textwidth}
		\centering
		\includegraphics[width=0.9\textwidth, height=0.9\textwidth]{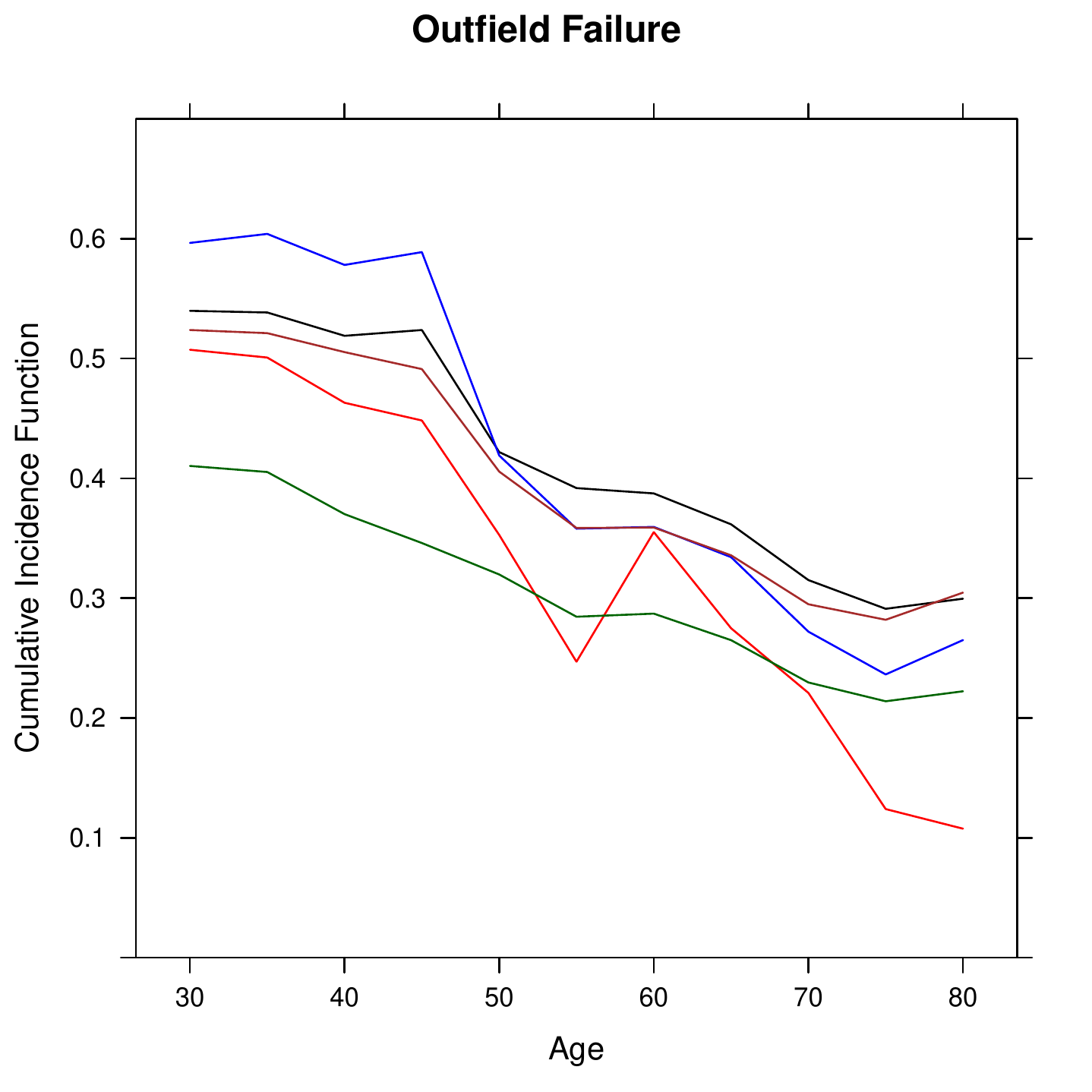}
	\end{minipage}%
	\begin{minipage}{0.5\textwidth}
		\centering
		\includegraphics[width=0.9\textwidth, height=0.9\textwidth]{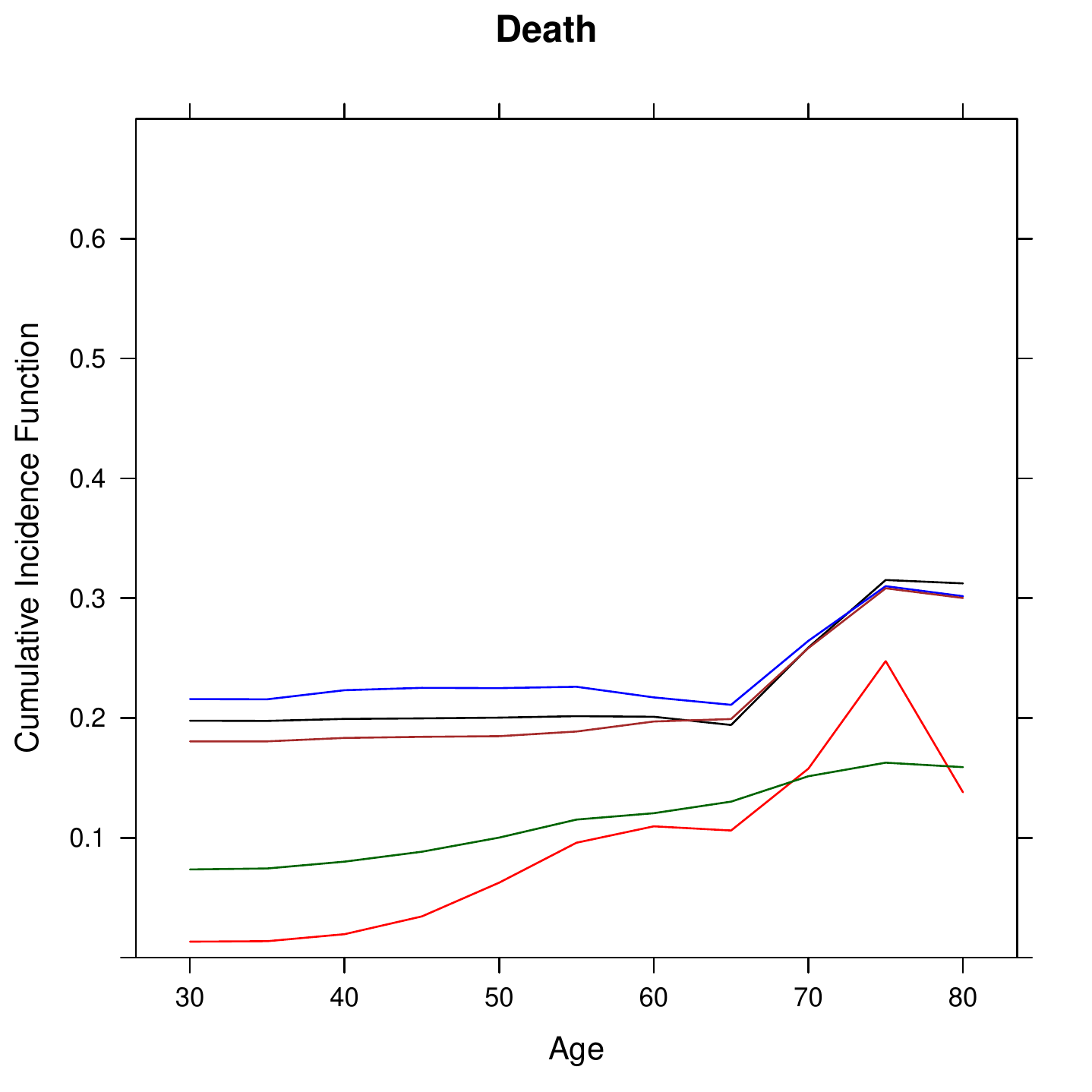}
	\end{minipage}
	\caption{Partial dependence plots with respect to KPS (top) and Age (bottom) from forests with outfield failure (left) and death (right) using artificially censored data and uncensored data for 50th percentile time. The 5 lines are as follows:  black = BJ; blue = DR; red = {\tt rfsrc}(censored); brown = uncensored data estimate using proposed methods in Section \ref{implement0};  
		dark green = uncensored data estimate using {\tt rfsrc}.}
	\label{figPDP}
\end{figure}

To summarize the results in a meaningful way, we created partial dependence plots (PDP) 
\citep[e.g.,][]{greenwell2017pdp} to characterize the influence of Age and KPS 
on the CIF.  For reference, the middle 50\% of patients in this dataset are aged 54-67, and
76\% of patients have KPS scores of 90 or above. 
The PDPs for the 50th percentile time point for the outfield failure and death outcomes are summarized in 
Figure \ref{figPDP}. For outfield failure, the CIFs do not demonstrate substantial changes across the levels of KPS, 
though a slight uptick in risk is observed for the healthiest patients when using both \textit{BJ-RF} and \textit{DR-RF}; however,
there is a decreasing risk with increasing age, which may be possibly due to patients dying before experiencing outfield failure. 
For death, all methods suggest a decreased risk of death for healthier patients, and
an increasing risk of death as patients age, particularly for the oldest patients. Similar trends are observed when looking at
the CIF values calculated at other time points; see Figures \ref{figPDPA1}-\ref{figPDPA4} in the Appendix. Other noteworthy
features from these plots include (i) the trends seen in the PDPs for all methods is generally comparable; and, 
(ii) the CIF estimates produced by {\tt rfsrc} are always smaller than those obtained using the proposed methods
for these data, even in the case where the data are not censored.

%

\begin{figure}[!htb]    
	\begin{minipage}{.5\textwidth}
		\centering
		\includegraphics[width=0.9\textwidth, height=0.9\textwidth]{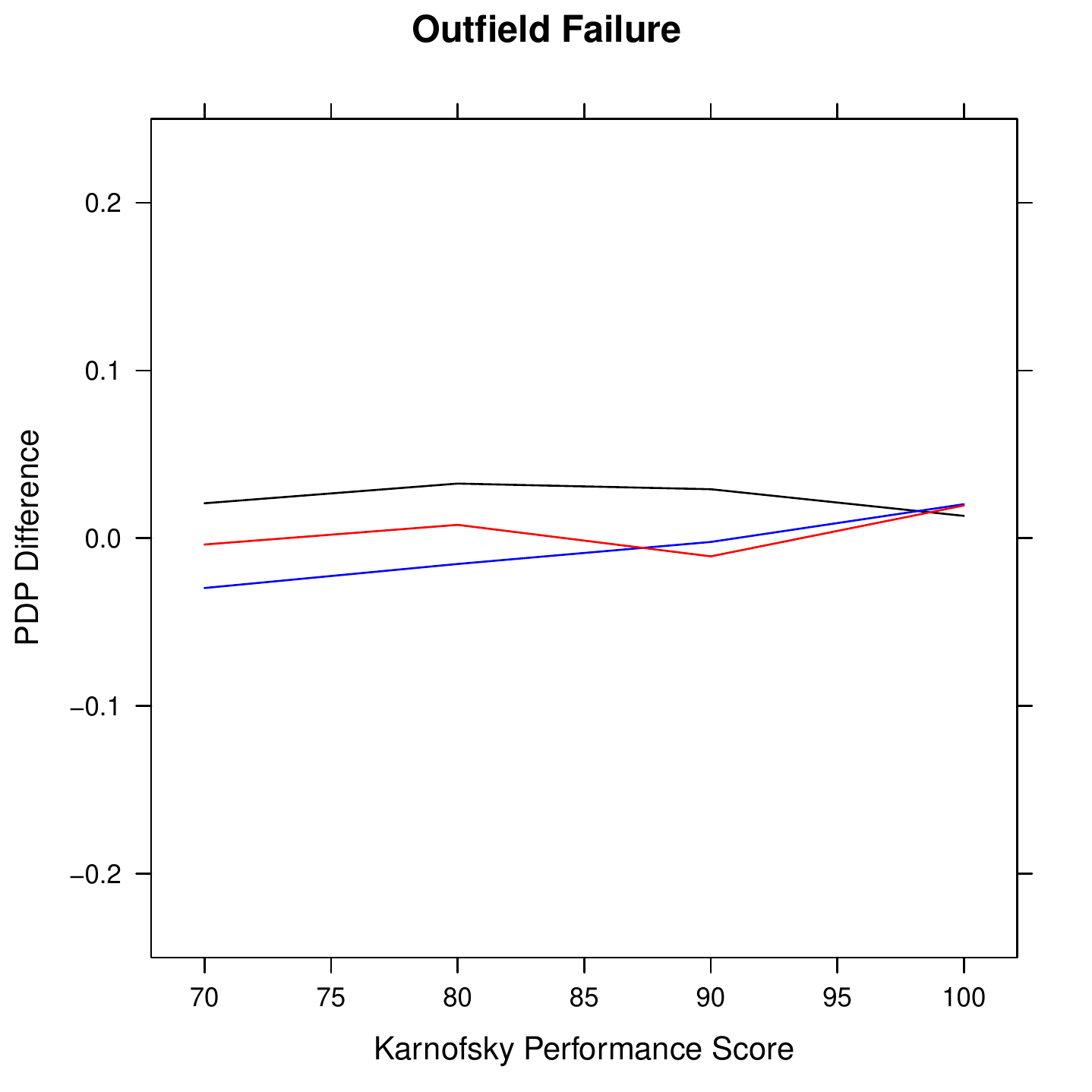}
	\end{minipage}%
	\begin{minipage}{0.5\textwidth}
		\centering
		\includegraphics[width=0.9\textwidth, height=0.9\textwidth]{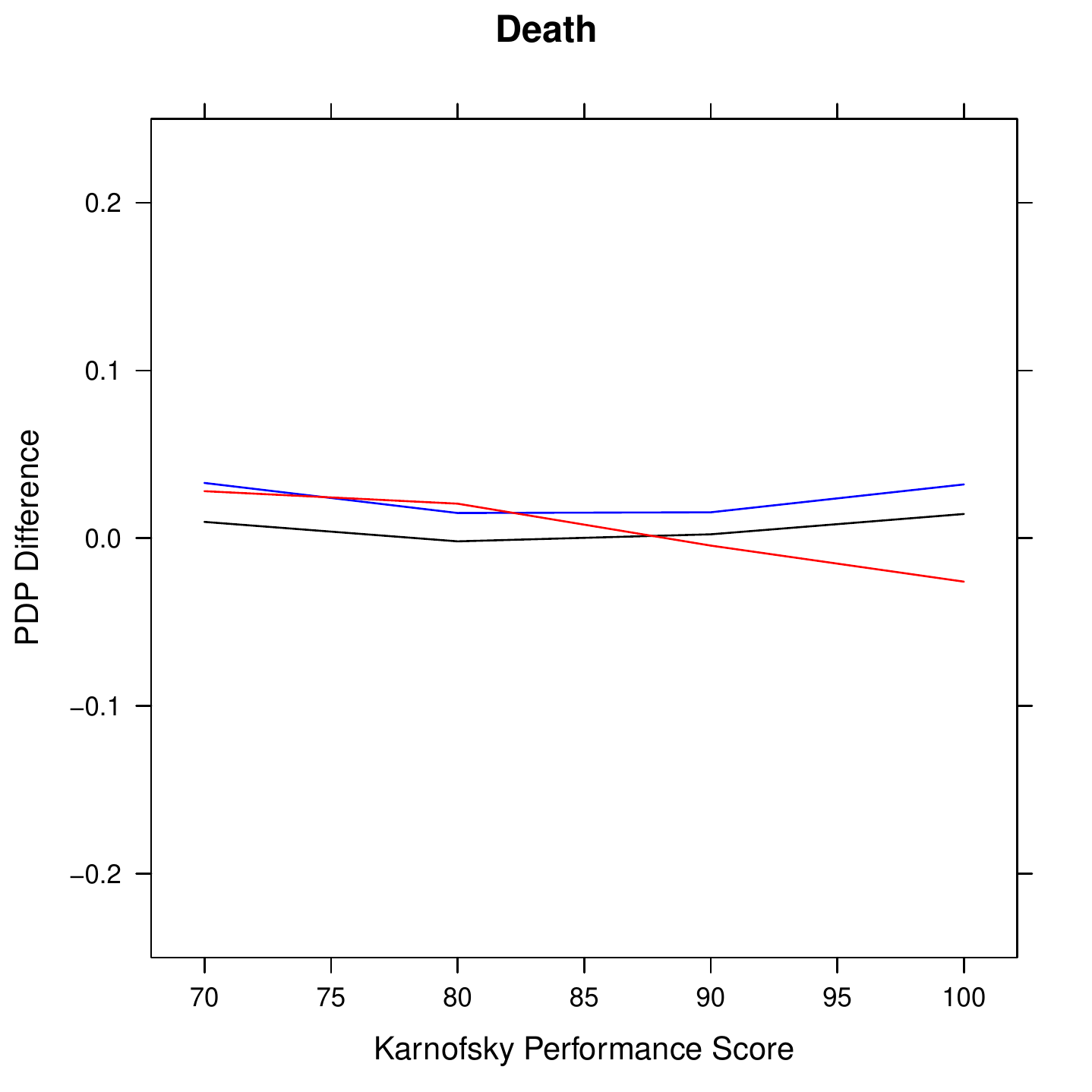}
	\end{minipage}
	\begin{minipage}{.5\textwidth}
		\centering
		\includegraphics[width=0.9\textwidth, height=0.9\textwidth]{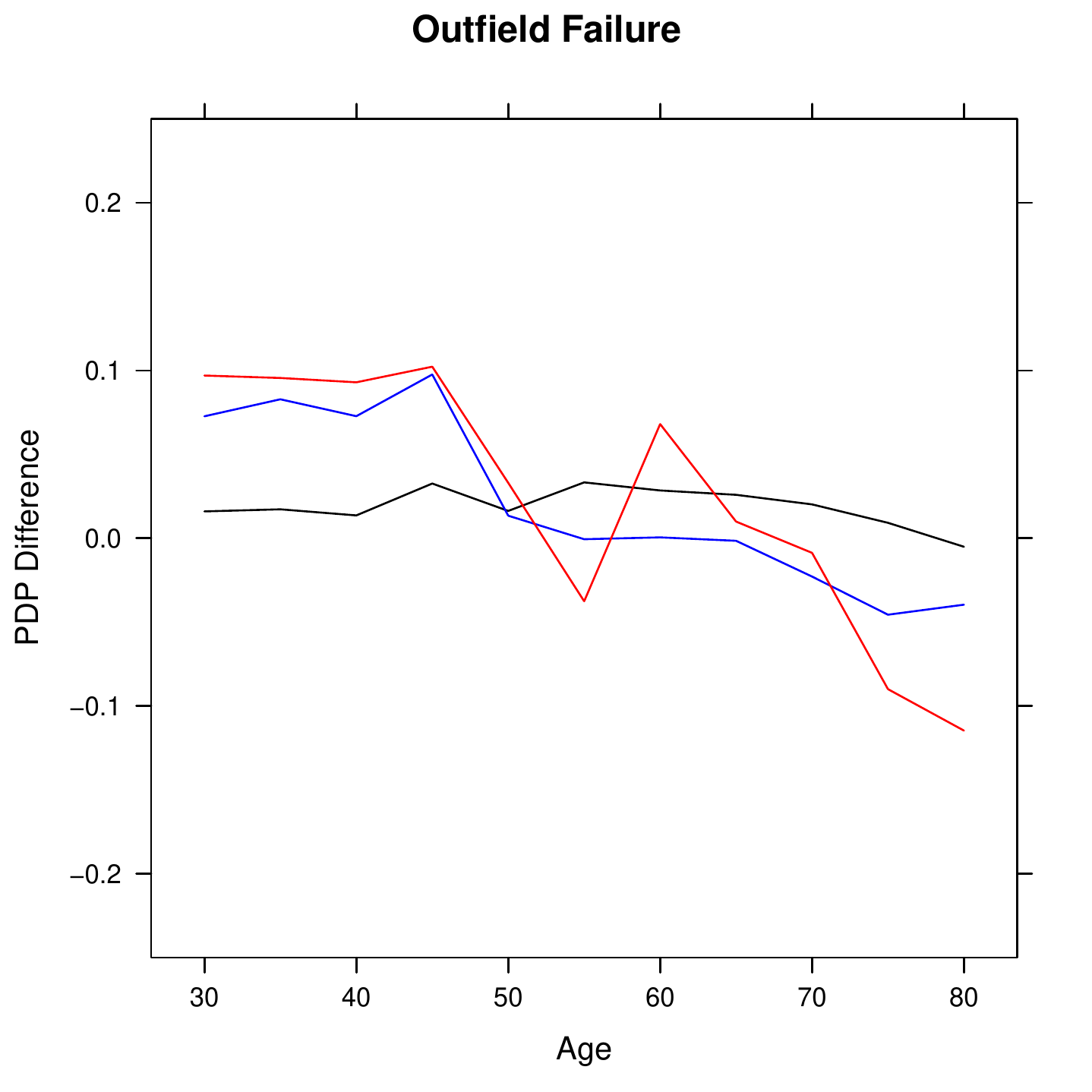}
	\end{minipage}%
	\begin{minipage}{0.5\textwidth}
		\centering
		\includegraphics[width=0.9\textwidth, height=0.9\textwidth]{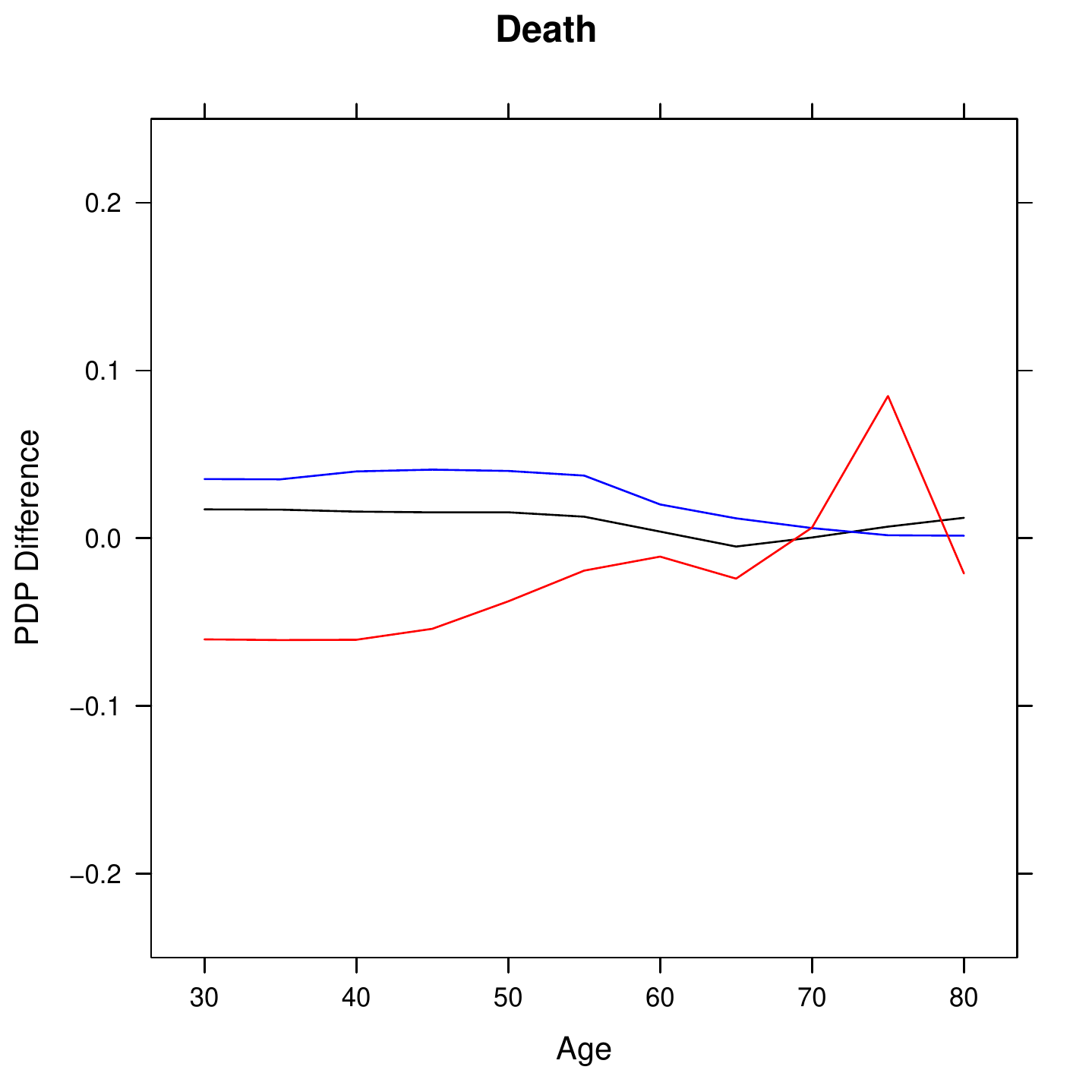}
	\end{minipage}
	\caption{Plots of the difference in partial dependence CIF estimates with respect to KPS (top) and Age (bottom) between uncensored and censored data with outfield failure (left) and death (right) for 50th percentile time (black = BJ; blue = DR; red =  {\tt rfsrc})}
	\label{figPDPdiff}
\end{figure}

In Figure \ref{figPDPdiff}, the difference in the PDPs for KPS and Age that are obtained using censored and uncensored data are compared. Specifically, for \textit{BJ-RF} and \textit{DR-RF}, we calculate the difference compared to the uncensored
estimator obtained using the methods described in Section \ref{implement0}; for {\tt rfsrc}, we compute the estimators obtained 
using the censored and uncensored outcomes and calculate the difference. In general it can be seen that censoring has a minimal
effect on the PDP estimates for all methods, though there is evidence of a somewhat more pronounced difference for outfield 
failure at the lowest and highest ages, particularly for {\tt rfsrc}.  Importantly, however, these results may reflect the 
comparatively small number of patients at these ages (i.e., only 5\% of the patients are under 45, and only 5\% are older than 74).
Overall, \textit{BJ-RF} tends to exhibit the smallest changes when comparing results for censored and uncensored data.

\clearpage

We also compute similar measures for the categorical variables and report those in Tables
\ref{tabA1} and \ref{tabA2} of the Appendix. Similarly to Age and KPS, the proposed methods 
tend to estimate CIFs that are larger than those estimated by {\tt rfsrc} for these data. In
all cases, the estimated CIF values demonstrate monotonicity in time; this is easier to 
see in Tables \ref{tabA1} and \ref{tabA2} than it is in the figures generated
for KPS and Age. In addition, we again observe that the impact of censoring is relatively
small when comparing the results for censored and uncensored data within each method
for estimating the CIF.

The difference in the estimates obtained between {\tt rfsrc} and the proposed methods 
persist whether or not there is censoring present. In the absence of censoring, 
both {\tt rfsrc} and the proposed methods use a bootstrap ensemble of trees; moreover,
for each tree in the ensemble, the CIF is estimated in each terminal node using the corresponding 
average cause-specific number of events in that node. Hence, the differences observed
here in the case of uncensored data, and consequently also in the case of censored data, 
appear to stem directly from the different splitting rules used to build the trees
that make up each ensemble.


\section{Discussion}
\label{discuss}

In our simulation studies, the proposed methods demonstrate similar or better performance 
(i.e., with respect to the chosen MSE measure) compared to the methods of  \cite{ishwaran2014random}, 
as implemented in the {\sf randomForestSRC} package. Of interest is the ease with which the proposed methods 
can be implemented using existing software. This ease extends to the case of regression trees for the CIF, which 
should be of interest to practitioners as (to our knowledge) there is currently no publicly available software package that 
directly implements tree-based regression methods for competing risks.

The proposed methods focus on building a tree or ensemble estimator for a single 
cause in the presence of other possible causes. When interest lies in multiple causes, the method can be applied to each 
cause in the same exact manner. However, this is probably inefficient, and one interesting direction for further research 
would be to extend both regression tree and ensemble procedures to the problem of 
simultaneous estimation of multiple CIFs. For example, when building a regression tree, one could easily adapt the 
approach taken earlier to accommodate multiple causes in addition to multiple time points; see
\cite{ishwaran2014random} for a similar proposal. However, such an approach is not ideal, for it makes the restrictive 
assumption that the predictor space is to be partitioned in the same way for all causes. This restriction may ultimately 
be less worrisome when building a predictor derived from ensembles of trees; nevertheless we conjecture that there may
be better ways in which one might proceed.  A second interesting area of possible extension would be to consider 
generalizations of the model-based recursive partitioning algorithm proposed in \cite{mob} to the setting of
CIF estimation, either as a tree or ensemble estimator.

The reliance of the proposed methods on the need to estimate a ``nuisance'' parameter that coincides with the target
of primary interest (i.e., the CIF) is an evident drawback of the proposed approach. Of course, this problem is inherent to using all augmented IPCW estimators. Currently, we use {\tt rfsrc} for this purpose, and 
it is therefore perfectly reasonable to ask whether the proposed approach has any advantages over the 
methods introduced in \cite{ishwaran2014random}.
We believe the answer to this question is affirmative. In particular, a splitting process that makes use of a
one-variable-at-a-time log-rank-type criteria may be subject to greater bias as a result of informative censoring, 
particularly so in the early stages of splitting; see, for example,
\cite{steingrimsson2016doubly, steingrimsson2019censoring} and especially \cite{cui2019consistency} for discussion
and results in the case of right-censored survival data. Although the validity of the proposed methods also requires that 
$(T,M) \bot C \mid W,$ the observed data loss functions used in Algorithms $M_0$ and $M_1$ are essentially unbiased 
both conditionally and unconditionally on $W$ and should be less susceptible to similar biases. We further believe it would be 
interesting to study the performance of iterated versions of the proposed algorithms in which the CIF required for computing the 
augmentation term is updated with each iteration of the proposed algorithm, possibly only being initialized with {\tt rfsrc}.

	\begin{acknowledgement}
		We thank the NRG Oncology Statistics and Data Management Center for providing de-identified RTOG 9410 clinical trial data
		under a data use agreement. 
	\end{acknowledgement}
	
	\begin{funding}
		This work was partially supported by the National Institutes of Health (R01CA163687: AMM, RLS, YC; U10-CA180822: CH).	
	\end{funding}
	
\bibliographystyle{apalike}
\bibliography{refsmain_revision2020}	

\newpage
\appendix 


\begin{center}
	{\Large \bf
		Supplementary Material for \\
		{\em Regression Trees and Ensembles for Cumulative Incidence Functions} \\[1ex]
		by Cho, Molinaro, Hu and Strawderman \\[2ex]
	}
\end{center}
\section{Appendix}

\setcounter{section}{0}
\renewcommand{\thesection}{A.\arabic{section}}

\numberwithin{equation}{section}
\setcounter{equation}{0} \renewcommand{\theequation}{A.\arabic{equation}} 

\setcounter{table}{0}
\renewcommand{\thetable}{A.\arabic{table}}

\setcounter{figure}{0}
\renewcommand{\thefigure}{A.\arabic{figure}}

References to figures and tables preceded by ``A.'' are internal to this appendix; all
other references refer to the main paper.

\section{Additional simulation results}
In this section, we summarize some additional simulation results using the same setting as one in the main paper,
except that $\boldsymbol{X} = (X_1,\ldots,X_{20}) \sim N(0,\Sigma)$, where $\Sigma$ has elements 
$\sigma_{ij}, \sigma_{ij} = 0.75^{|i-j|}, i \ne j$. Here we only display results for Event 1; the results for both events $m=1$ and $m=2$ 
are very similar to case of independent covariates. See the corresponding captions in Figures 1-8 of the main paper 
to interpret the labels on the x-axis that denote the methods used.

\begin{figure}[!htb]
	\centering
	\includegraphics[width=0.8\textwidth, height=0.4\textheight]{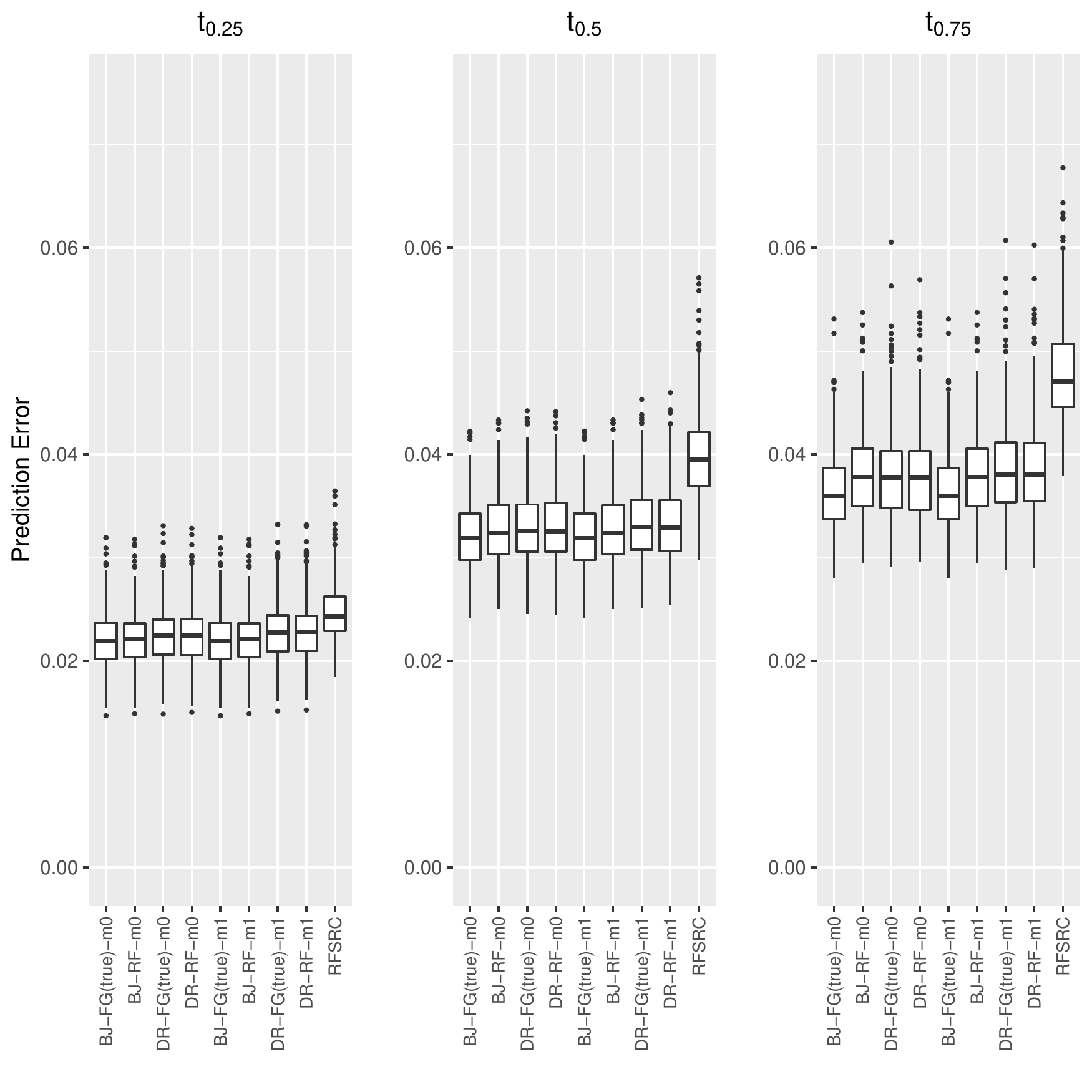}
	\caption{Comparing Algorithms $M_0$ and $M_1$ with Fixed Tuning Parameters for Event 1 with dependent covariates.}
	\label{Afig:figm1}
\end{figure}
\hspace*{-10mm}
\begin{figure}[!htb]
	\centering
	\includegraphics[width=0.8\textwidth, height=0.4\textheight]{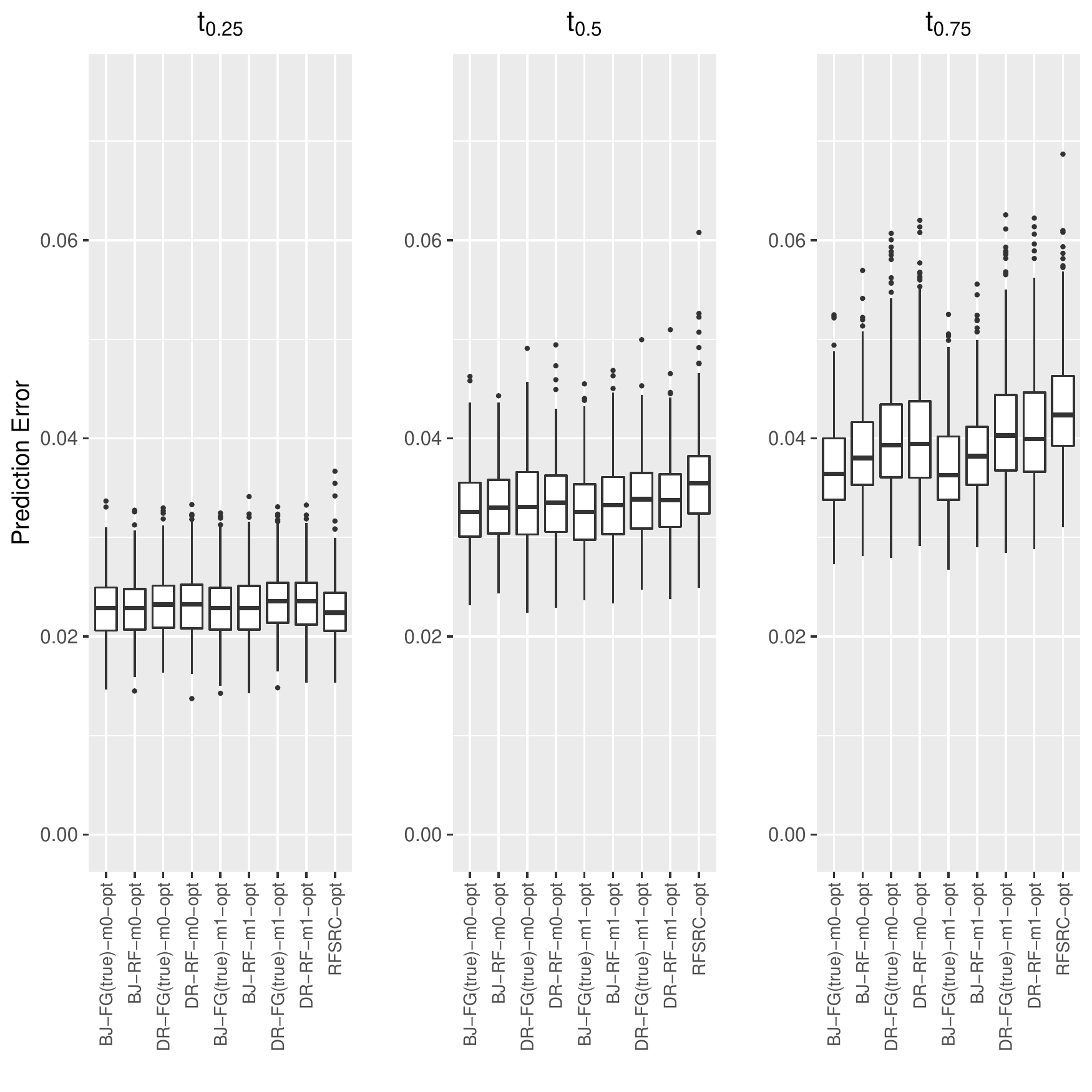}
	\caption{Comparing Algorithms $M_0$ and $M_1$ with Optimized Tuning Parameters for Event 1 with dependent covariates.}
	\label{Afig:figm2}
\end{figure}

\begin{figure}[!htb]
	\centering
	\includegraphics[width=0.8\textwidth, height=0.4\textheight]{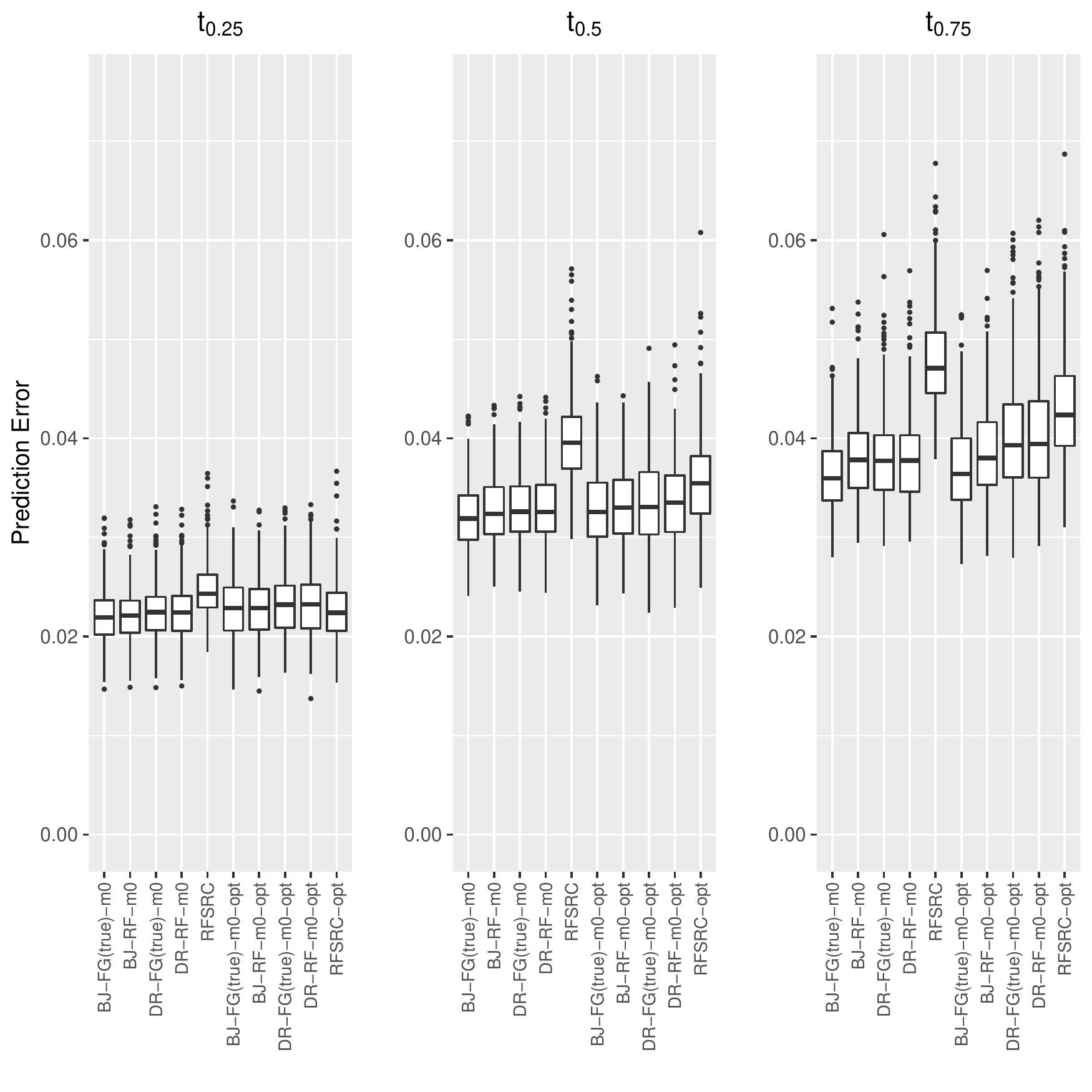}
	\caption{Comparing Fixed and Optimized Tuning Parameters for Algorithm $M_0$ for Event 1 with dependent covariates.}
	\label{Afig:figm3}
\end{figure}
\hspace*{-10mm}
\begin{figure}[!htb]
	\centering
	\includegraphics[width=0.8\textwidth, height=0.4\textheight]{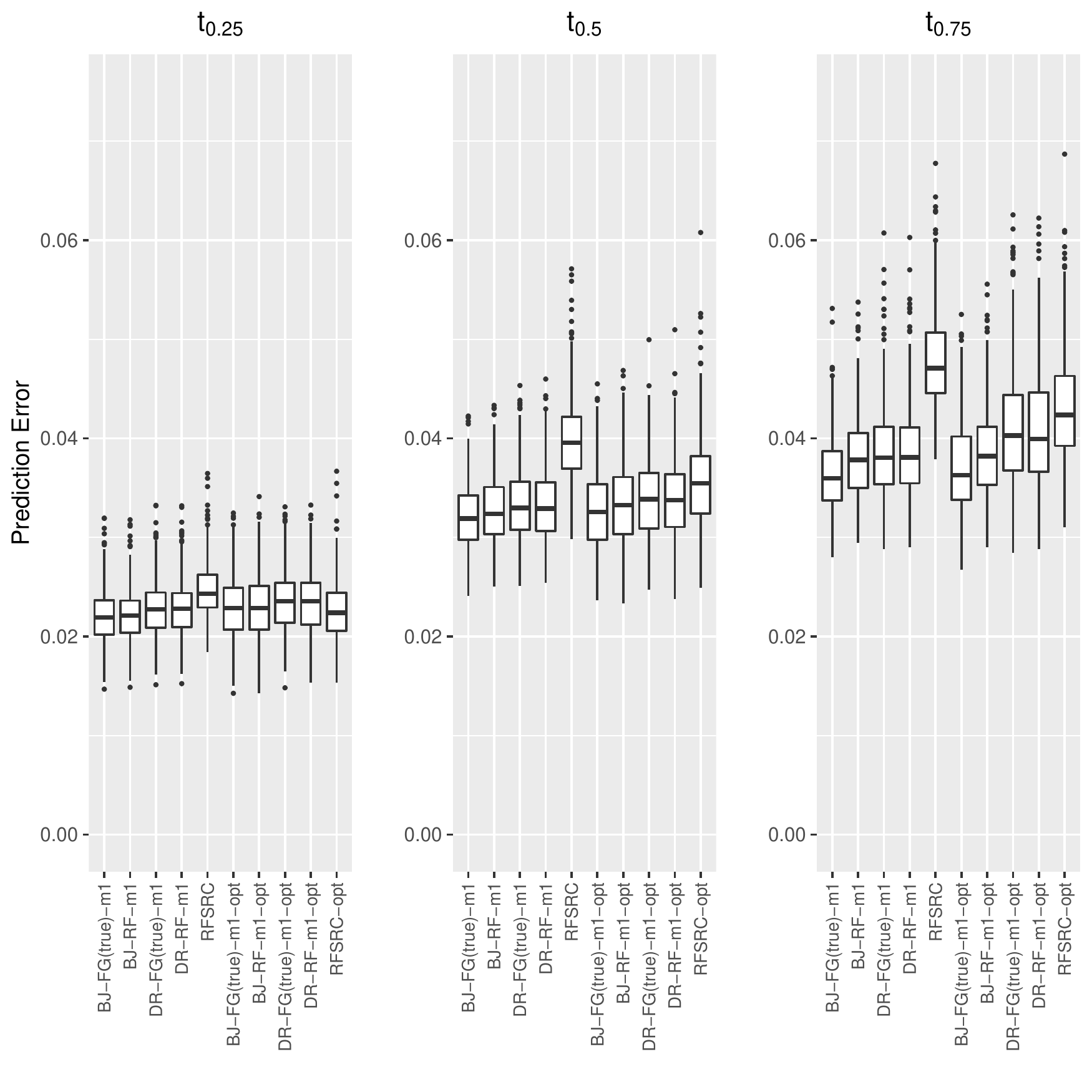}
	\caption{Comparing Fixed and Optimized Tuning Parameters for Algorithm $M_1$ for Event 1 with dependent covariates.}
	\label{Afig:figm4}
\end{figure}

\clearpage

\section{Additional data analysis}
Partial dependence plots for KPS and Age at the 25th and 75th percentile of (marginal) failure time, as well as difference plots, were also created and show a similar pattern to 50th percentile of failure time as summarized in the main paper. 

\begin{figure}[!htb]
	\centering
	\begin{minipage}{.5\textwidth}
		\centering
		\includegraphics[width=0.9\textwidth, height=0.9\textwidth]{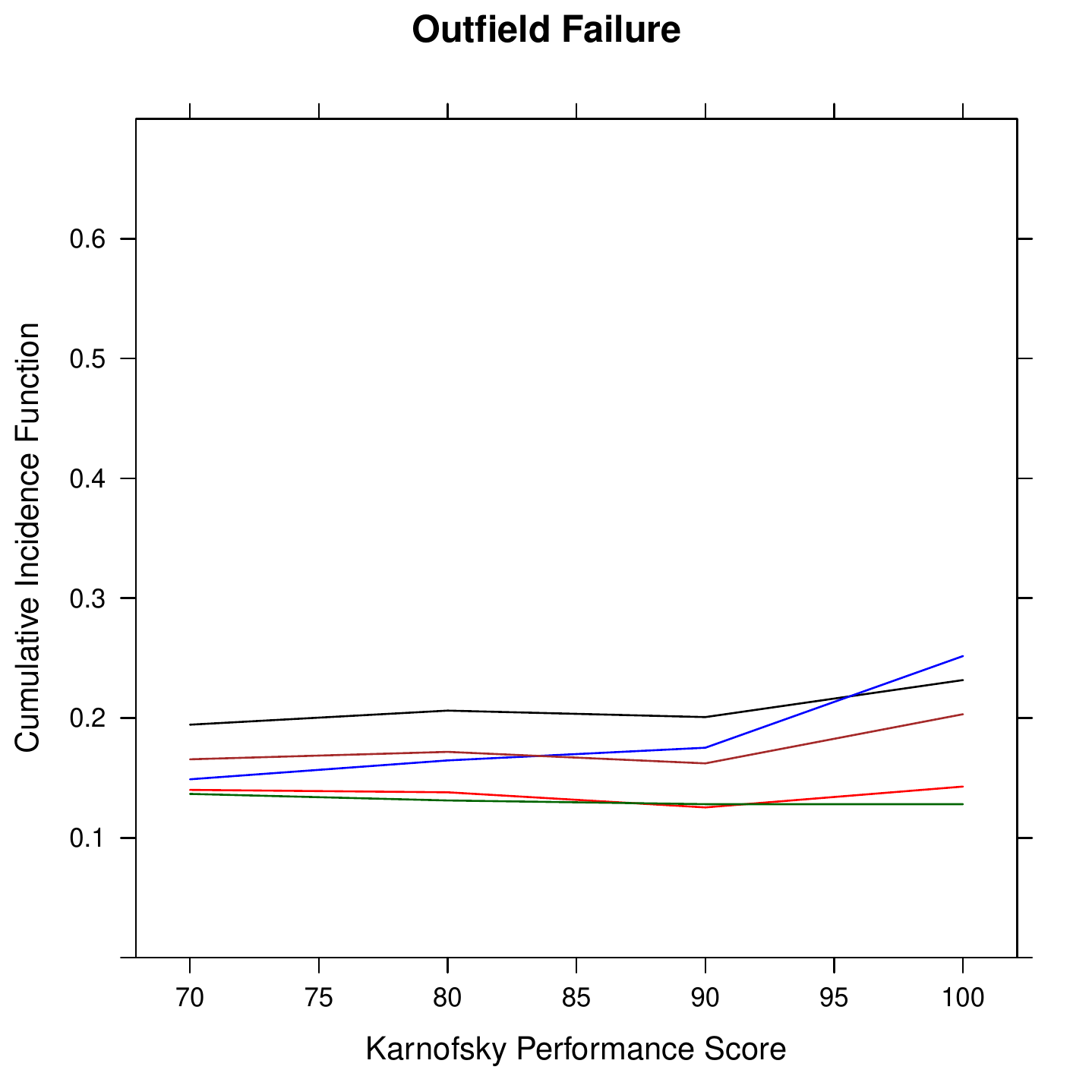}
	\end{minipage}%
	\begin{minipage}{0.5\textwidth}
		\centering
		\includegraphics[width=0.9\textwidth, height=0.9\textwidth]{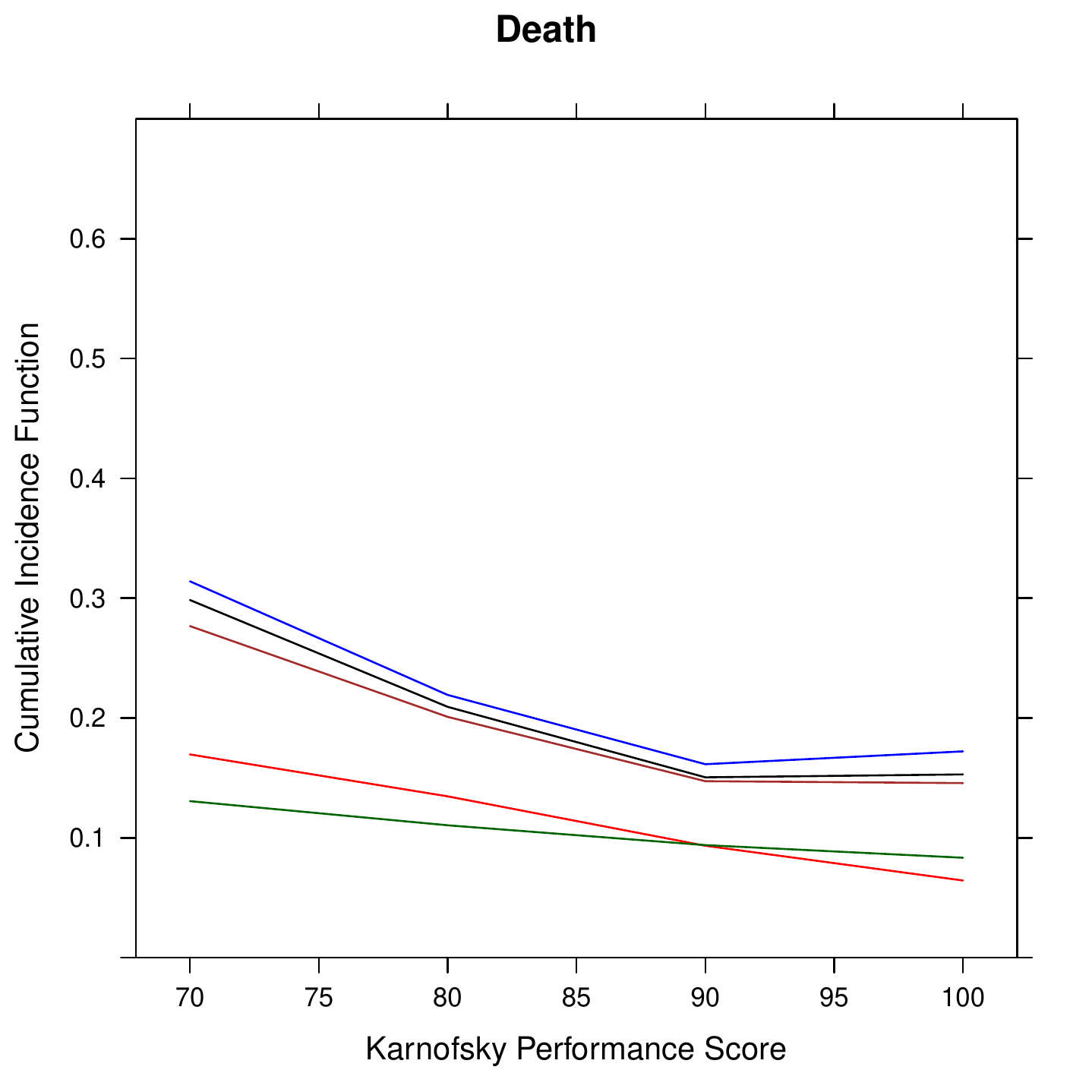}
	\end{minipage}
	\begin{minipage}{.5\textwidth}
		\centering
		\includegraphics[width=0.9\textwidth, height=0.9\textwidth]{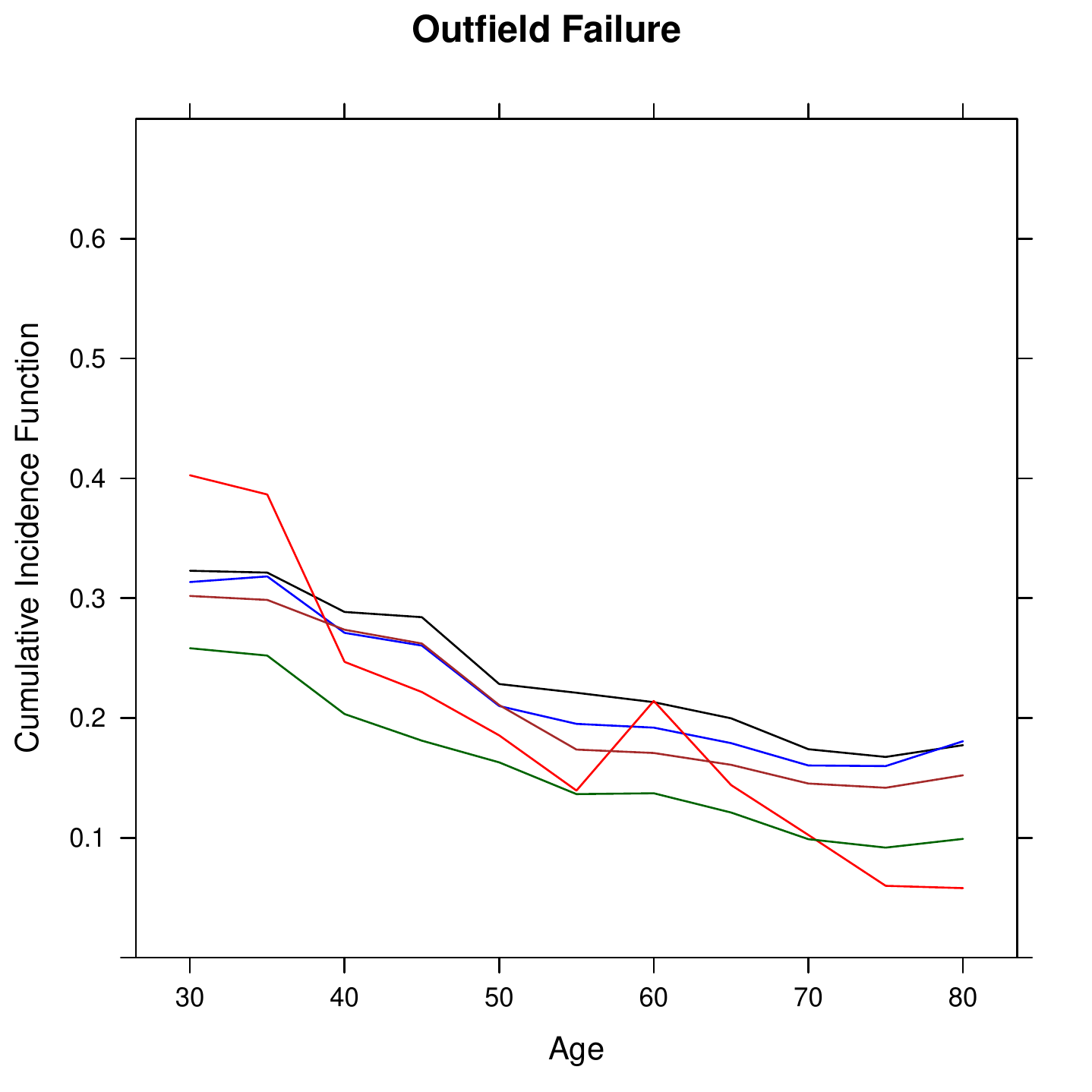}
	\end{minipage}%
	\begin{minipage}{0.5\textwidth}
		\centering
		\includegraphics[width=0.9\textwidth, height=0.9\textwidth]{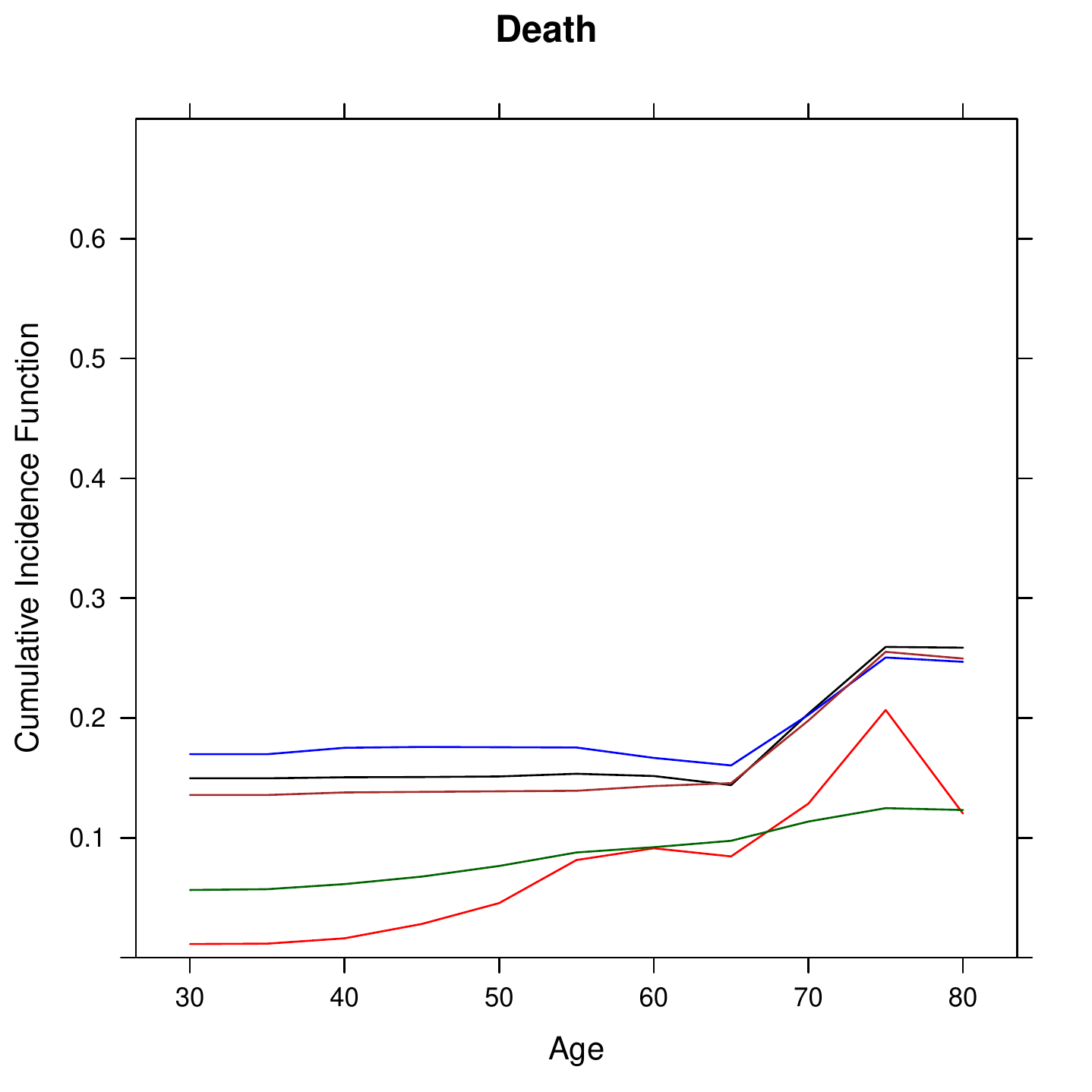}
	\end{minipage}
	\caption{Partial dependence plots with respect to KPS (top) and Age (bottom) from forests with outfield failure (left) and death (right) using artificially censored data and uncensored data for 25th percentile time. The 5 lines are as follows:  black = BJ; blue = DR; red = {\tt rfsrc}(censored); brown = uncensored data estimate using proposed methods in Section \ref{implement0};  
		dark green = uncensored data estimate using {\tt rfsrc}.}
	\label{figPDPA1}
\end{figure}    

\begin{figure}[!htb]
	\centering
	\begin{minipage}{.5\textwidth}
		\centering
		\includegraphics[width=0.9\textwidth, height=0.9\textwidth]{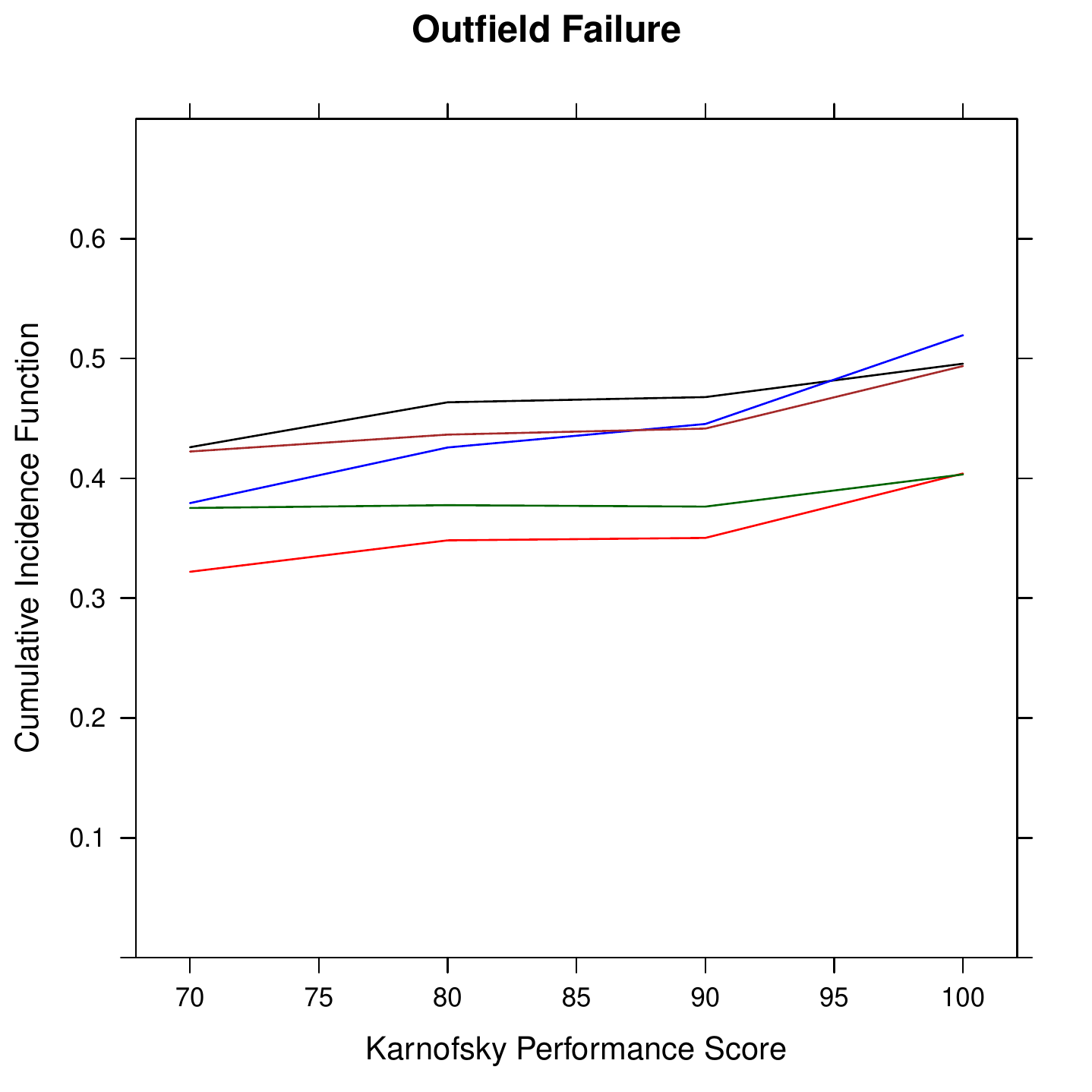}
	\end{minipage}%
	\begin{minipage}{0.5\textwidth}
		\centering
		\includegraphics[width=0.9\textwidth, height=0.9\textwidth]{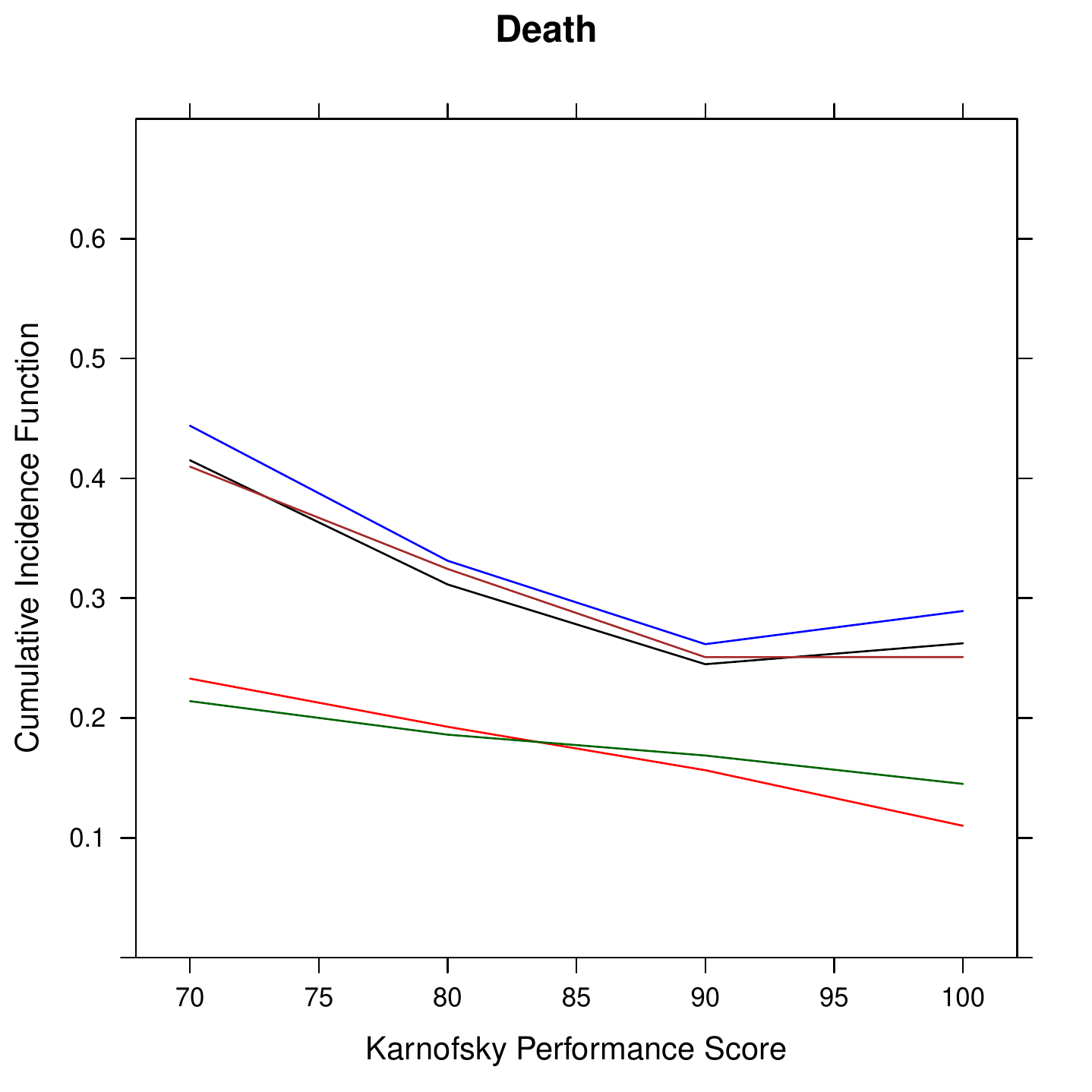}
	\end{minipage}
	\begin{minipage}{.5\textwidth}
		\centering
		\includegraphics[width=0.9\textwidth, height=0.9\textwidth]{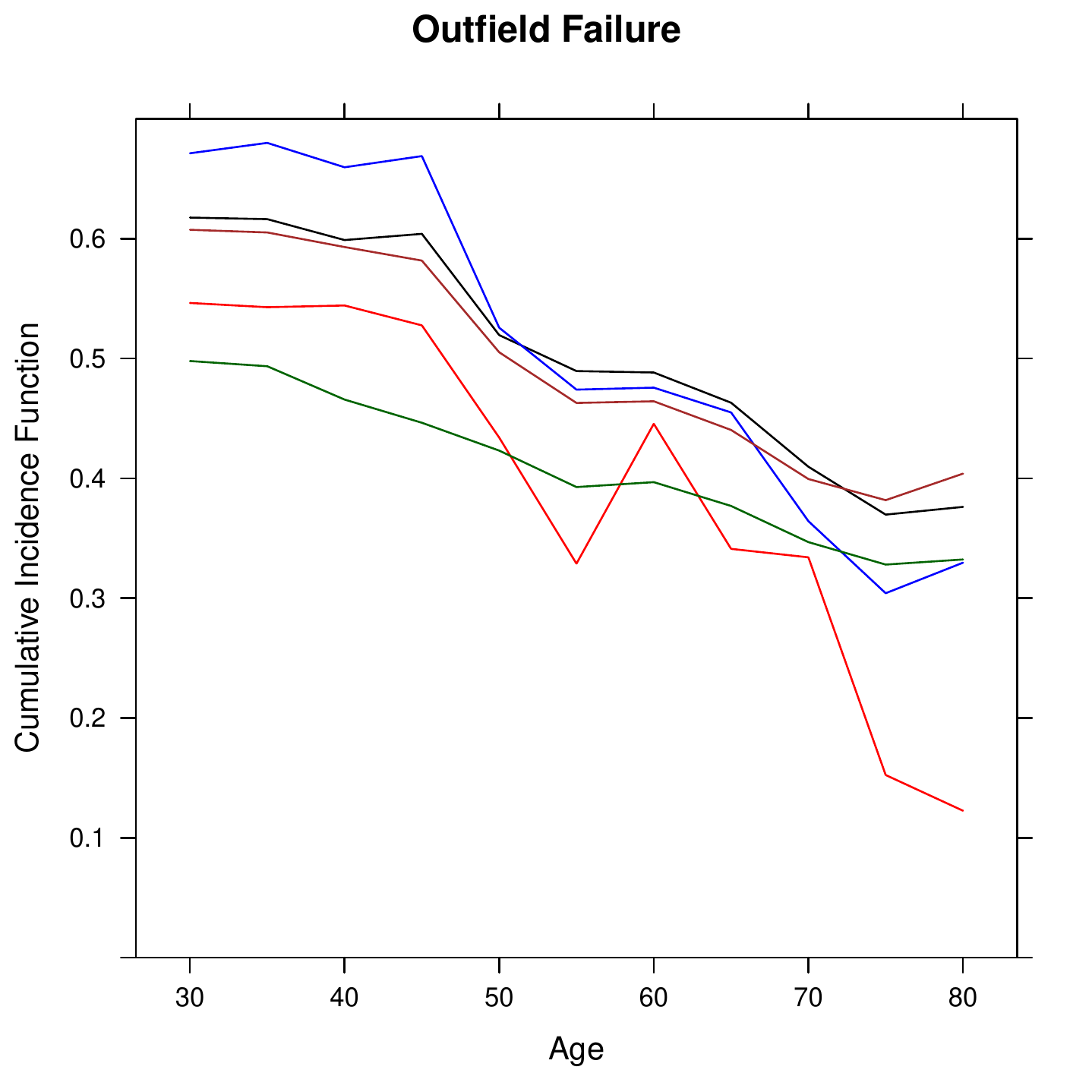}
	\end{minipage}%
	\begin{minipage}{0.5\textwidth}
		\centering
		\includegraphics[width=0.9\textwidth, height=0.9\textwidth]{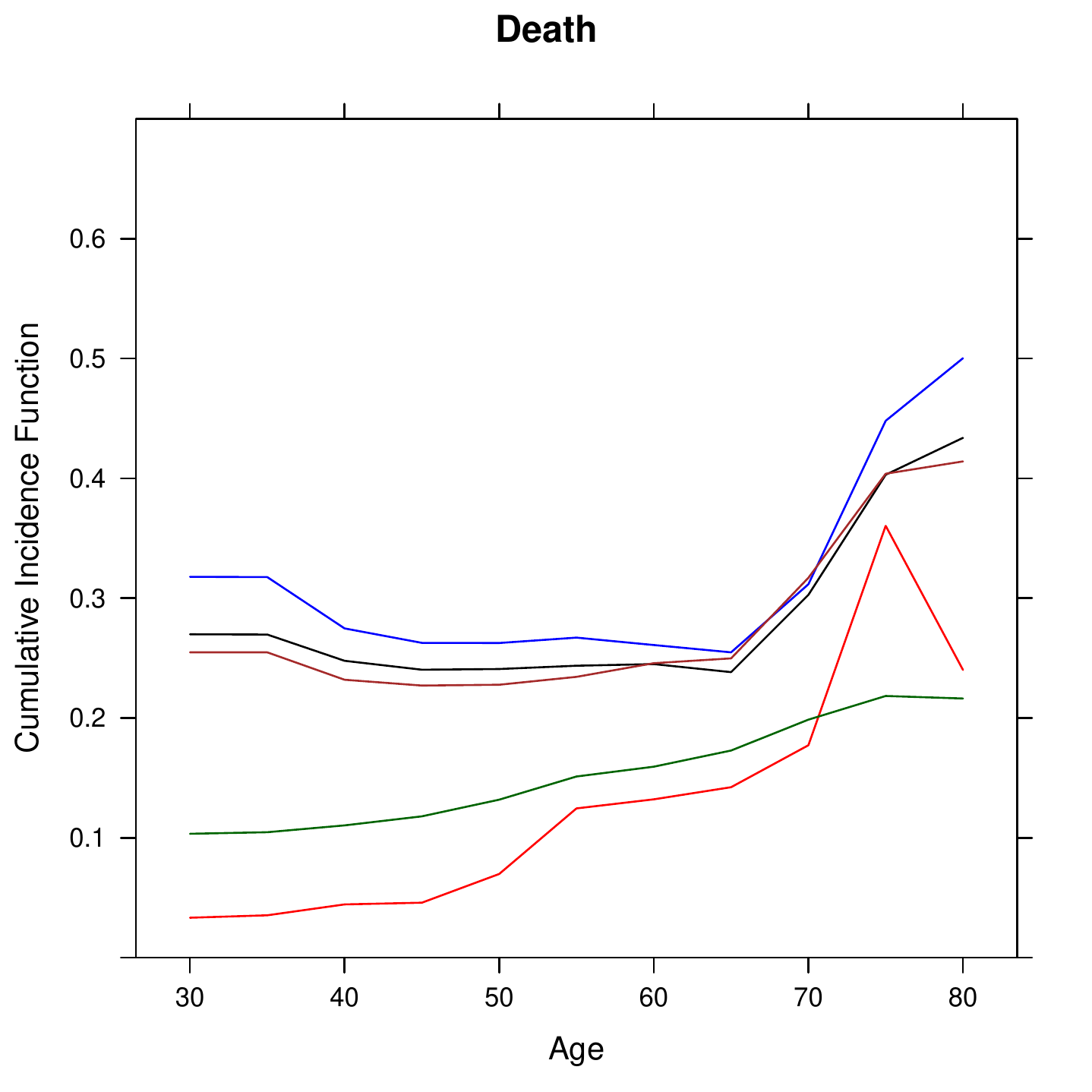}
	\end{minipage}
	\caption{Partial dependence plots with respect to KPS (top) and Age (bottom) from forests with outfield failure (left) and death (right) using artificially censored data and uncensored data for 75th percentile time. The 5 lines are as follows:  black = BJ; blue = DR; red = {\tt rfsrc}(censored); brown = uncensored data estimate using proposed methods in Section \ref{implement0};  
		dark green = uncensored data estimate using {\tt rfsrc}.}
	\label{figPDPA2}
\end{figure}    

\begin{figure}[!htb]    
	\begin{minipage}{.5\textwidth}
		\centering
		\includegraphics[width=0.9\textwidth, height=0.9\textwidth]{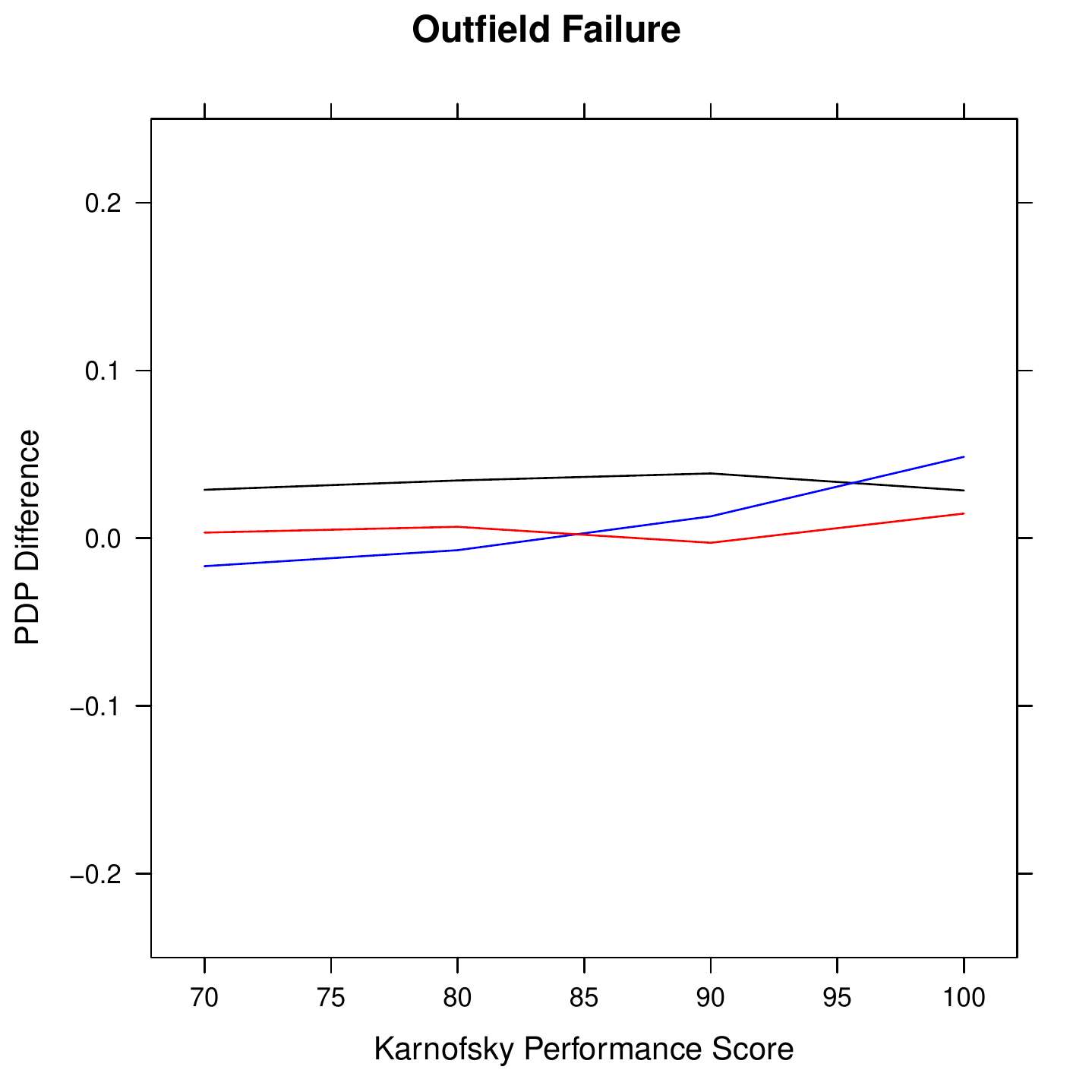}
	\end{minipage}%
	\begin{minipage}{0.5\textwidth}
		\centering
		\includegraphics[width=0.9\textwidth, height=0.9\textwidth]{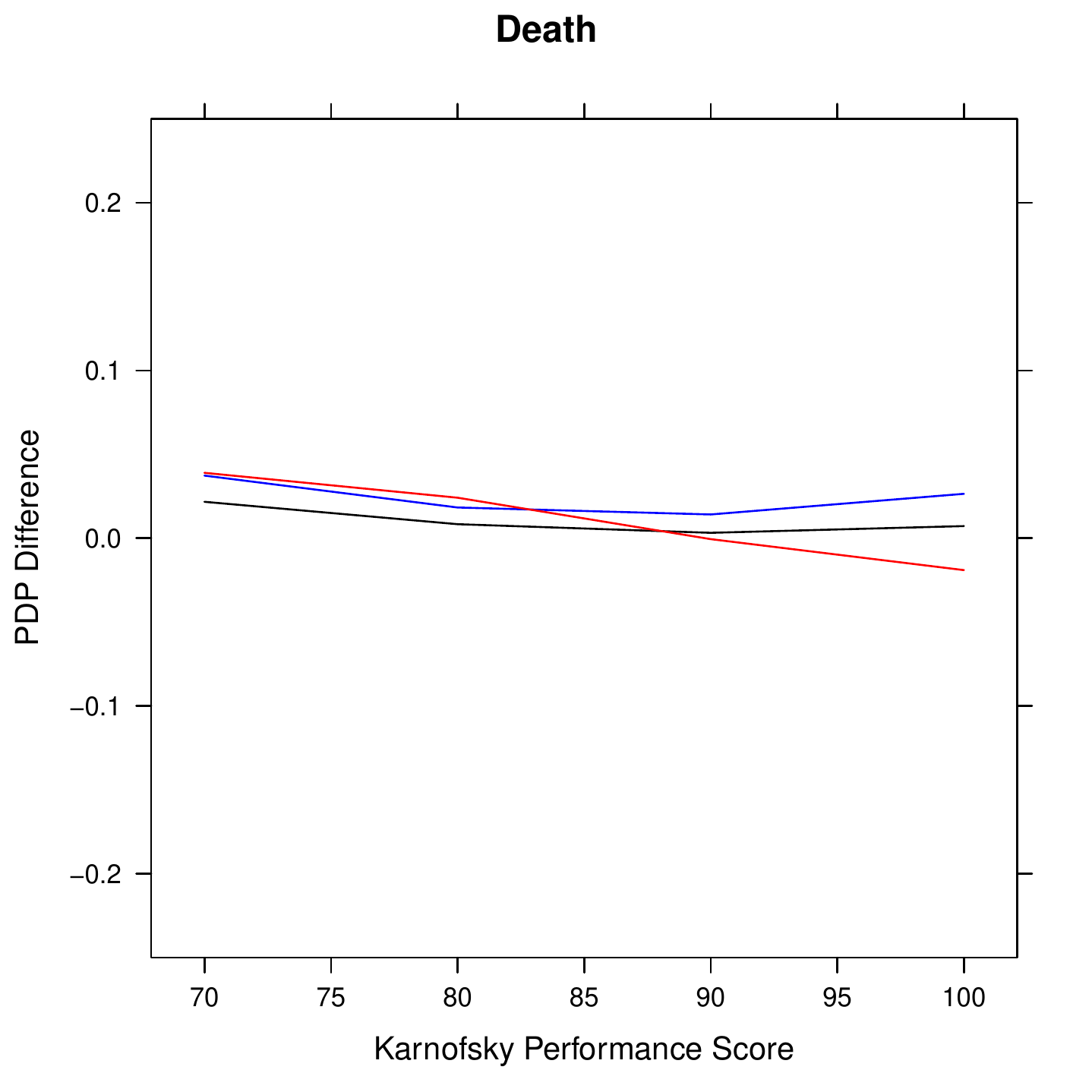}
	\end{minipage}
	\begin{minipage}{.5\textwidth}
		\centering
		\includegraphics[width=0.9\textwidth, height=0.9\textwidth]{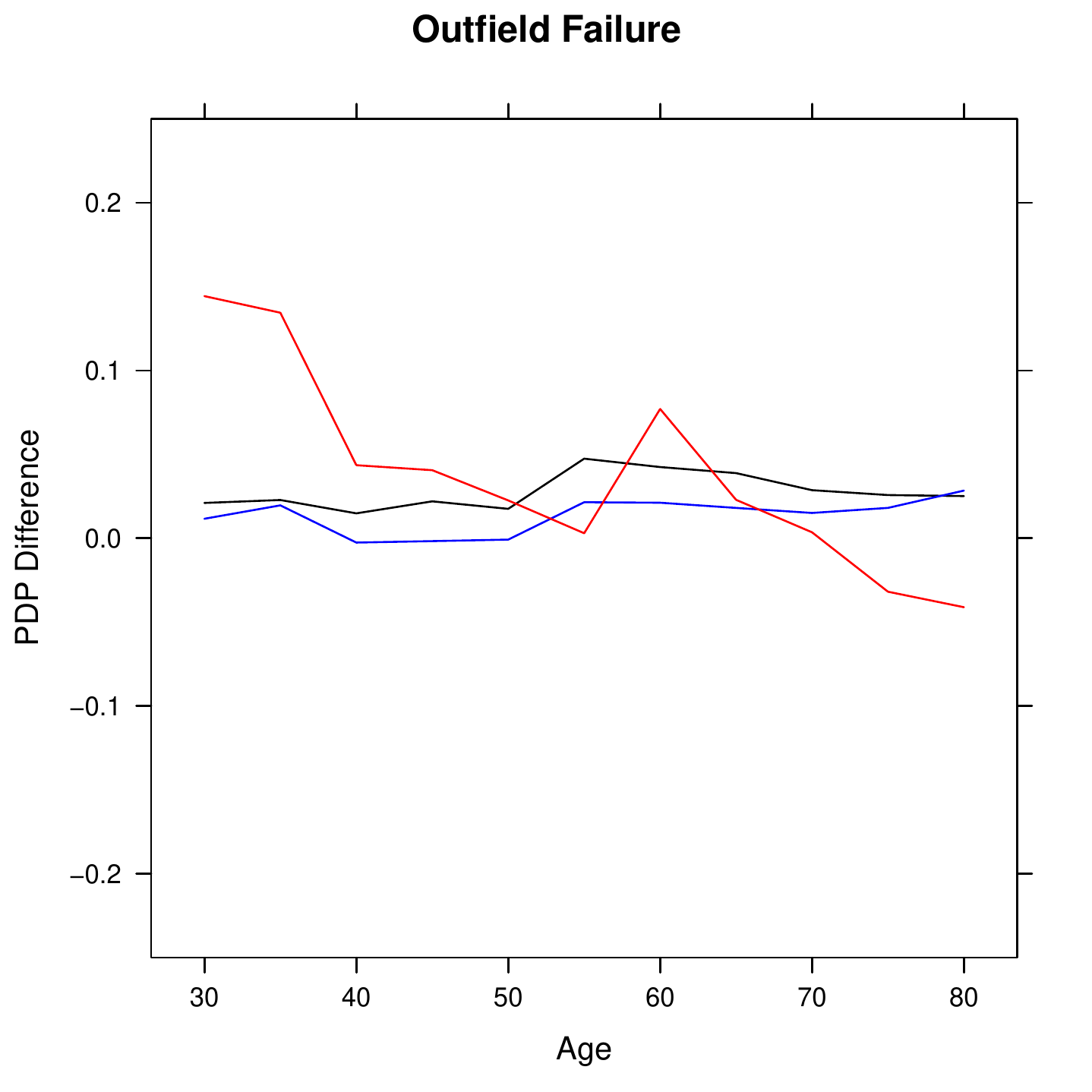}
	\end{minipage}%
	\begin{minipage}{0.5\textwidth}
		\centering
		\includegraphics[width=0.9\textwidth, height=0.9\textwidth]{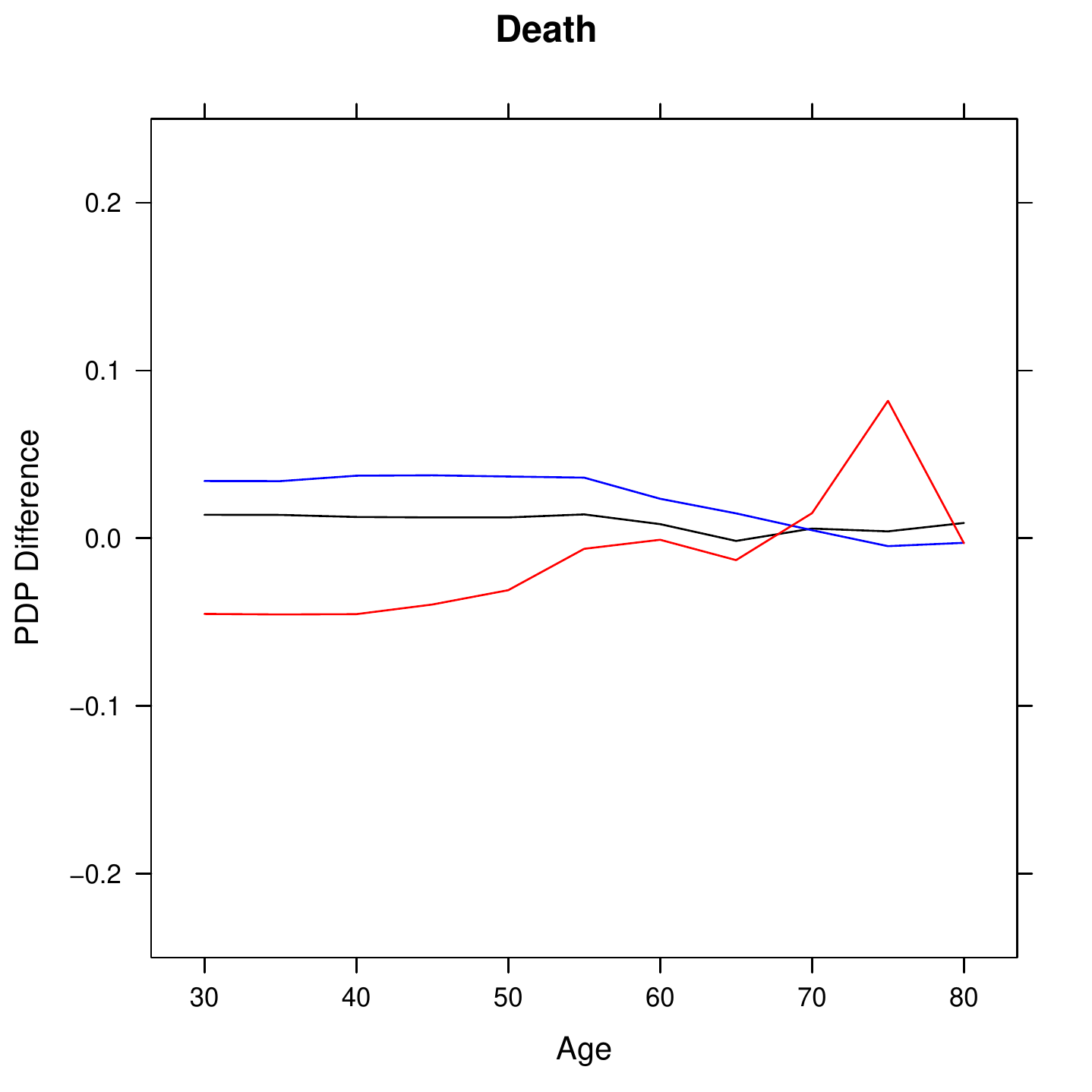}
	\end{minipage}
	\caption{Plots for the difference in partial dependence CIF estimates with respect to KPS (top) and Age (bottom) between uncensored and censored data with outfield failure (left) and death (right) for 25th percentile time (black : BJ, blue : DR, red : {\tt rfsrc})}
	\label{figPDPA3}
\end{figure}

\begin{figure}[!htb]    
	\begin{minipage}{.5\textwidth}
		\centering
		\includegraphics[width=0.9\textwidth, height=0.9\textwidth]{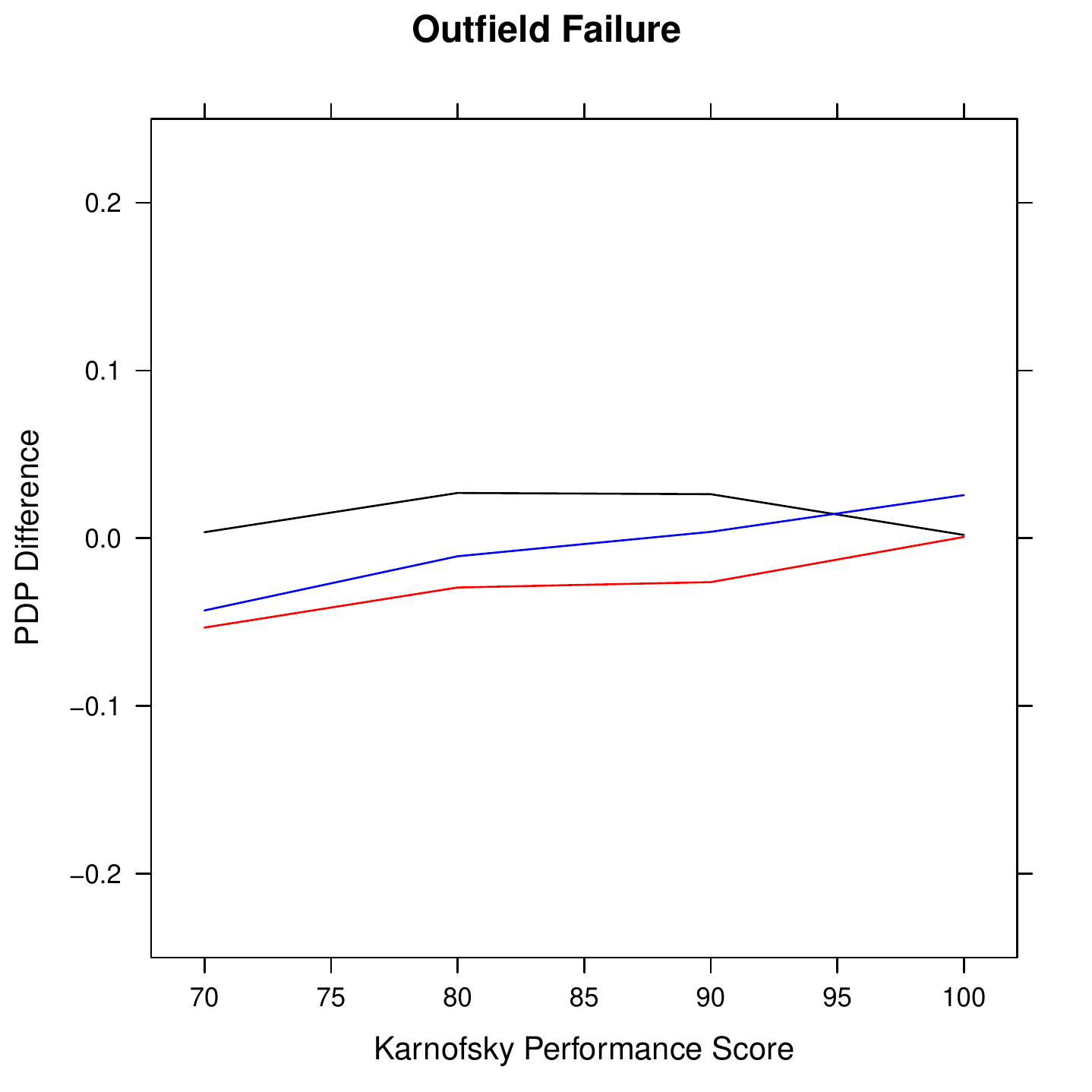}
	\end{minipage}%
	\begin{minipage}{0.5\textwidth}
		\centering
		\includegraphics[width=0.9\textwidth, height=0.9\textwidth]{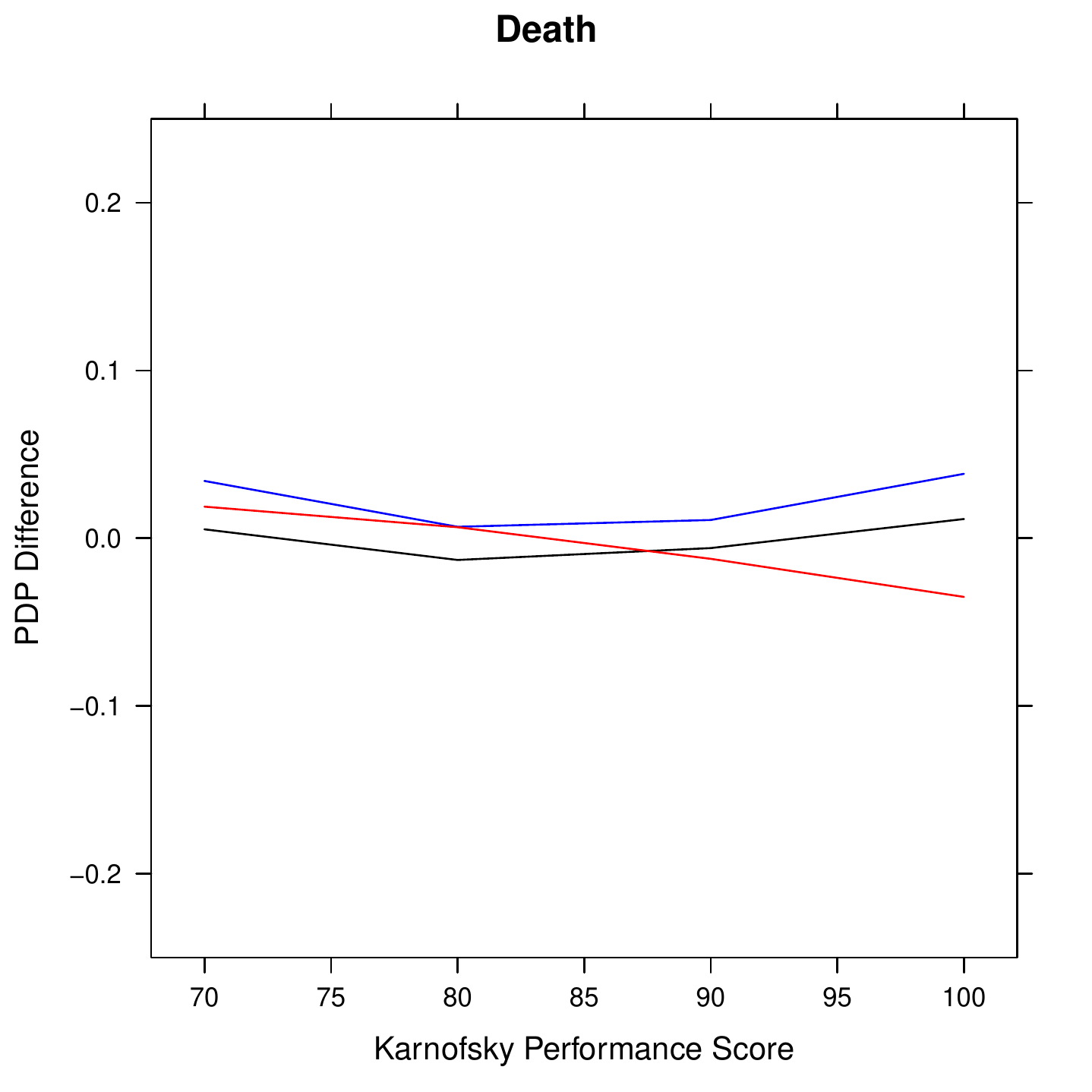}
	\end{minipage}
	\begin{minipage}{.5\textwidth}
		\centering
		\includegraphics[width=0.9\textwidth, height=0.9\textwidth]{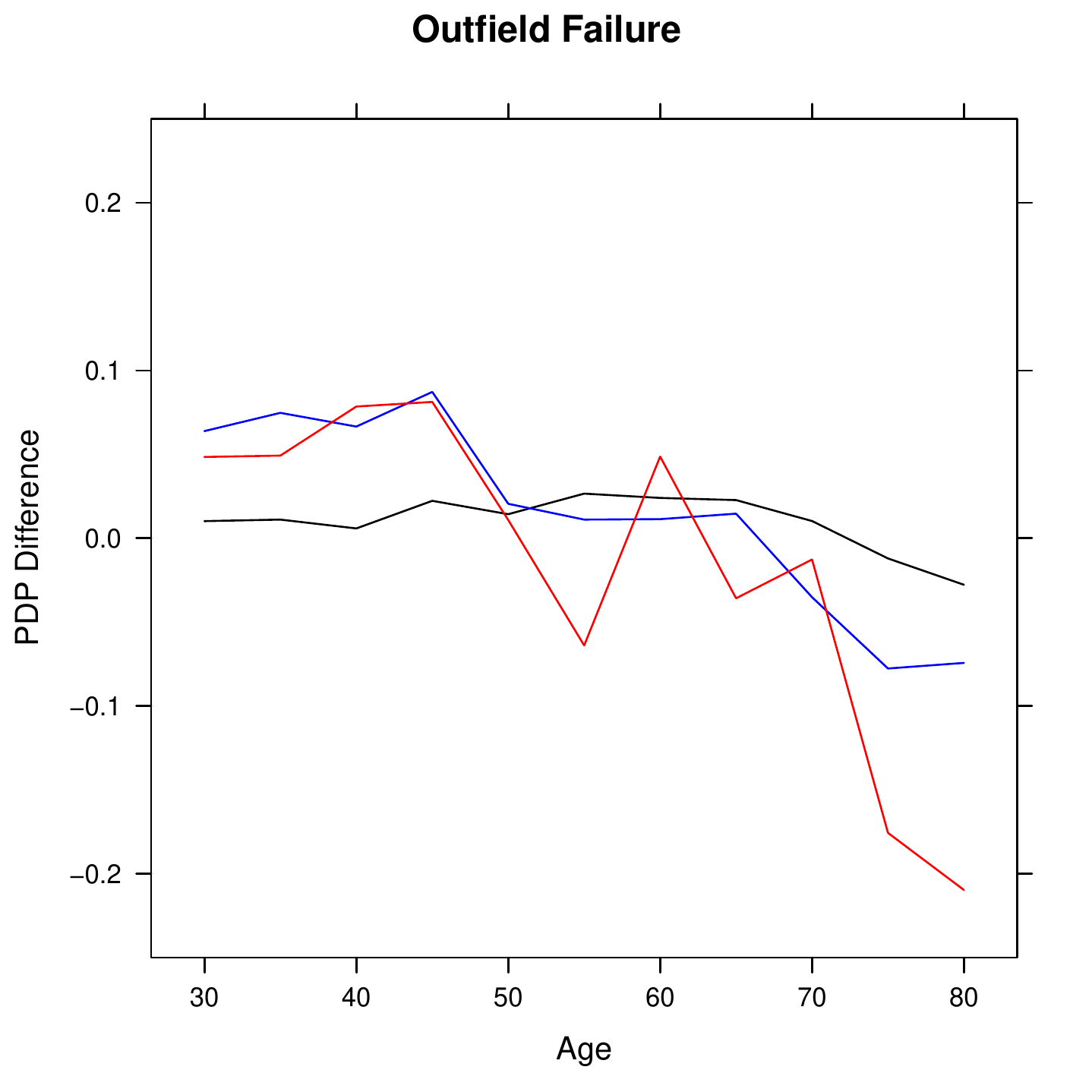}
	\end{minipage}%
	\begin{minipage}{0.5\textwidth}
		\centering
		\includegraphics[width=0.9\textwidth, height=0.9\textwidth]{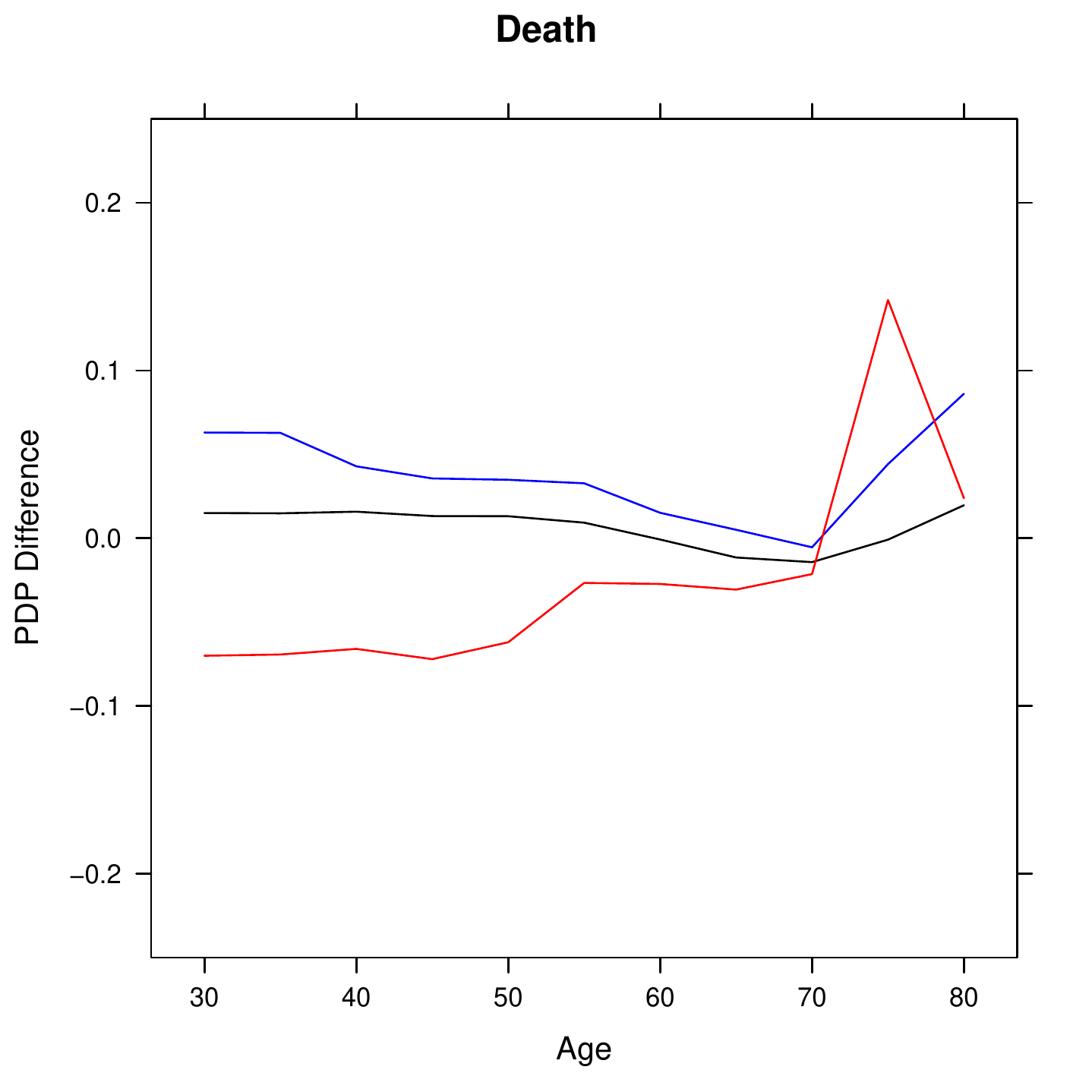}
	\end{minipage}
	\caption{Plots for the difference in partial dependence CIF estimates with respect to KPS (top) and Age (bottom) between uncensored and censored data with outfield failure (left) and death (right) for 85th percentile time (black : BJ, blue : DR, red : {\tt rfsrc})}
	\label{figPDPA4}
\end{figure}
\clearpage

Below, we summarize the PDPs for all categorical variables in Tables \ref{tabA1} and \ref{tabA2}. 

\begin{table}[!htb]
	{\small
		\centering
		\begin{tabular}{rrrrrrrrrrrrrr}
			\hline
			&& \multicolumn{3}{c}{RX} & \multicolumn{2}{c}{AJCC} & \multicolumn{2}{c}{Gender} & \multicolumn{2}{c}{Race} & \multicolumn{2}{c}{Histology} \\
			&& \multicolumn{1}{c}{1} & 
			\multicolumn{1}{c}{2} &
			\multicolumn{1}{c}{3} &
			\multicolumn{1}{c}{0} &
			\multicolumn{1}{c}{1} &
			\multicolumn{1}{c}{0} &
			\multicolumn{1}{c}{1} &
			\multicolumn{1}{c}{0} &
			\multicolumn{1}{c}{1} &
			\multicolumn{1}{c}{0} &
			\multicolumn{1}{c}{1} \\ \hline
			& $t_1$ & 0.197 & 0.223 & 0.207 & 0.190 & 0.223 & 0.226 & 0.198 & 0.185 & 0.212 & 0.217 & 0.194 \\ 
			BJ-RF & $t_2$ & 0.363 & 0.393 & 0.372 & 0.353 & 0.394 & 0.400 & 0.362 & 0.333 & 0.383 & 0.386 & 0.359 \\ 
			& $t_3$ & 0.456 & 0.494 & 0.467 & 0.452 & 0.487 & 0.501 & 0.456 & 0.417 & 0.481 & 0.479 & 0.460 \\  \hline
			&$t_1$ & 0.178 & 0.206 & 0.195 & 0.168 & 0.208 & 0.189 & 0.191 & 0.154 & 0.196 & 0.195 & 0.182 \\ 
			DR-RF &$t_2$ & 0.336 & 0.373 & 0.347 & 0.316 & 0.374 & 0.383 & 0.332 & 0.269 & 0.362 & 0.365 & 0.325 \\ 
			&$t_3$ & 0.436 & 0.484 & 0.455 & 0.427 & 0.477 & 0.502 & 0.433 & 0.351 & 0.473 & 0.471 & 0.433 \\ \hline
			& $t_1$ & 0.140 & 0.144 & 0.122 & 0.102 & 0.153 & 0.146 & 0.124 & 0.112 & 0.132 & 0.155 & 0.097 \\ 
			{\tt rfsrc}  & $t_2$ & 0.286 & 0.279 & 0.259 & 0.235 & 0.298 & 0.305 & 0.254 & 0.239 & 0.273 & 0.303 & 0.223 \\ 
			& $t_3$ & 0.368 & 0.364 & 0.354 & 0.337 & 0.377 & 0.392 & 0.345 & 0.304 & 0.367 & 0.394 & 0.307 \\ \hline \hline
			& $t_1$ & 0.160 & 0.192 & 0.171 & 0.164 & 0.181 & 0.174 & 0.174 & 0.151 & 0.178 & 0.183 & 0.158 \\ 
			RF & $t_2$ & 0.333 & 0.361 & 0.359 & 0.331 & 0.365 & 0.364 & 0.344 & 0.299 & 0.359 & 0.365 & 0.327 \\ 
			& $t_3$ & 0.436 & 0.464 & 0.460 & 0.433 & 0.467 & 0.473 & 0.443 & 0.408 & 0.460 & 0.467 & 0.430 \\ \hline
			& $t_1$ & 0.133 & 0.133 & 0.122 & 0.114 & 0.140 & 0.138 & 0.125 & 0.123 & 0.130 & 0.148 & 0.098 \\ 
			{\tt rfsrc}  & $t_2$ & 0.277 & 0.276 & 0.266 & 0.251 & 0.290 & 0.292 & 0.264 & 0.254 & 0.276 & 0.296 & 0.237 \\ 
			& $t_3$ & 0.388 & 0.376 & 0.385 & 0.365 & 0.397 & 0.401 & 0.375 & 0.373 & 0.385 & 0.405 & 0.349 \\ \hline
		\end{tabular}
		\caption{Summary of PDPs for categorical variables for out-field failure. For AJCC, 1 = Stage IIB; Gender, 1 = Male; 
			Race, 1 = White; and Histology, 1 = Squamous. Results above the double horizontal line correspond to 
			artificially censored data; results below are for uncensored data.}
		\label{tabA1}
	}
\end{table}

\begin{table}[ht]
	{\small
		\centering
		\begin{tabular}{rrrrrrrrrrrrrr}
			\hline
			&& \multicolumn{3}{c}{RX} & \multicolumn{2}{c}{AJCC} & \multicolumn{2}{c}{Gender} & \multicolumn{2}{c}{Race} & \multicolumn{2}{c}{Histology} \\
			&& \multicolumn{1}{c}{1} & 
			\multicolumn{1}{c}{2} &
			\multicolumn{1}{c}{3} &
			\multicolumn{1}{c}{0} &
			\multicolumn{1}{c}{1} &
			\multicolumn{1}{c}{0} &
			\multicolumn{1}{c}{1} &
			\multicolumn{1}{c}{0} &
			\multicolumn{1}{c}{1} &
			\multicolumn{1}{c}{0} &
			\multicolumn{1}{c}{1} \\ \hline
			%
			& $t_1$ & 0.152 & 0.161 & 0.195 & 0.167 & 0.172 & 0.122 & 0.194 & 0.160 & 0.171 & 0.135 & 0.224 \\ 
			BF-RF & $t_2$ & 0.200 & 0.216 & 0.247 & 0.215 & 0.227 & 0.165 & 0.250 & 0.211 & 0.223 & 0.175 & 0.295 \\ 
			& $t_3$ & 0.241 & 0.263 & 0.306 & 0.263 & 0.275 & 0.223 & 0.294 & 0.257 & 0.272 & 0.219 & 0.353 \\  \hline
			& $t_1$ & 0.147 & 0.182 & 0.221 & 0.180 & 0.186 & 0.128 & 0.213 & 0.172 & 0.186 & 0.148 & 0.239 \\ 
			DR-RF & $t_2$ & 0.196 & 0.242 & 0.272 & 0.227 & 0.245 & 0.168 & 0.273 & 0.225 & 0.240 & 0.181 & 0.326 \\ 
			& $t_3$ & 0.240 & 0.300 & 0.336 & 0.280 & 0.299 & 0.234 & 0.320 & 0.279 & 0.294 & 0.226 & 0.396 \\  \hline
			& $t_1$ & 0.092 & 0.071 & 0.120 & 0.106 & 0.084 & 0.067 & 0.110 & 0.085 & 0.101 & 0.075 & 0.126 \\ 
			{\tt rfsrc}  & $t_2$ & 0.121 & 0.099 & 0.147 & 0.126 & 0.115 & 0.089 & 0.140 & 0.123 & 0.127 & 0.099 & 0.159 \\ 
			& $t_3$ & 0.146 & 0.128 & 0.185 & 0.165 & 0.141 & 0.130 & 0.166 & 0.153 & 0.159 & 0.127 & 0.204 \\ \hline \hline
			& $t_1$ & 0.138 & 0.150 & 0.204 & 0.155 & 0.171 & 0.121 & 0.186 & 0.156 & 0.166 & 0.138 & 0.204 \\ 
			RF & $t_2$ & 0.187 & 0.204 & 0.261 & 0.200 & 0.231 & 0.163 & 0.244 & 0.207 & 0.219 & 0.182 & 0.271 \\ 
			& $t_3$ & 0.234 & 0.266 & 0.320 & 0.254 & 0.289 & 0.221 & 0.300 & 0.261 & 0.276 & 0.235 & 0.334 \\ \hline
			& $t_1$ & 0.095 & 0.079 & 0.113 & 0.101 & 0.092 & 0.078 & 0.104 & 0.090 & 0.097 & 0.079 & 0.123 \\ 
			{\tt rfsrc}  & $t_2$ & 0.127 & 0.109 & 0.146 & 0.124 & 0.129 & 0.106 & 0.137 & 0.123 & 0.128 & 0.107 & 0.159 \\ 
			& $t_3$ & 0.165 & 0.149 & 0.189 & 0.168 & 0.168 & 0.152 & 0.175 & 0.168 & 0.168 & 0.144 & 0.208 \\ 
			\hline
		\end{tabular}
		\caption{Summary of PDPs for categorical variables for death. For AJCC, 1 = Stage IIB; Gender, 1 = Male; 
			Race, 1 = White; and Histology, 1 = Squamous. Results above the double horizontal line correspond to 
			artificially censored data; results below are for uncensored data.}
		\label{tabA2}
	}
\end{table}

\end{document}